\documentclass[12pt,showpacs,twocolumn, preprintnumbers,aps,pre, reprint,superscriptaddress]{revtex4-1}

\usepackage{graphicx}
\usepackage{subfigure}
\usepackage{color}
\usepackage{float}
\usepackage{amsmath}
\usepackage[linktocpage,colorlinks=true,linkcolor=blue,citecolor=blue,breaklinks=true]{hyperref}
\usepackage{verbatim}
\usepackage{breakcites}

\begin{document}

\preprint{}

\title{EXOPLANETARY DETECTION BY MULTIFRACTAL SPECTRAL ANALYSIS}

\author{Sahil Agarwal}
\email[]{sahil.agarwal@yale.edu}
\affiliation{Program in Applied Mathematics, Yale University, New Haven, USA}

\author{Fabio Del Sordo}
\affiliation{Department of Astronomy, Yale University, New Haven, USA}
\affiliation{Departments of Geology \& Geophysics,  Yale University, New Haven, USA}
\affiliation{Nordita, Royal Institute of Technology and Stockholm University, SE-10691 Stockholm, Sweden}

\author{John S. Wettlaufer}

\affiliation{Program in Applied Mathematics, Yale University, New Haven, USA}
\affiliation{Departments of Geology \& Geophysics,  Yale University, New Haven, USA}
\affiliation{Nordita, Royal Institute of Technology and Stockholm University, SE-10691 Stockholm, Sweden}
\affiliation{Departments of Mathematics, Yale University, New Haven, USA}
\affiliation{Departments of Physics, Yale University, New Haven, USA}
\affiliation{Mathematical Institute, University of Oxford, Oxford, UK}

\date{\today}

\begin{abstract}

Owing to technological advances, the number of exoplanets discovered has risen dramatically in the last few years.   
However, when trying to observe Earth analogs, it is often difficult to test the veracity of detection.
We have developed a new approach to the analysis of exoplanetary spectral observations based on temporal multifractality, which identifies time scales that characterize planetary orbital motion around the host star, and those that arise from stellar features such as spots. Without fitting stellar models to spectral data, we show how the planetary signal can be robustly detected from noisy data using noise amplitude as a source of information. 
For observation of transiting planets, combining this method with simple geometry allows us to relate the time scales obtained to primary and secondary eclipse of the exoplanets. Making use of data obtained with ground-based and space-based observations we have tested our approach on HD 189733b. 
Moreover, we have investigated the use of this technique in measuring planetary orbital motion via Doppler shift detection. Finally, we have analyzed  synthetic spectra obtained using the SOAP 2.0 tool, which simulates a stellar spectrum and the influence of the presence of a planet or a spot on that spectrum over one orbital period. We have demonstrated that,  so long as the signal-to-noise-ratio $\ge$ 75, our approach reconstructs the planetary orbital period, as well as the rotation period of a spot on the stellar surface.

\end{abstract}

\pacs{}

\maketitle



\section{Introduction} \label{sec:intro}

The last three decades have seen the birth of exo-planetary science. With the advent of various techniques, which include, but are not limited to, pulsar timing \citep{Wolszczan:1992}, Doppler measurements \citep{Mayor:1995}, transit photometry \citep{Charbonneau:2000}, micro-lensing \citep{Beaulieu:2006} and direct imaging \citep{Chauvin:2004}, thousands of planets have been detected orbiting distant stars.  A central focus is the detection of so-called \emph{Exo-Earths}, 
 Earth-like planets in terms of mass and radius, 
orbiting around a star at a distance that, given sufficient atmospheric pressure, would allow for the existence of liquid water on its surface \citep[e.g][and references therein]{Kopparapu:2013}.   The techniques that are most commonly used in discovering other planets are transit photometry \citep[e.g.][and references therein]{Lissauer:2014aa} or Doppler measurements \citep[e.g.][and references therein]{Mayor:2014aa}. Recently, using the transit method, detection of nine candidates for habitable planets was announced that may fall within the habitable zones of their host stars \citep{Anglada:2016,Morton:2016aa}.
 
Whilst the combination of these approaches have provided an impressive range of observations, 
the detection of Earth analogs is a challenging problem. Indeed, the presence of instrumental and astrophysical noise are sources of uncertainty for such discoveries \citep[e.g.][]{Fischer:2016aa}. The fingerprints of an exo-Earth could easily be hidden in stellar noise, or stellar signals might mimic the presence of an exoplanet.  Moreover, when such noise is modeled, there is a risk of introducing spurious signals in the analysis of data \citep[e.g.][]{Dumusque:2015aa, Rajpaul:2016aa}.

Contemporary studies aim to understand and correctly evaluate the imprint of stellar activity on exoplanet detections \citep[e.g.][]{Lanza:2007,Lanza:2011,Aigrain:2012,Korhonen:2015}. Nonetheless, radial velocity and transit photometry methods can still produce false detections, especially when dealing with Earth-like planets, or planets characterized by a signal $\leq$1m s$^{-1}$ \citep{Dumusque:2016,Morton:2016aa}, 
which is, in part, due to the models upon which these methods are based.  Such exoplanet evolution and stellar models over-fit parameters to the data, which turns out to be crucial when one aims to detect terrestrial-like planets \citep{Dumusque:2012aa,Rajpaul:2016aa}. These parameters include, but are not limited to, planet and stellar radius, their masses, eccentricity, impact factor, transit duration, transit ingress/egress duration, transit depth, orbital inclination, distance of the planet from the star, limb-darkening parameters (which themselves may vary with the law used), and shape of the transit curve. In some cases, instrumental signals such as the wobble of the instrument cluster aboard the \emph{Spitzer} satellite or the latent charge build-up in the pixels (ramp) are also modeled \citep{Grillmair:2007aa,Todorov:2014aa}. Additionally, noise sources such as granulation over the stellar surface employ model fitting to estimate the effect of that noise on the data \citep{Dumusque:2012aa}, stellar activity concurrent with the stellar rotation are modeled by fitting sine waves through the radial velocity data, and long term stellar activity, light contamination from near by stars, among others are parametrically modeled. This fitting of models has led to the introduction of spurious signals in the observed data and hence to false detections.

In order to fit these stellar models to data, one must begin by considering a particular system, for example an unblended eclipsing binary or a transiting planet. Hence, the list of models that the data can represent needs to be complete, which constitutes a weakness for these types of fitting schemes \citep{Morton:2016aa}. Moreover, these methods cannot always distinguish whether the signal is instrumental or from stellar activity \citep{Morton:2016aa}. Finally, because it affects the details of the fitting and hence the results, the observations must have a high signal-to-noise ratio (S/N).

A key aspect of the analysis of exo-planetary systems is the identification of periodic timescales. This is also central to the study of multi-planetary systems \citep[e.g.][]{Millholland:2016aa}, and often requires the combined use of the two most successful approaches in exoplanet searches; the transit and radial velocity methods. The most widely used method  for identifying periodic timescales is the Lomb--Scarge periodogram, which is based on the assumption that the data can be interpreted as a sum of periodic signals. However, this technique is used on data that have been ``filtered'' using models for stellar noise and terrestrial atmospheric contamination. Hence, it is again possible to accumulate artifacts in the data through the use of such models. Many approaches have been developed by the exoplanetary science community to address these matters. For example, \cite{Dumusque:2016} have compared state-of-the-art methods on simulated radial velocity data and concluded that the detection of planets below the 1 m s$^{-1}$ threshold is still controversial. 
Moreover, the extraction of radial velocities from spectra relies on cross-correlation techniques that are based on the use of spectral templates based on stellar models. 
Therefore, the determination of the Doppler shift remains a method-dependent challenge.
For example, in the recent radial velocity study of Proxima Centauri \citep{Anglada:2016}, the use of the TERRA algorithm  \citep{Anglada:2012} rather than the HARPS pipeline, has been crucial for the quantification of the signal.

Here, to extract the timescales that characterize a planetary system and stellar features without {\em a priori} assumptions about the data itself,
we introduce a new approach to spectral analysis.  Namely, in the spirit of the Langevin theory of Brownian motion, we quantify a signal coming from a star as the combination of a deterministic dynamics and stochastic noise, but make no {\em a priori} assumptions about the nature of these processes, and thereby examine an unfiltered time series $X_i$ of a stellar spectral signal.  If the star hosts a planetary system, the time scales associated with stellar rotation and activity, as well as with planetary motions must be present. We need not (a) make assumptions regarding the combination of periodic signals, or (b) use stellar models. The goal is to identify the dominant timescales of the observed system as agnostically as possible, and then use elementary geometry to reveal the underlying dynamics of the system.
The flexibility of this approach, based on temporal multifractality, allows one to identify stellar signals that would otherwise be missed by fitting sine waves to the data.
 
We begin with a description of the method in \S \ref{sec:method}. The proposed approach can be used both for transiting planets and for radial velocity measurements. We show two examples  in \S \ref{sec:data}; one for a transiting planet, and the other for a simulated observation of a planet detectable only via radial velocity measurements. Finally, we discuss the results and their robustness in \S \ref{sec:discuss}, and conclude in \S \ref{sec:conc}.

\section{Method} \label{sec:method}

We analyze a series of spectra taken at an approximately constant time intervals. If each of these spectra spans a wavelength range of $L$ wavelengths, we construct $L$ time series, each of which consists of the flux measured at a given wavelength. Hence, if we have $m$ equally spaced-in-time spectra, we have $L$ time series of length $m$. The time delay between spectral observations corresponds to the best resolution obtainable. 

At each wavelength the variability of flux in time can arise from, among other things, the Doppler shift, photometric effects, atmospheric/telluric effects, and instrumental noise, each with characteristic time scales we aim to extract with our method. Importantly, with no \textit{a priori} knowledge of the dynamics of the system, and without the use of model-fitting, we can extract the time scales associated with either a transiting planet or the Doppler shift underlying planetary motion.  In the former case we use elementary geometry  to reconstruct time scales connected with the transit.  In the latter case, it is known that a Doppler shift for stellar spectra can be caused both by an orbiting planet as well as by intrinsic stellar features, such as spots. We shall show that we can identify both of them, but to distinguish between them is the subject of future work.

\subsection{Multi-fractal Temporally Weighted Detrended Fluctuation Analysis}
\label{Sec:method}
We analyze spectral time series using Multi-fractal Temporally Weighted Detrended Fluctuation Analysis (MF-TWDFA) \cite[e.g.,][and references herein]{Sahil:MF}, which does not \textit{a priori} assume anything about the temporal structure of the data.
The approach has four stages, which we describe briefly below.

\begin{enumerate}
\item We construct a non-stationary {\em profile} $Y(i)$ of the original time series $X_i$ as, 
\begin{equation}
Y(i )\equiv \sum_{k=1}^{i} \left(X_k - \overline{X_k}~ \right), \qquad \text{where}\qquad  i = 1, ... , N.  
\label{eq:profile}
\end{equation}
The profile is the cumulative sum of the time series and $\overline{X_k}$ is the average of the time series $X_1 ... X_k$.

\item 
This non-stationary profile is divided into $N_s = \text{int}(N/s)$ non-overlapping segments of equal length $s$, where $s$ is an integer and varies in the interval $1<s\le N/2$.  Each value of $s$ represents a time scale $s \times \Delta t$, where $\Delta t$ is the temporal resolution of the time series.
The time series has a length that is rarely an exact multiple of $s$, which is handled by repeating the procedure from the end of the profile and returning to the beginning, thereby creating $2 N_s$ segments.  

\item A point by point approximation $\hat{y}_{\nu}(i)$ of the profile is made using a moving window, smaller than $s$ and weighted by separation between the points $j$ to the point $i$ in the time series such that $\vert i - j \vert \le s$.\footnote{In regular MF-DFA, rather than using temporal-weighting,  $n$th order polynomials $y_{\nu}(i)$ are used to approximate $Y(i)$ {\em within} a fixed window, without reference to points in the profile outside that window.} A larger (or smaller) weight $w_{ij}$ is given to $\hat{y}_{\nu}(i)$ according to whether $\vert i - j \vert $ is small (large) \citep[][]{Sahil:MF}. This approximated profile is then used to compute the variance spanning up ($\nu = 1,...,N_s$) and down ($\nu = N_s + 1,...,2 N_s$) the profile as
\begin{eqnarray}
\hspace{-12mm}\text{Var}(\nu, s) \equiv & \hspace{-20mm}\frac{1}{s}  \sum_{i=1}^{s} \{ Y([\nu - 1]s + i) - {\hat{y}}([\nu-1]s +i) \}^2 \nonumber \\
&\hspace{-15mm} \text{for $\nu = 1,...,N_s$ ~~~~and} \nonumber \\
\hspace{-12mm}\text{Var}(\nu, s)  \equiv & \frac{1}{s}  \sum_{i=1}^{s} \{ Y(N-[\nu - N_s]s + i) - {\hat{y}}(N-[\nu-N_s]s +i)\}^2\nonumber \\
&\hspace{-15mm} \text{for $\nu =  N_s + 1,...,2 N_s$.}
\label{eq:varTW}
\end{eqnarray}

\item Finally, a {\em generalized fluctuation function} is obtained and written as
\begin{equation}
F_q (s) \equiv \left[ \frac{1}{2 N_s} \sum_{\nu=1}^{2 N_s} \{ \text{Var}(\nu, s)\}^{q/2} \right]^{1/q}.
\label{eq:fluct}
\end{equation}
\end{enumerate}

The behavior of $F_q (s)$ depends on the choice of time segment $s$ for a given order $q$ of the moment taken. The principal focus is to study the scaling of  $F_q (s)$ as characterized by a generalized Hurst exponent $h(q)$ viz., 
\begin{equation}
F_q (s) \propto s^{h(q)} .  
\label{eq:power}
\end{equation}
When $h(q)$ is independent of $q$ the time series is said to be monofractal, in which case $h(q)$ is  equivalent to the classical Hurst exponent $H$.  
For $q$ = 2, regular MF-DFA and DFA are equivalent \citep[][]{Kantelhardt:2002}, and  $h(2)$ can also be related to the decay of the power spectrum $S$. If $S(f) \propto f^{- \beta}$, with frequency $f$ then $h(2) = (1 + \beta)/2$ \citep[e.g.,][]{Ding}.  For white noise $\beta$ = 0 and hence $h(2) = 1/2$, whereas for Brownian or red noise $\beta = 2$ and hence $h(2) = 3/2$. The dominant timescales in the data set are the points where the fluctuation function $\ell{\text{og}}_{10}F_2(s)$ changes slope with respect to $\ell{\text{og}}_{10}s$.
At each wavelength a crossover in the slope of a fluctuation function is calculated if the change in slope of the curve exceeds a set threshold, $C_{th}$.
Because the window length is constrained as $1 < s \le N/2$ \citep{Zhou:2010}, this analysis is limited to time scales of $t \leq t_\textrm{up} =  N \Delta t/2$.
\cite{Dobson:1990} studied stellar features by focusing on time-series observations of H and K lines of Ca II.  Their method constructs diagrams based on the concept of pooled variance, and has been subsequently used in other analyses of stellar activity \citep{Donahue:1997,Scholz:2004,Lanza:2006,Messina:2016}. Pooled variance is 
defined as the mean variance at a particular time scale, $\tau_p$, by stepping through a time series in consecutive bins of size $\tau_p$ and calculating the variance within each bin.  A pooled variance diagram (PVD) plots this mean variance versus $\tau_p$, thereby examining the time-scale dependence of the mean variance.

The key differences between a PVD and our method are as follows.  First, while a PVD computes a mean variance of the data itself, MF-TWDFA examines the variability relative to the temporally weighted fit to the profile of the data, thereby exploiting the intuition that points closer in time are more likely to be related than more distant points.  Second, because  the procedure of MF-TWDFA produces a smooth profile of the data, which is the core time series analyzed, it does not suffer from the intrinsic noise present in the data \cite[see Fig. 4 of][]{Sahil:MF}.  Third, in MF-TWDFA  the fluctuation functions are calculated for an arbitrary range of moments \cite[up to ten in geophysical data;][]{Sahil:MF}, both positive and negative, thereby providing a rich tapestry of the temporal dynamics underlying variability, as well as explicit information about the processes producing them (e.g, pink or Brown noise processes).  Finally, to the best of our knowledge PVD's have not been used to quantify exoplanetary timescales.

\section{Demonstrating the Method with three types of Data}\label{sec:data}

To test the method described in \S \ref{sec:method} we apply it to three different kinds of data.
The first two are spectral observations of HD 189733b, a well known transiting exoplanet discovered in 2005 \citep{Bouchy:2005aa}.
The third is a set of simulated stellar spectra affected by the Doppler shift induced by an orbiting planet. These simulated data are obtained with the Spot Oscillation And Planet (SOAP) 2.0 tool \citep[][]{2014ApJ...796..132D}.

\subsection{HD 189733b}

First detected in 2005 \citep{Bouchy:2005aa}, HD 189733b is a hot Jupiter orbiting the star HD 189733A in the constellation Vulpecula, approximately 63 light years away from Earth. We use spectral data for HD 189733b to test our method of extracting time scales related to the effects of a transiting planet. We employ both ground-based observations (high-resolution in wavelength) and space-based observations (low-resolution in wavelength) to test if and how the analysis is affected by resolution, instrumental noise, and the terrestrial atmosphere. The ground-based observations are obtained from the High Accuracy Radial velocity Planet Searcher \emph{HARPS} spectrograph  \citep{2003Msngr.114...20M}, and the space-based observations are from the NASA \emph{Spitzer} space mission.

\subsubsection{HARPS Data}

These are reduced 1D spectral data from the High Accuracy Radial velocity Planet Searcher at the European Southern Observatory (ESO) La Silla 3.6m telescope for planet HD 189733b. Programs 072.C-0488(E), 079.C-0127(A), and 079.C-0828(A) are used from the ESO archive (see Table \ref{T:harps}). These data cover four nights, but that from the fourth night has been removed from this analysis as it is known to have been affected by severe weather \citep{Triaud:09, Wyttenbach:15}. Each night is treated as an independent data set. On night 1, these spectra are taken approximately every 10.5 minutes, and on nights 2 and 3 every 5.5 minutes. 

\begin{table}[]
\centering
\caption{\bf HARPS Data Analyzed for HD 189733b}
\label{T:harps}
\begin{tabular}{@{}llllll@{}}
\toprule
        && \begin{tabular}[c]{@{}l@{}}Observation\\ Date\end{tabular} && Program ID    \\ \hline
Night 1 &\hspace{10mm}& 2006-09-07                                                 &\hspace{10 mm}& 072.C-0488(E) \\
Night 2 & & 2007-07-19                                                 & & 079.C-0828(A) \\
Night 3 & & 2007-08-28                                                 & & 079.C-0127(A) \\ \hline
\end{tabular}
\end{table}

\subsubsection{\emph{Spitzer} Data}

The \emph{Spitzer} space mission observed secondary eclipses of HD 189733b.  We use Basic Calibrated Data (.bcd) files from the \emph{Spitzer} pipeline version 18.18.0, obtained with the optimal extraction tool in the SPICE software that employs the Optimal Extraction Algorithm of \citet{Horne:86} to obtain the reduced 1D spectra from the observed images (see Table \ref{T:Spitzer}). 

\begin{table}[t]
\centering
\caption{\bf \emph{Spitzer} Data Analyzed for HD 189733b}
\label{T:Spitzer}
\begin{tabular}{@{}llllll@{}}
\toprule
\begin{tabular}[c]{@{}l@{}}AOR\\ Key\end{tabular} && \begin{tabular}[c]{@{}l@{}}Observation\\ Date\end{tabular} && \begin{tabular}[c]{@{}l@{}}Wavelength\\ Range ($\mu m$)\end{tabular} \\ \hline
18245632                                          &\hspace{10mm} & 2006-10-21                                                 &\hspace{10mm}& 7.4-14.0                                                           \\
20645376                                          & & 2006-11-21                                                 & & 7.4-14.0                                                           \\
23437824                                          & & 2008-05-24                                                 & & 7.4-14.0                                                           \\
23438080                                          & & 2008-05-26                                                 & & 7.4-14.0                                                           \\
23438336                                          & & 2008-06-02                                                 & & 7.4-14.0                                                           \\
23438592                                          & & 2008-05-31                                                 & & 7.4-14.0                                                           \\
23438848                                          & & 2007-10-31                                                 & & 7.4-14.0                                                           \\
23439104                                          & & 2007-11-02                                                 & & 7.4-14.0                                                           \\
23439360                                          & & 2007-06-26                                                 & & 7.4-14.0                                                           \\
23439616                                          & & 2007-06-22                                                 & & 7.4-14.0                                                           \\
23440384                                          & & 2008-06-09                                                 & & 5.0-7.5                                                            \\
23440640                                          & & 2008-06-04                                                 & & 5.0-7.5                                                            \\
23440896                                          & & 2007-12-07                                                 & & 5.0-7.5                                                            \\
23441152                                          & & 2007-11-06                                                 & & 5.0-7.5                                                            \\
23441408                                          & & 2007-11-11                                                 & & 5.0-7.5                                                            \\
23441664                                          & & 2007-11-09                                                 & & 5.0-7.5                                                            \\
23441920                                          & & 2007-11-24                                                 & & 5.0-7.5                                                            \\
23442176                                          & & 2007-11-15                                                 & & 5.0-7.5                                                            \\
23439872                                          & & 2007-11-04                                                 & & 13.9-21.3                                                          \\
23440128                                          & & 2007-06-17                                                 & & 13.9-21.3                                                          \\
23442432                                          & & 2007-12-10                                                 & & 19.9-39.9                                                          \\
23442688                                          & & 2007-06-20                                                 & & 19.9-39.9                                                          \\ \hline
\end{tabular}
\end{table}

\begin{figure*}[htbp]
    \centering
    \begin{subfigure}[]
    \centering
    \includegraphics[trim = 0 0 20 0, clip, width = 0.31\textwidth]{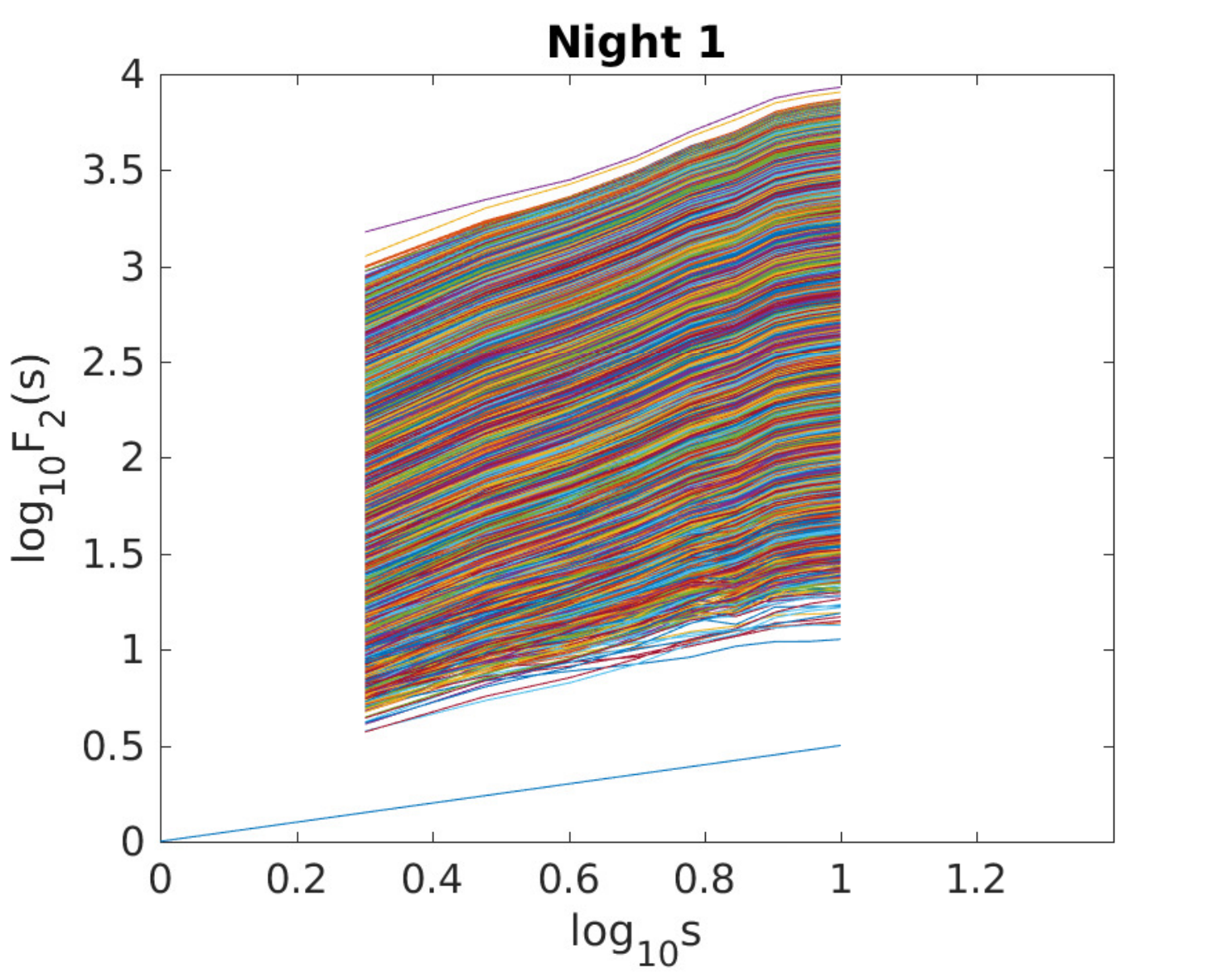}
    \end{subfigure}
    ~
    \begin{subfigure}[]
    \centering
    \includegraphics[trim = 0 0 20 0, clip, width = 0.31\textwidth]{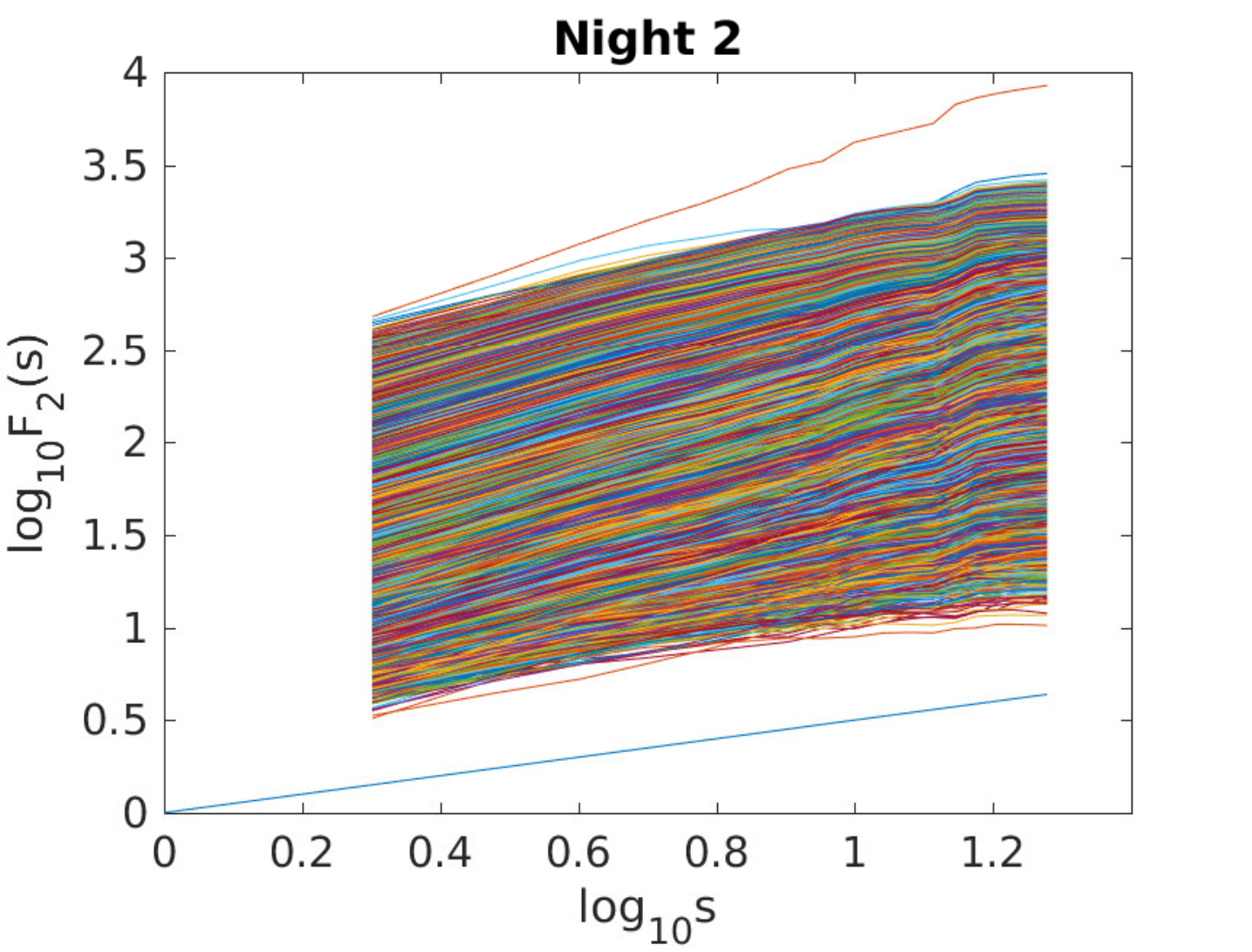}	  	
    \end{subfigure}
    ~
    \begin{subfigure}[]
    \centering
    \includegraphics[trim = 0 0 20 0, clip, width = 0.31\textwidth]{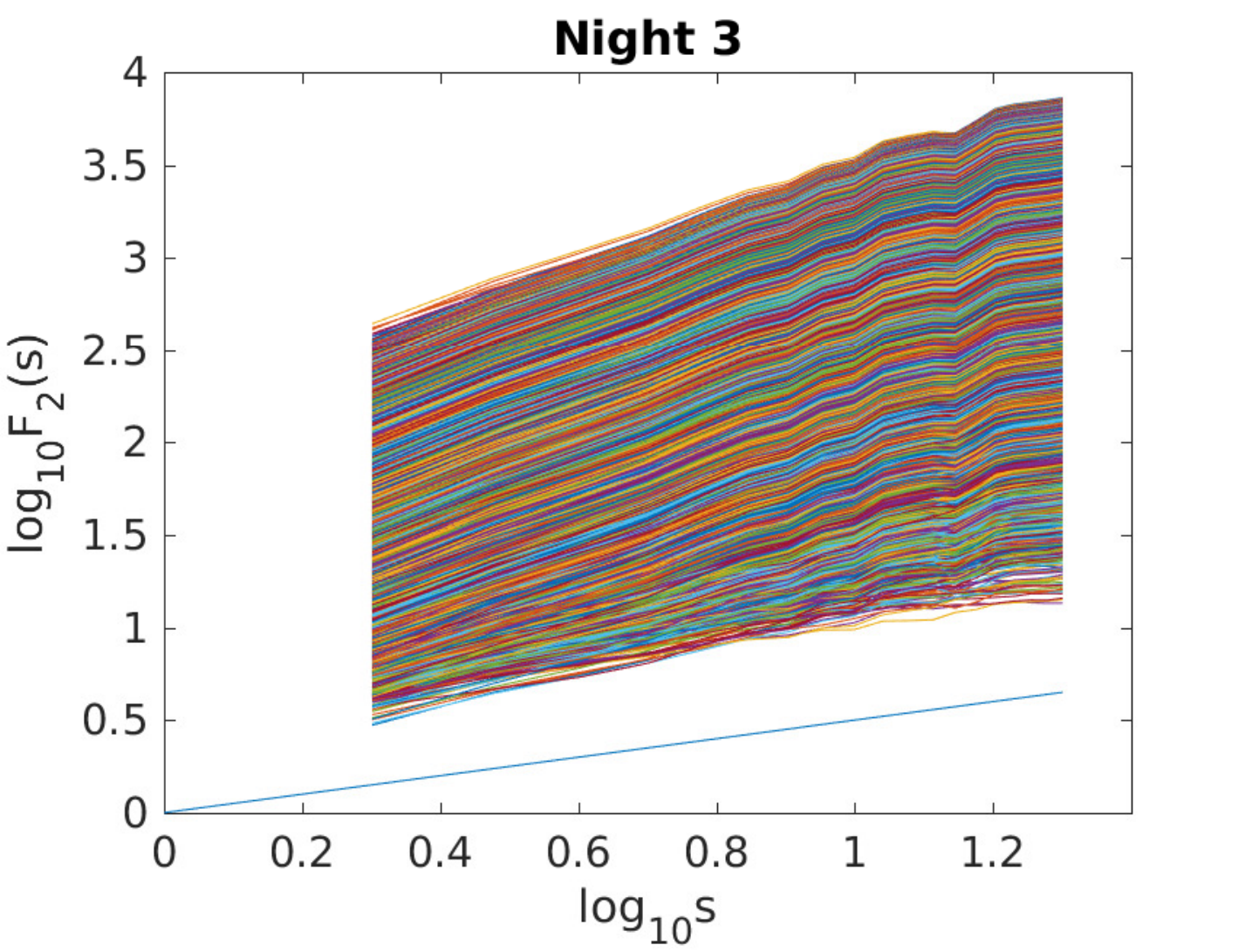}	  	
    \end{subfigure}	
\caption{The second moment of the fluctuation functions for (a) Night 1 (time resolution $10.602988 \pm 0.072236$ minutes), (b) Night 2 (time resolution $5.517202 \pm 0.005683$ minutes), and (c) Night 3 (time resolution $5.523740 \pm 0.004415$ minutes), with original wavelength resolution (plotted at alternate wavelengths). The abscissa $(\ell{\text{og}}_{10}\text{s})$ is measured using the number of data points in the time series, $1<s\le N/2$, where $s$ represents a timescale $s \times \Delta t$, with $\Delta t$ being the temporal resolution of the time series. Different colors represent different wavelengths. In all of the fluctuation functions, wavelength generally increases in the positive y--direction.
The straight blue line has a slope of 0.5, the slope of fluctuation function denoting white noise. Shorter wavelengths show higher amounts of noise, which may be associated with atmospheric turbulence and/or telluric effects on Earth or the exoplanet. \label{fig:N1_complete}}
\end{figure*}

\subsection{Simulations}

The use of simulated data provides a good test and application of our method to radial velocity measurements, because we can control the input and thereby facilitate a clear interpretation of the results. We use data produced with the SOAP 2.0 tool \citep[][]{2014ApJ...796..132D}.
These spectra are generated by filling a grid of cells with a telluric cleaned solar spectrum from the National Solar Observatory, simulating a rotating star. Depending on the grid cell position, the spectrum is Doppler shifted to account for stellar rotation, and the intensity of each cell is weighted by a limb-darkening law.  To simulate the presence of a spot, a typical spectrum of a solar spot is enclosed in a cell.  Additionally, the intensity is reduced in the presence of a dark spot or increased in the presence of a bright plage. Such features follow the rotation of the star. The final integrated spectrum is the sum of the spectra from all of the cells.  

First, we use stellar spectra in the absence of stellar spots and thus influenced solely by an orbiting planet.  This allows us to focus on, for example, the ``bare'' effect of the Doppler shift on the spectra.  Because we would like to examine orbital periods, we stack the 25 spectra corresponding to the orbital period of the planet to provide  8 orbital periods worth of data. The time units are arbitrary, so the analysis can be compared to a wide range of observations, provided they are evenly spaced in time.
Moreover, since we are interested in the effect of noise on our data analysis, Gaussian white noise with a specific S/N per resolution element is then added at each time to obtain data sets with different S/Ns. This is important for the comparison of results obtained with this simulated data set to real observations from spectrographs characterized by different S/Ns.

Second, we use stellar spectra in the absence of a planet and thus influenced solely by a stellar spot.  This allows us to focus on the ``bare'' effect of a spot, which can in principle mimic the Doppler shift in radial velocity measurements induced by a spot.  Here again, as we have 25 spectra corresponding to one full rotation period of the star, we stack these to give 8 rotation periods worth of data, and then add Gaussian white noise as in the first (planet only) case discussed above. The spot covers 5\% of the star, has a contrast of 663K and it is set to rotate on the stellar equator (i.e., a latitude of $0^{\circ}$). The star rotates "equator-on", meaning the rotational axis is at $90^{\circ}$ relative to the line of sight. We have selected such values in order to deal with the simplest possible realistic observational case.  For each of these spectra the continuum has been subtracted.

\section{Discussion} \label{sec:discuss}


\subsection{HARPS analysis of HD 189733b}\label{sec:HARPS}

The 1-D spectra from the HARPS instrument provide the time series for flux at each of the wavelengths for each night separately. These are analyzed using MF-TWDFA as described in \S\ref{sec:method} above.  The second moment of the associated  fluctuation functions are shown in Figures \ref{fig:N1_complete}(a)--(c) for Nights 1, 2, and 3 respectively. In all of the fluctuation functions, wavelength generally increases in the positive y--direction.

\begin{figure}[htbp]
    \centering
    \includegraphics[trim = 20 0 20 10, clip, width = 0.5\textwidth]{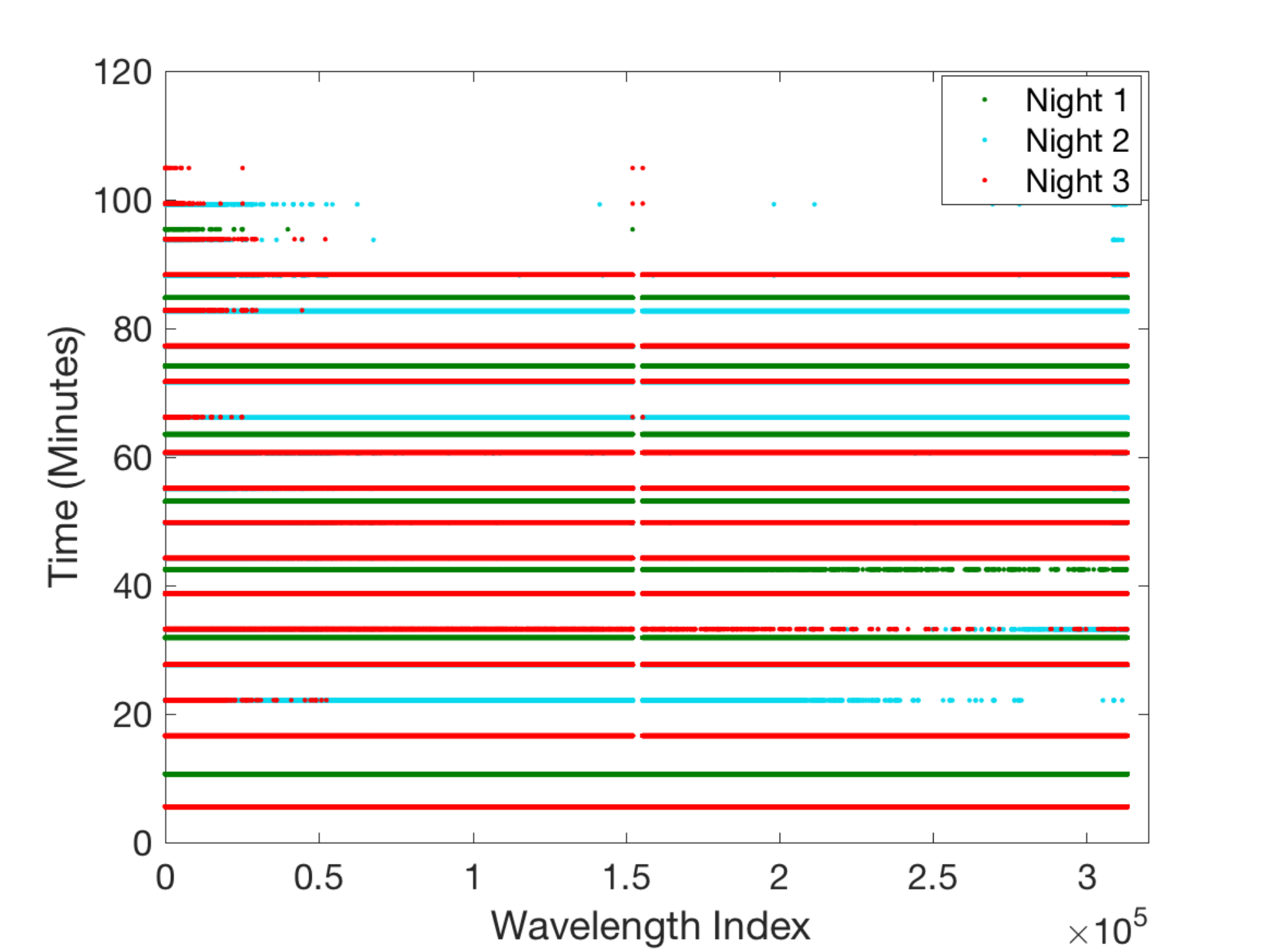}	  	
    \caption{The crossover times plotted for all wavelengths, for all three nights for HD 189733b. Only the $85$ minute timescale is present for all three nights.}
    \label{fig:CP_Harps}
\end{figure}

The key aspects of the fluctuation functions are as follows.  
First, nearly all of these curves are parallel to each other, demonstrating that the spectra evolve in time with a similar noisy behavior at all wavelengths.
Second, those fluctuation functions that show deviations, do so principally at smaller wavelengths and can thus be ascribed to atmospheric interference from the Earth or the exoplanet, such as variability of air masses, telluric contamination and/or turbulent effects.  
Third, for all nights and all the wavelengths we see a timescale of $85$ minutes.   Fourth, for times longer than 85 minutes, the dynamics for all wavelengths exhibit a white noise structure.

In Figure \ref{fig:CP_Harps} we plot all of the times at which the slope of the fluctuation functions changes----the crossover times----for the three nights for all of the wavelengths. The robustness of the 85 minute timescale is reflected by its presence on all nights, whereas other timesscales are present for only one or two nights. Lines at shorter times only appear to be continuous due to the large number of points, but are in fact quite noisy, as shown by the fluctuation functions. Possible origins of the other time scales include turbulent and convective processes in Earth's and the exoplanet's atmosphere, stellar activity or instrumental noise. It is important that we have identified these scales here and this provides a foundation for systematic examination of them in future studies as a possible means of systematically filtering them out.  

\begin{figure}[htbp!]
    \centering
    \includegraphics[trim = 0 0 0 0, clip, width = 0.5\textwidth]{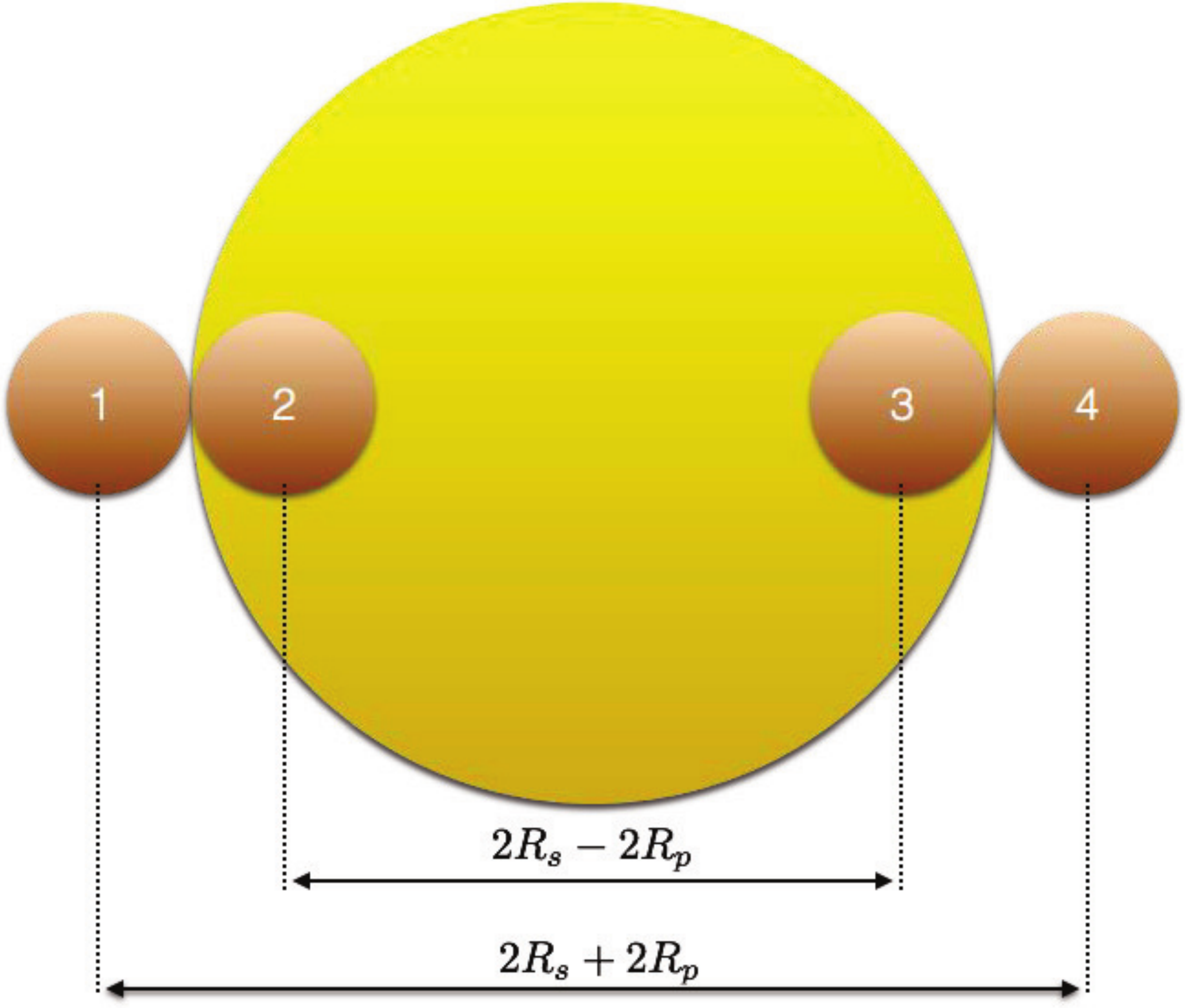}	  	
    \caption{A schematic showing the planet of radius $R_p$ transiting its parent star of radius $R_s$.}
    \label{fig:ExoTransit}
\end{figure}

To understand this 85 minute  time scale, we study the following parameters of the planet HD 189733b, as found from previous studies:  (1) the ratio of the planetary ($R_p$) to the stellar ($R_s$) radii, $\left ({R_p}/{R_s} \right )^2 = 0.02391 \pm 0.00007$ \citep{Torres:2008aa}, and (2) the duration of the transit of the planet in front of the star $\tau_{14} = 0.07527 \pm 0.00037 \textnormal{days} \approx 108.4 \textnormal{minutes}$, or the time between first and last ``contact'' of the planet and the star \citep{Triaud:09} (see Figure \ref{fig:ExoTransit}).

\begin{figure}[htbp]
    \centering
    \includegraphics[trim = 20 0 0 0, clip, width = 0.55\textwidth]{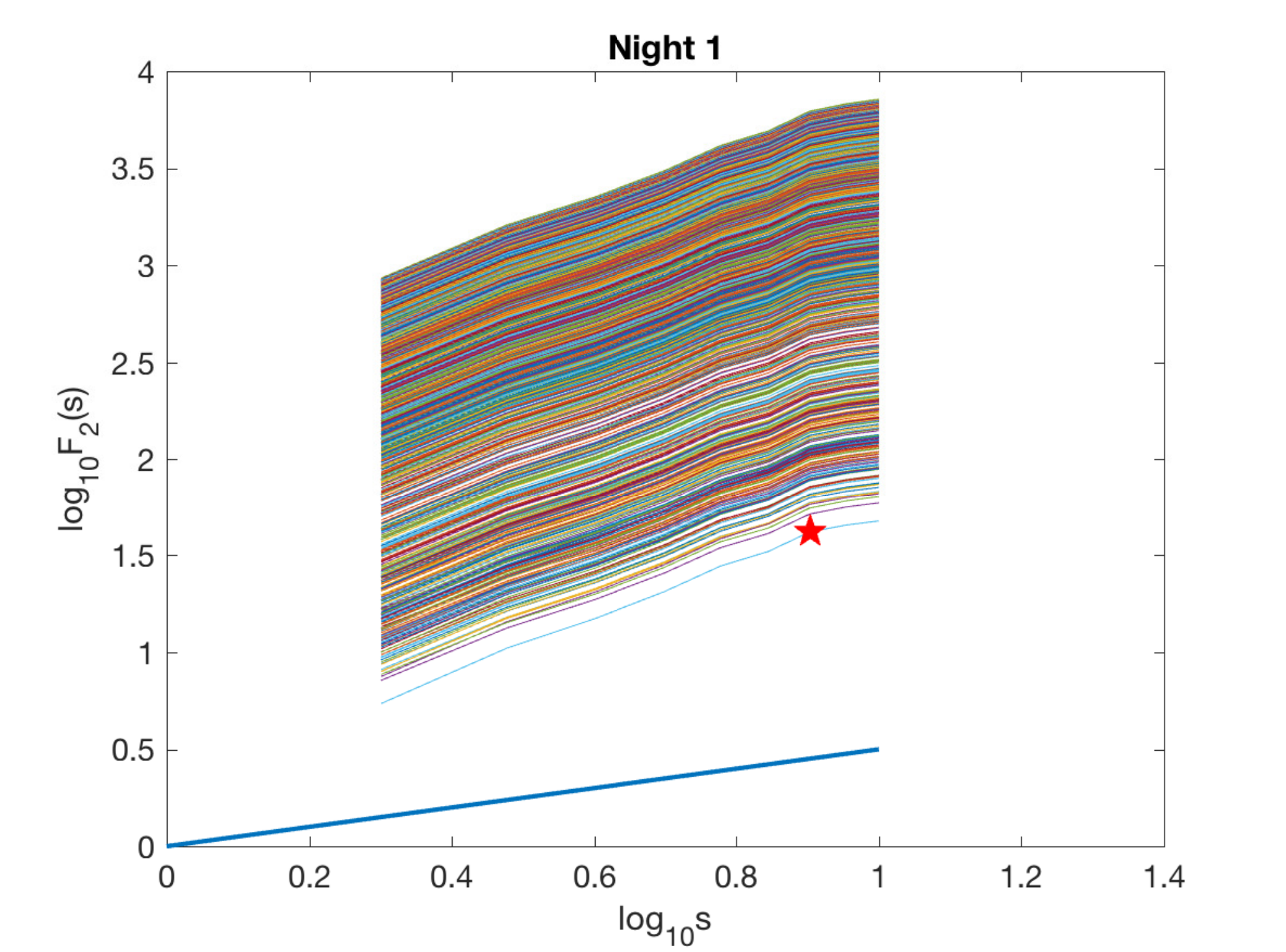}	  	
    \caption{The second moment of the fluctuation functions for Night 1 (approximate time resolution $10.5$ minutes), with a wavelength resolution degraded by a factor 100. As in figure \ref{fig:CP_Harps}, the straight blue line has a slope of 0.5, the slope of the fluctuation function denoting white noise. The red star shows the location of the change of slope of the fluctuation functions that corresponds to the detected timescale of 85 minutes.}
    \label{fig:N1_lowres}
    \end{figure}

Due to the time-resolution of the data for each night, as well as the total length of these time series, we are not able to observe the shortest and the longest time scales in the system. 
The $85$ minute time scale  corresponds to the interval between the second and third contact of the planet, $\tau_{23}$, i.e., the period during which the planet is completely between the observer and the star (Figure \ref{fig:ExoTransit}). If we assume that the planet is transiting in the equatorial plane of the star, and hence the impact parameter is $b=0$, we calculate this time by tracking the center of the planetary disk across the stellar disk as
\begin{equation}
\frac{\tau_{14}}{2(R_s + R_p)} = \frac{\tau_{23}}{2(R_s - R_p)}, 
\label{eq:transit}
\end{equation}
from which we obtain $\tau_{23} \approx 0.0551~\textnormal{days} = 79.659~\textnormal{minutes}$, using $\frac{R_p}{R_s}$ from \citet{Torres:2008aa}.  Within the resolution of the time series, this value of  $\tau_{23}$ is consistent with the 85 minute time scale we have found. 
The timescale of the full transit $\tau_{14}$ would generally be the dominant timescale of the system. Because the Rossiter--McLaughlin effect for this system operates on the same timescale \citep{Winn:2006}, it might underlie the spectral modifications detected here. Our method only works up to $N/2$, namely half the duration of the observational time, $t_\textrm{up} = N \Delta t / 2$, and thus due to the length of the time series from the HARPS data, the method does not allow $\tau_{14}$ to be detected.
To test how wavelength-resolution affects our results, we degrade the resolution of the spectra in wavelength and repeat our analysis.  We decrease the resolution by a factor 100 by using the average of each 100 points, and plot the fluctuation functions for the degraded spectra from Night 1 in Figure \ref{fig:N1_lowres}. Clearly, the structure at longer time scales is unchanged, whereas the noisy behavior at short wavelengths is less evident than in the analysis of high resolution spectra, showing that the averaging acts as a crude high-pass filter.  Importantly,  the $85$ minute scale persists as does the general behavior of the fluctuation functions for this reduction in resolution.
 
\subsection{\emph{Spitzer} analysis of HD 189733b}

Because stars can only be observed at night, all Earth--based instruments provide a strong constraint on the detection of time scales.  Moreover, telluric contamination is a well known problem that is also evident in our method at the shorter timescales in the HARPS data. To bypass these limitations, we examine data from the \emph{Spitzer} mission, which observed HD189733b for 22 nights, looking at the secondary eclipses of the planet, i.e., when the planet is behind the star.  

Although the data from \emph{Spitzer} is of low resolution in wavelength space, this does not affect the identification of robust time scales, which we have demonstrated in the case of the HARPS data by comparing the results using full resolution data with those from degraded data.
Regardless of the wavelength resolution, the time resolution as well as the total duration make the \emph{Spitzer} data sets compelling. 
 
\begin{figure}[h!]
    \centering
    \includegraphics[trim = 0 0 0 0, clip, width = 0.5\textwidth]{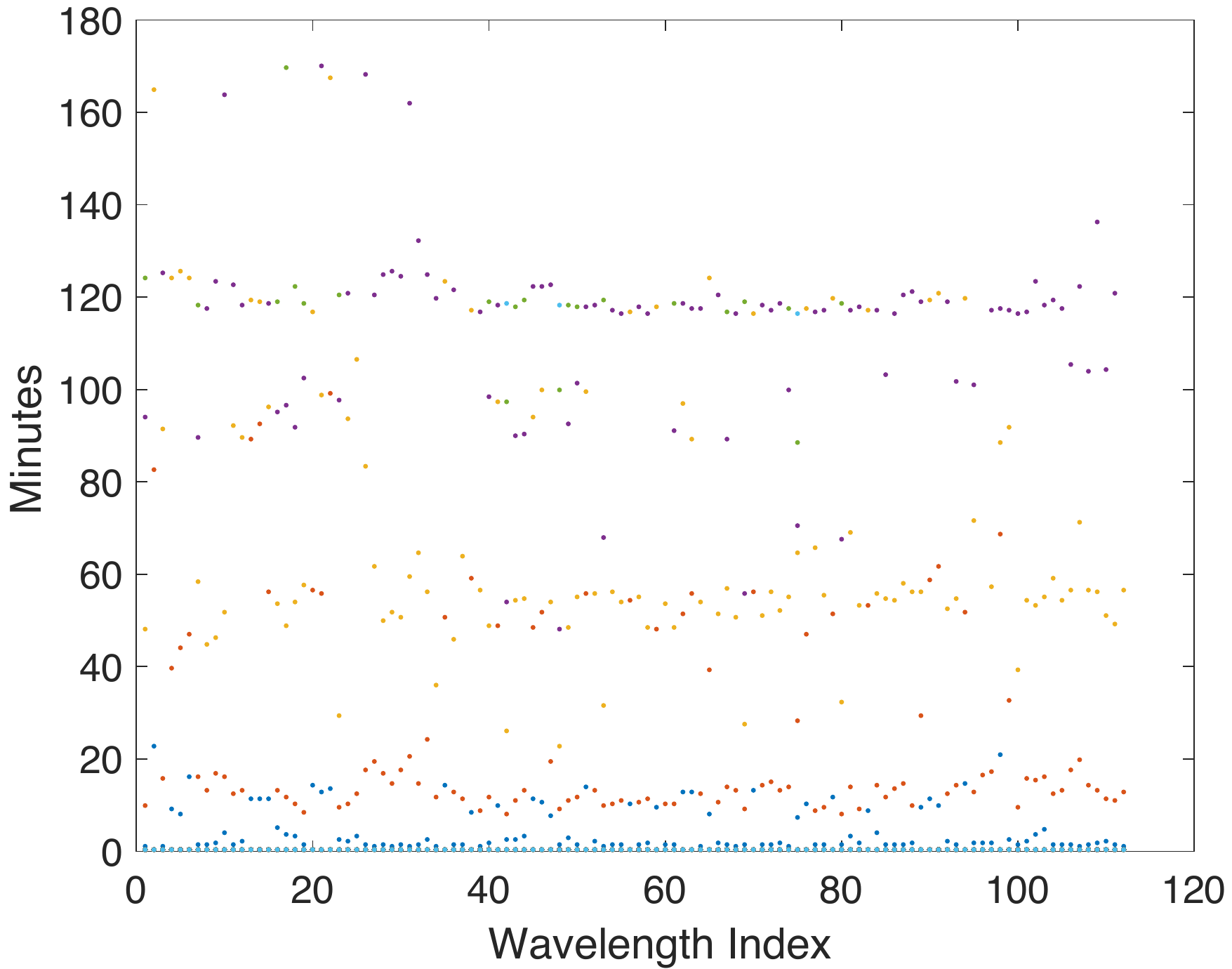}	  	
    \caption{The \emph{Spitzer} based crossover times plotted for all wavelengths, for Night 2 (AOR-20645376, see Table \ref{T:Spitzer}) for HD 189733b. All four significant timescales are robustly extracted using our method. }
    \label{fig:SpitzerD1}
\end{figure}

As was done in Figure \ref{fig:CP_Harps}, we plot these crossovers as a function of wavelength in Figure \ref{fig:SpitzerD1}.  It is clear that relative to the HARPS data the \emph{Spitzer} data is substantially more noisy.    We note that typically the raw data are filtered to remove the noisy characteristics, but  given experience with other systems \citep{Sahil:MF, Sahil:SUB} we take the perspective that noise can be an essential source of information.  

The four prominent timescales in this data are: (1) $\tau_{12} = $ 15.4804 $\pm$ 3.7660 minutes, (2) 55.0966 $\pm$ 5.8851 minutes, (3) $\tau_{23} = $ 87.0947 $\pm$ 2.7888 minutes, and (4) $\tau_{14} = $ 118.9671 $\pm$ 5.5764 minutes, where the uncertainty is one standard deviation about the mean.  First, the 55.0966 $\pm$ 5.8851 minute timescale is the pointing wobble in the Infrared Array Camera of the \emph{Spitzer} telescope, which is due to the battery heater in the telescope  \citep{Grillmair:2007aa, IRAC:10}. Secondly, the $85$ minute timescale obtained from the HARPS data is within one standard deviation of $\tau_{23}$ and hence is robust.  The other two timescales are related to the transit of the planet behind the star, namely the secondary eclipse. 

The necessary and sufficient condition for these timescales to represent the transit (Figure \ref{fig:ExoTransit}) is 
\begin{equation}
\tau_{14} = \tau_{23} + 2\times\tau_{12}
\label{eq:duration}
\end{equation}
It is evident that, within one standard deviation, equation \ref{eq:duration} is satisfied for the above timescales. 

\begin{figure}[htbp!]
    \centering
    \includegraphics[trim = 0 0 0 0, clip, width = 0.5\textwidth]{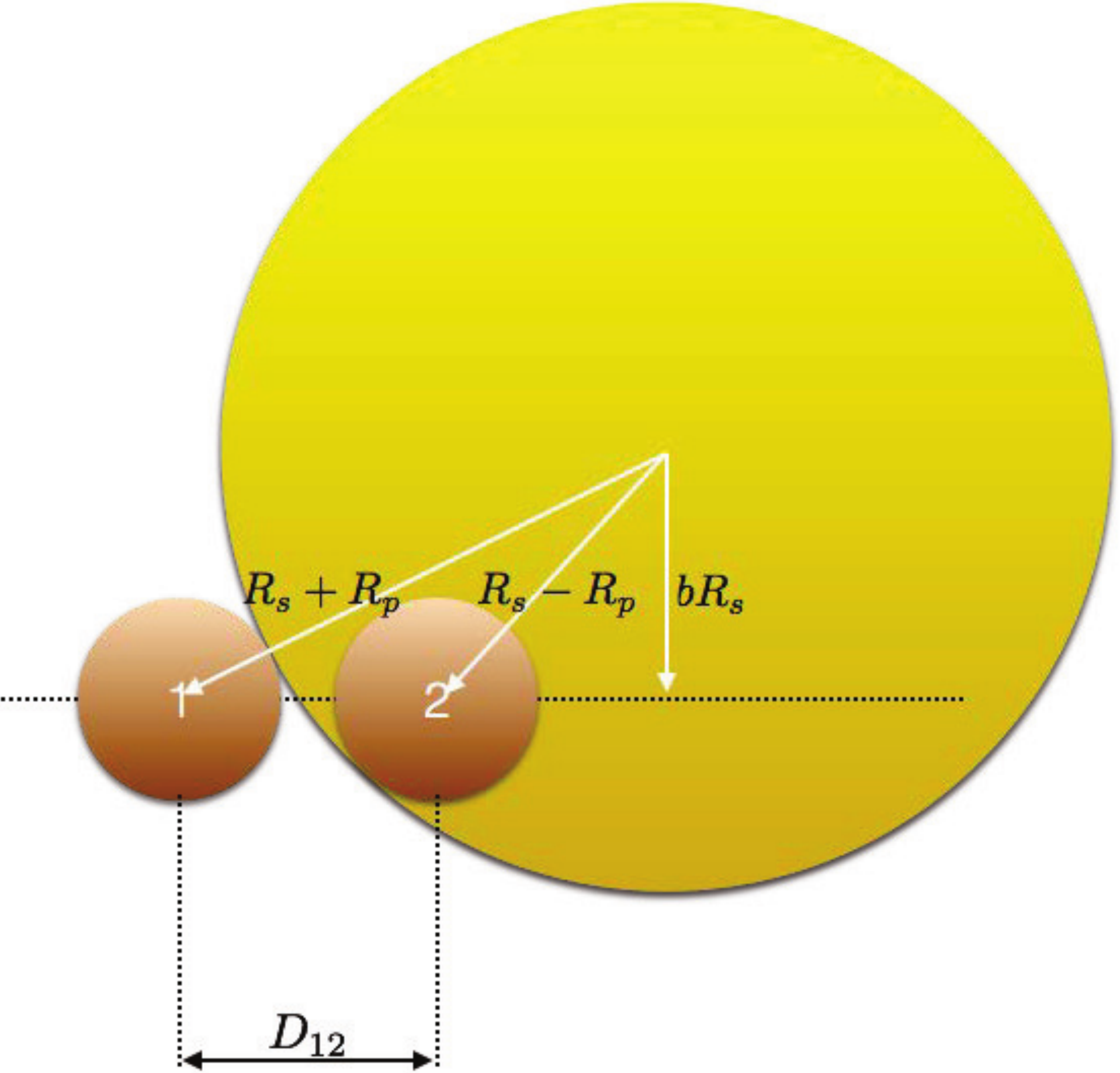}	  	
    \caption{A schematic showing a planet transiting its parent star at any general latitude measured by the impact parameter \emph{b}.}
    \label{fig:IP}
\end{figure}

Because $\tau_{14} < t_\textrm{up}$, the \emph{Spitzer} data sets span a sufficient time to allow the detection of the secondary eclipse time scale $\tau_{14}$.
Using the values for $\tau_{14}$ and $\tau_{23}$, we can then estimate the ratio of the radius of the planet to that of the star by rearranging Equation \ref{eq:transit} to obtain
\begin{equation}
\frac{R_p}{R_s} = \frac{\tau_{14} - \tau_{23}}{\tau_{14} + \tau_{23}} = 0.1543 \pm 0.0283.
\label{eq:ratio}
\end{equation}
Therefore, we have shown here that without the use of any fitting of model parameters such as the epoch of mid-transit, orbital period, fractional flux deficit, total duration of transit, the impact parameter of the planet's path across the stellar disc, the transit depth, the shape parameter for transit, the transit ingress/egress times, and many more \citep{Collier-Cameron:2007aa, Morton:2016aa}, we can now calculate most of these from the time scales alone \citep{Seager:2003aa}.  We note that, to within the precision of $\pm 0.0283$, the result in equation \ref{eq:ratio} is the same as that found by \cite{Bouchy:2005aa} and the average of that found by \cite{Pont:2007aa}.  

In this analysis, we did not take into account the possibility that the planet may be transiting with a non-zero impact parameter ($b \neq 0$, where $b =0$ implies the planet is transiting the equator of the star from the observer's perspective). This would add another unknown in equation \ref{eq:ratio}. 
By using the simple geometry shown in Figure \ref{fig:IP}, another equation can be derived that also involves $\tau_{12}$. Each of these two second order equations can be solved to determine the relationship between the ratio of radii and the impact parameter, as follows.

\begin{figure}[htbp]
    \centering
    \includegraphics[trim = 20 0 20 10, clip, width = 0.5\textwidth]{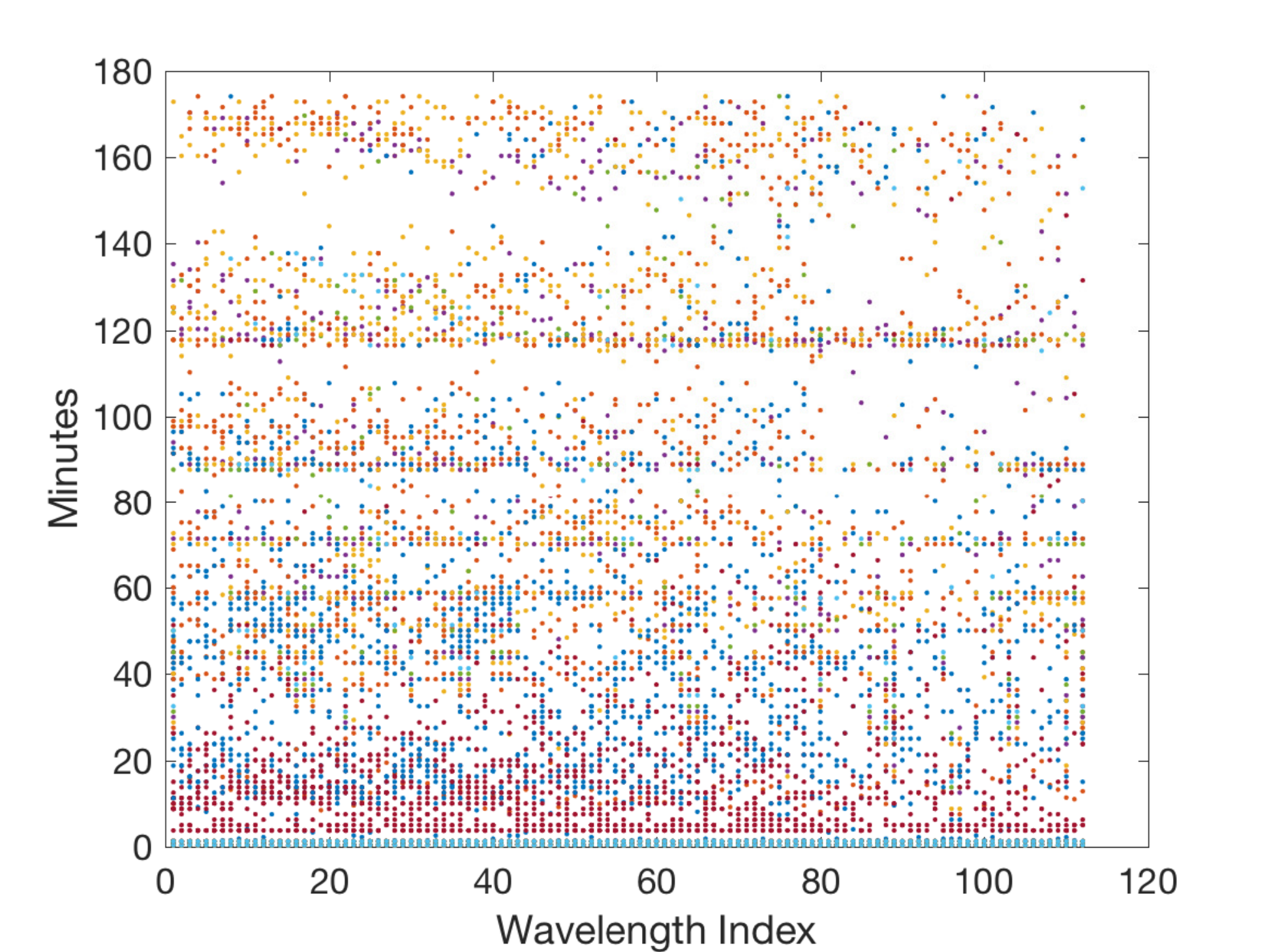}	  	
    \caption{The crossover times plotted for all wavelengths, for those nights that have high S/N \citep{Todorov:2014aa} for HD 189733b. All four significant timescales are robustly extracted using our method.}
    \label{fig:Spitzer_Safe}
\end{figure}
\begin{figure}[htbp]
    \centering
    \includegraphics[trim = 20 0 20 10, clip, width = 0.5\textwidth]{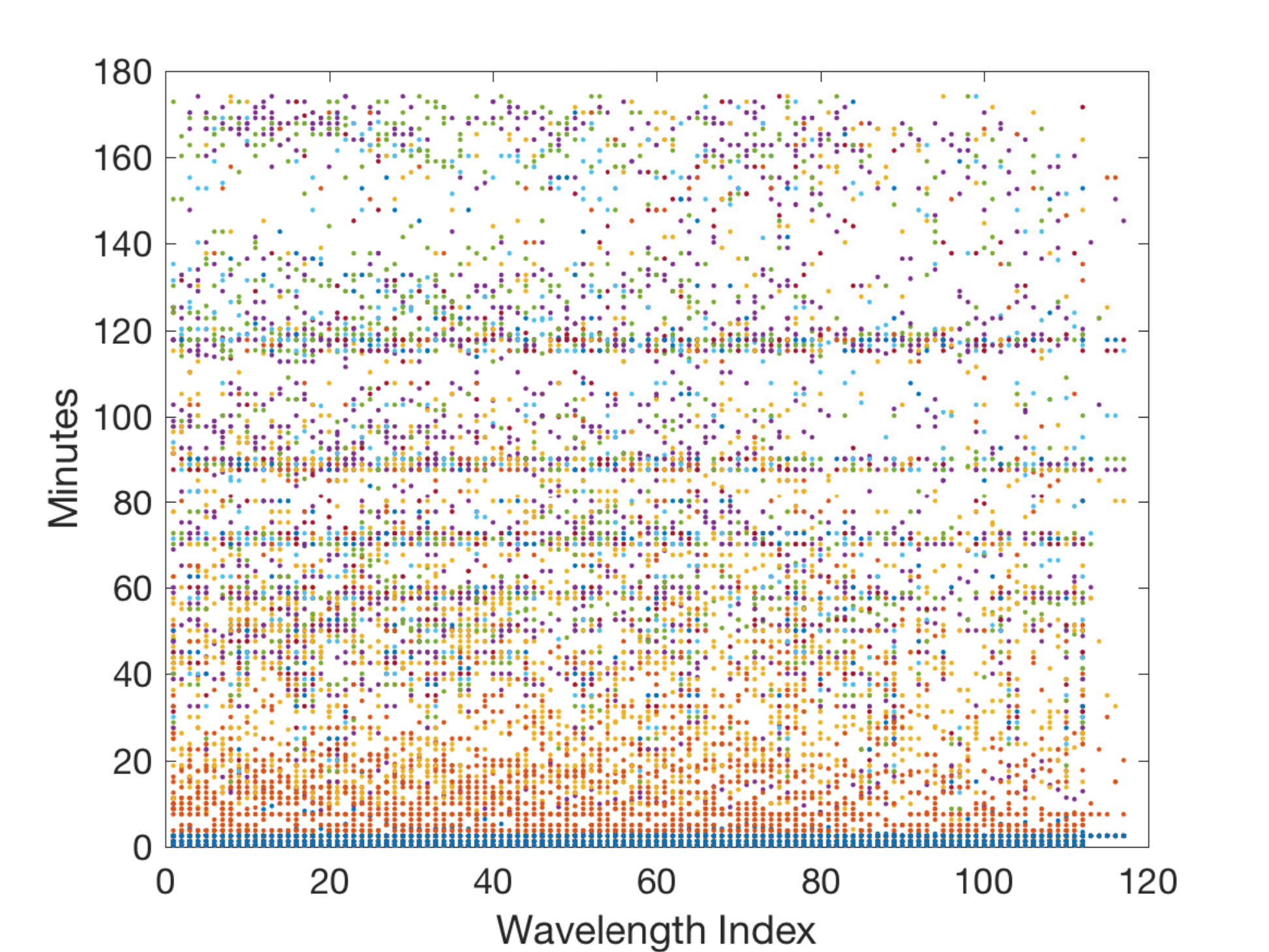}	  	
    \caption{The crossover times plotted for all wavelengths, for all nights for HD 189733b. All four significant timescales are robustly extracted using our method.}
    \label{fig:Spitzer}
\end{figure}

First, we can write $D_{12} = D_{34}$, the distance between the first-second or third-fourth contact from the Pythagorean theorem as
\begin{equation}
D_{12} = D_{34} = \sqrt{(R_s + R_p)^2 - b^2 R_{s}^2} - \sqrt{(R_s - R_p)^2 - b^2 R_{s}^2}.
\label{eq:D12}
\end{equation}
Second, $D_{23}$, the distance between the second and third contact is 
\begin{equation}
D_{23} = 2\sqrt{(R_s - R_p)^2 - b^2 R_{s}^2},
\label{eq:D23}
\end{equation}
and finally we have
\begin{equation}
D_{14} = 2\sqrt{(R_s + R_p)^2 - b^2 R_{s}^2}.
\label{eq:D14}
\end{equation}
Therefore, 
\begin{equation}
\frac{\tau_{14}}{D_{14}} = \frac{\tau_{23}}{D_{23}},
\label{eq:D14_23}
\end{equation}
which, upon substitution of  Eqs \ref{eq:D23} and \ref{eq:D14}, gives
\begin{equation}
\tau_{23} = \tau_{14} \times \sqrt{\frac{1 + \rho^2 - 2\rho - b^2}{1 + \rho^2 + 2\rho - b^2}},
\label{eq:ratio1}
\end{equation}
where $\rho = \frac{R_p}{R_s}$.  Similarly, Eqs \ref{eq:D12} and \ref{eq:D14} give
\begin{equation}
\tau_{12} = \tau_{14} \times \frac{1}{2}\left ( 1 - \sqrt{\frac{1 + \rho^2 - 2\rho - b^2}{1 + \rho^2 + 2\rho - b^2}} \right ), 
\label{eq:ratio2}
\end{equation}
and thus Eqs. \ref{eq:ratio1} and \ref{eq:ratio2} can be used to calculate $\rho$ and $b$, simultaneously. As a check,  Equations \ref{eq:ratio1} and \ref{eq:ratio2} combine to give Equations \ref{eq:transit}, and Equation \ref{eq:ratio1} becomes Equation \ref{eq:ratio} in the case $b=0$.

\begin{figure}[htbp]
    \centering
    \includegraphics[trim = 20 0 20 10, clip, width = 0.5\textwidth]{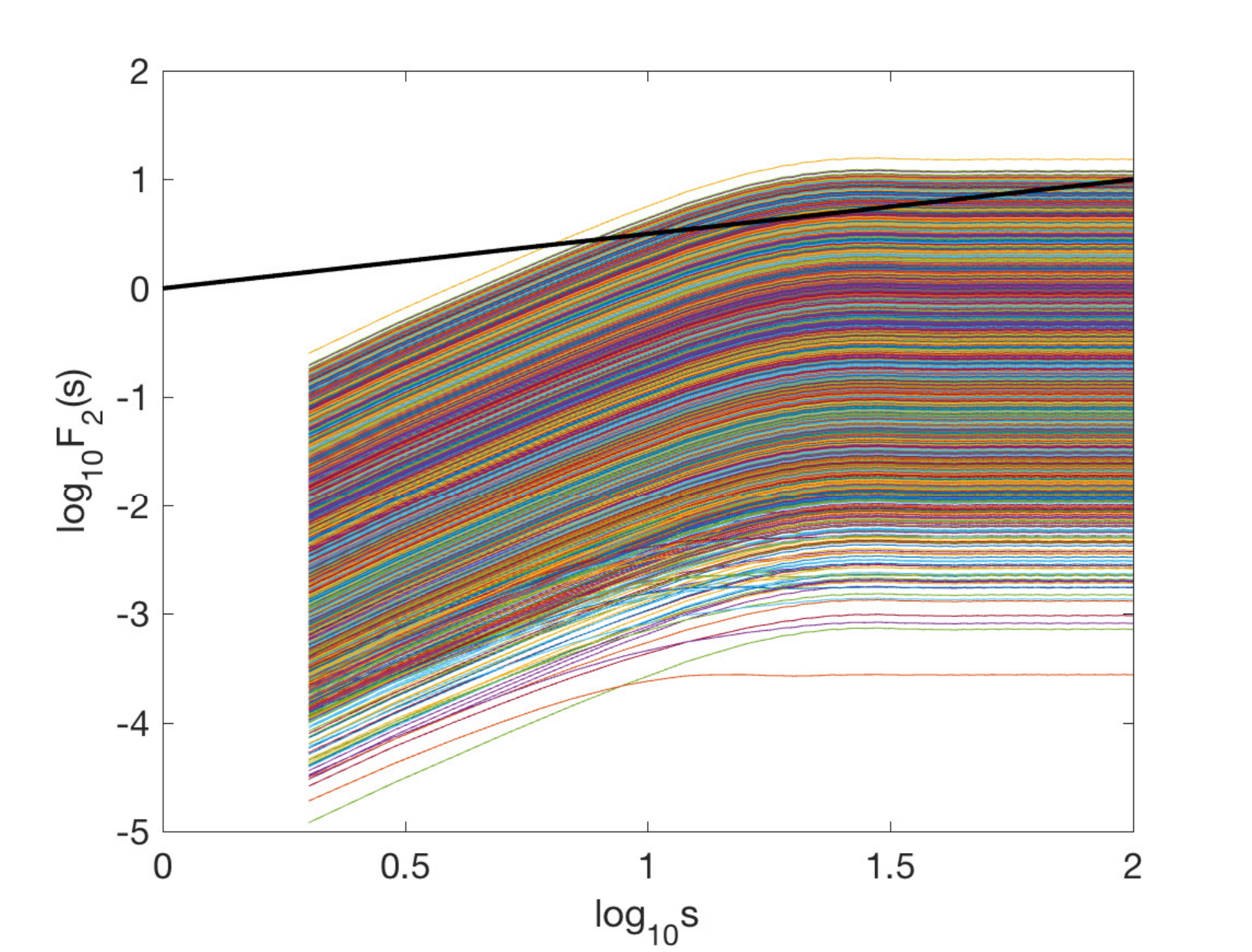}	  	
    \caption{The second moment of the fluctuation functions are shown for all of the wavelengths for the simulated planet without noise in the spectra. The straight black line has a slope of 0.5, which denotes white noise dynamics.}
    \label{fig:Soap_MF_planet}
\end{figure}

In Figures \ref{fig:Spitzer_Safe} and \ref{fig:Spitzer} we show the crossovers from all the available data sets, as shown in Table \ref{T:Spitzer}.  Additionally, Figure \ref{fig:Spitzer} includes those data sets that are not used in other studies due to low S/N, or other issues \citep{Todorov:2014aa}. Moreover, we see the clear emergence of the significant time scales discussed above from {\em all } of these data sets. Importantly, we note the amount of noise in these data sets, which as we will see below can be a source of information for the robust estimation of these time scales.

\subsection{SOAP Simulated Data}

\begin{figure*}[htbp]
    \centering
    \begin{subfigure}[]
        \centering
        \includegraphics[trim = 15 0 20 0, clip, width = 0.3\textwidth]{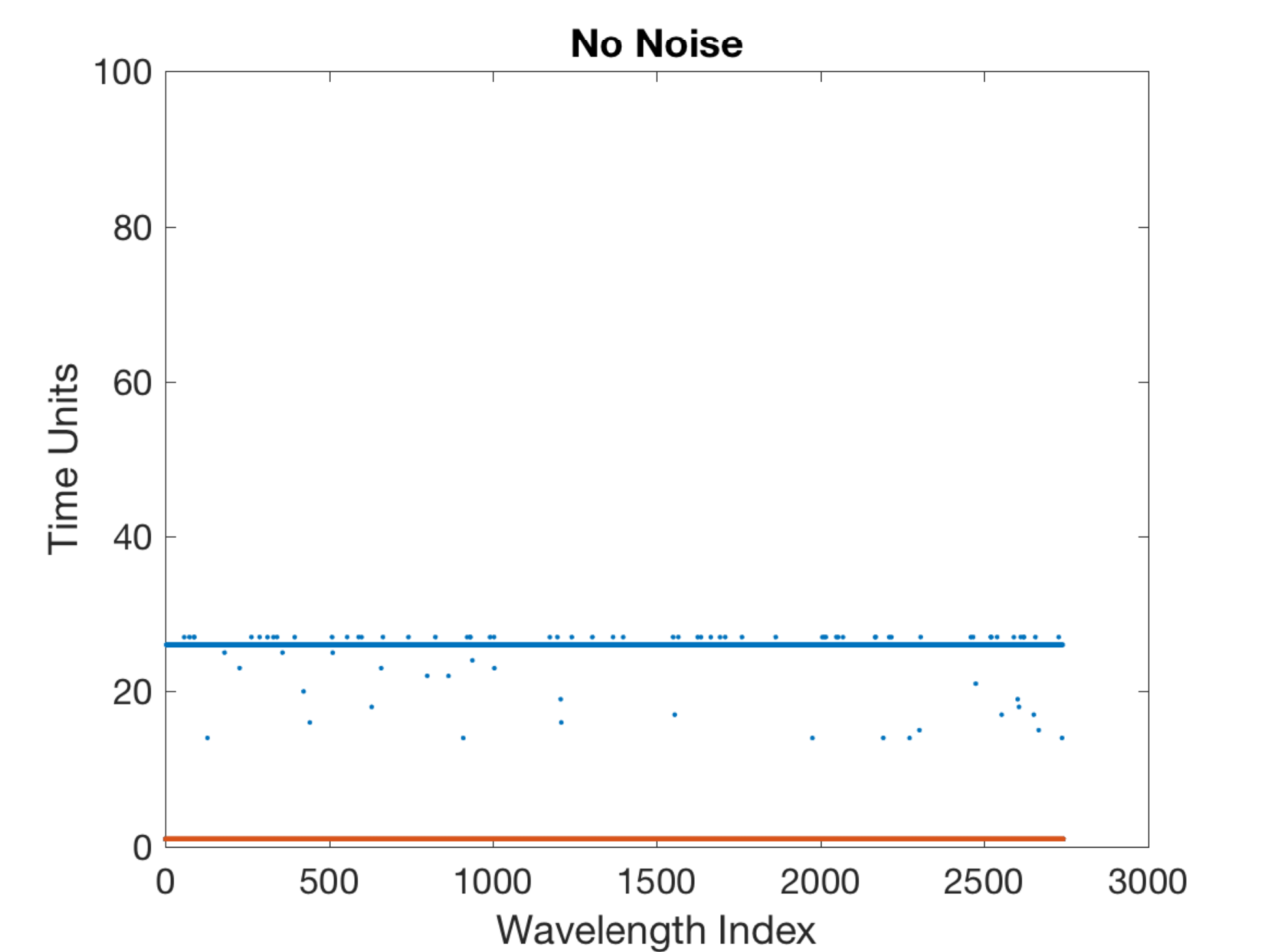}
            \end{subfigure}%
    ~ 
    \begin{subfigure}[]
        \centering
        \includegraphics[trim = 15 0 20 0, clip, width = 0.3\textwidth]{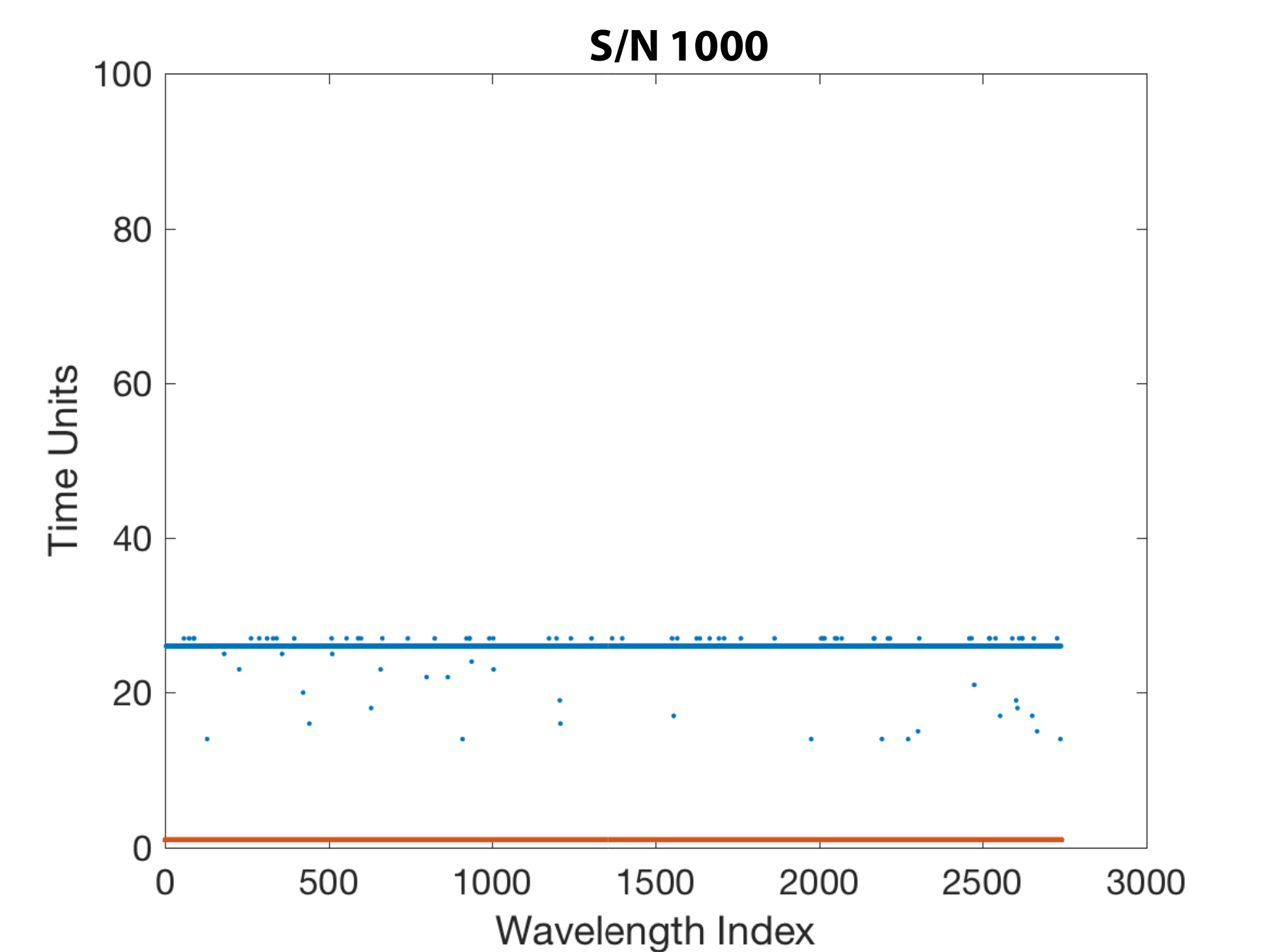}	
            \end{subfigure}
    ~
        \begin{subfigure}[]
        \centering
        \includegraphics[trim = 15 0 20 0, clip, width = 0.3\textwidth]{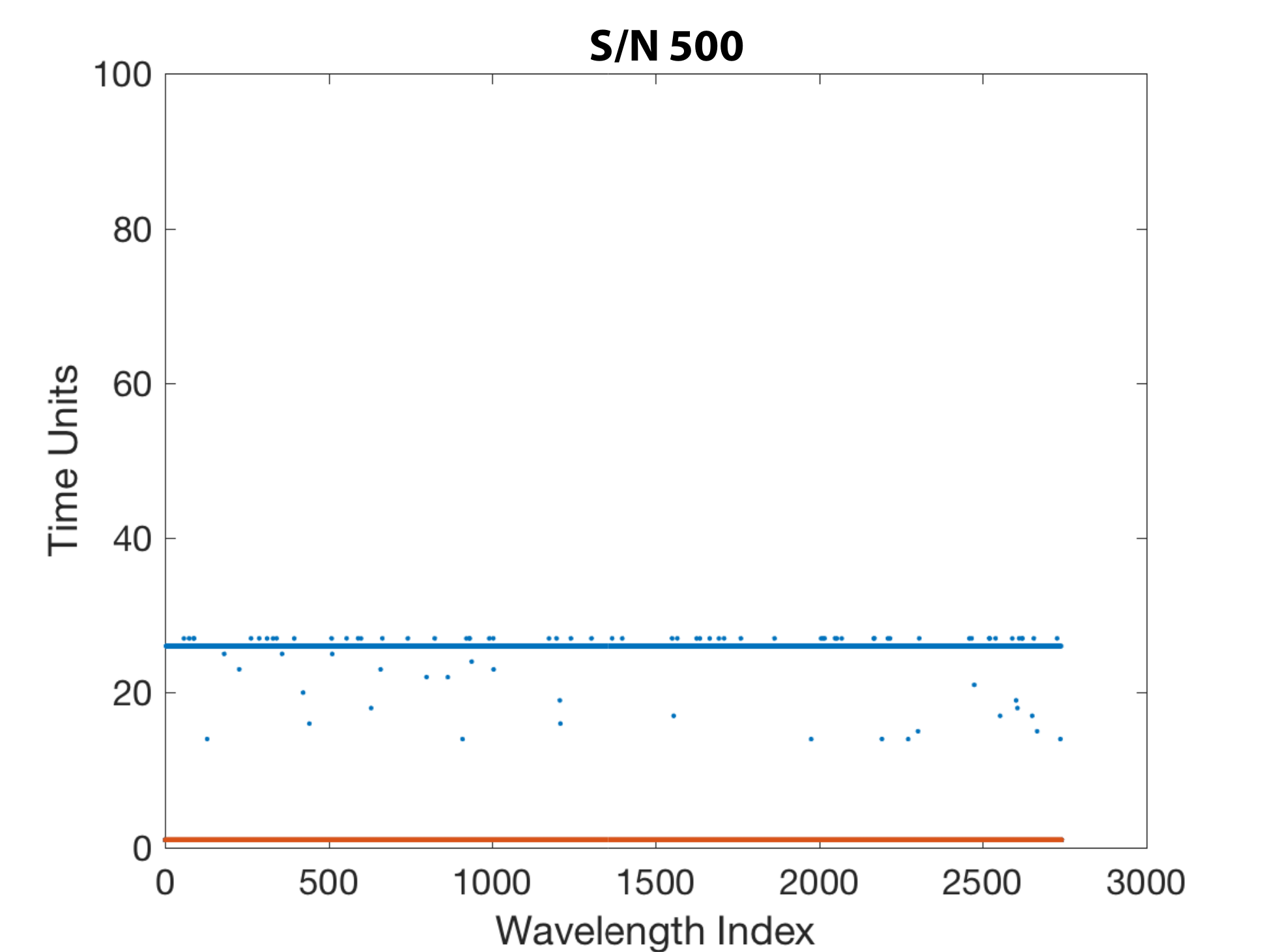}	
            \end{subfigure}%
   \\
    \begin{subfigure}[]
        \centering
        \includegraphics[trim = 15 0 20 0, clip, width = 0.3\textwidth]{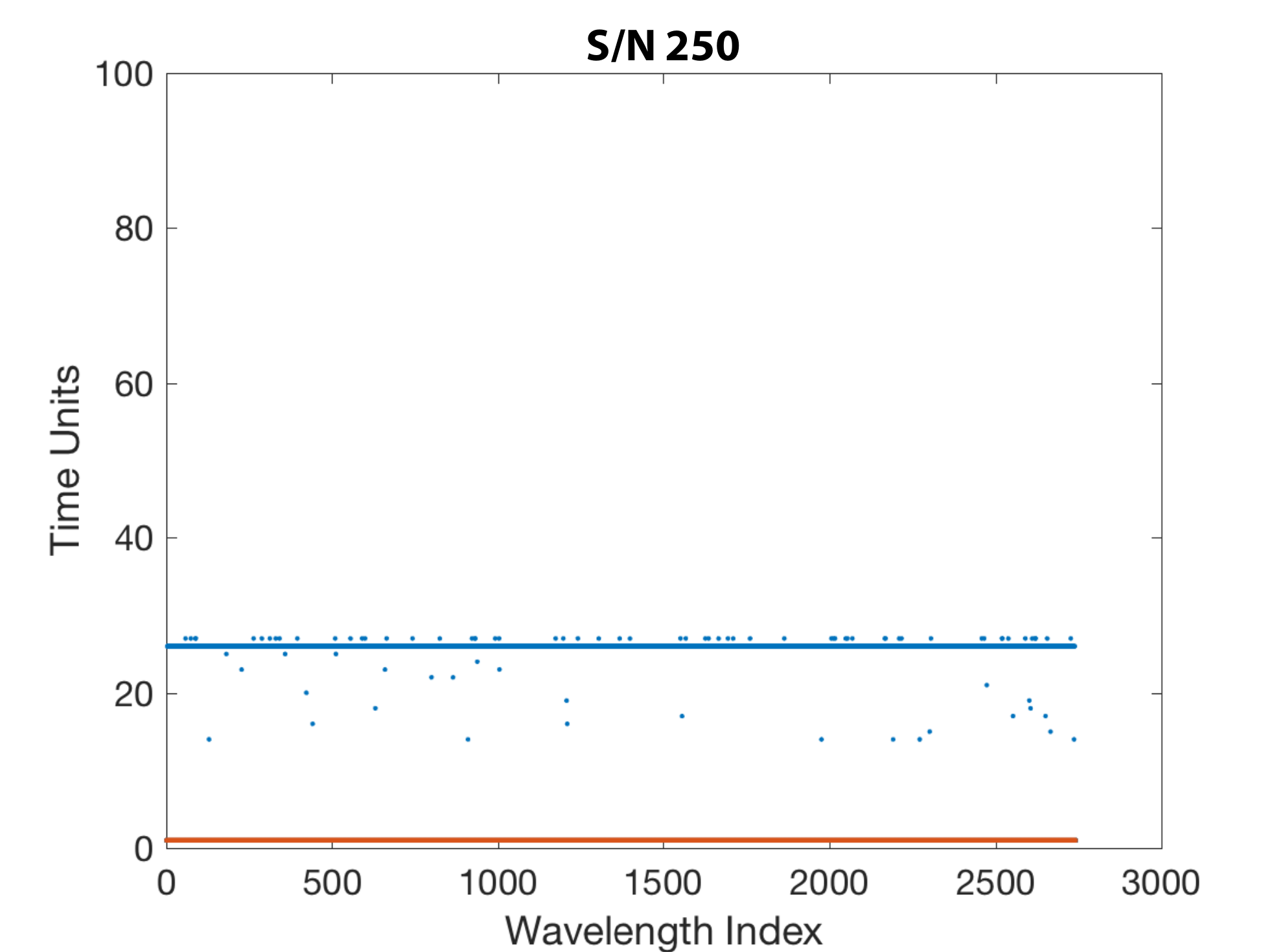}	
    \end{subfigure}
    ~
        \begin{subfigure}[]
        \centering
        \includegraphics[trim = 15 0 20 0, clip, width = 0.32\textwidth]{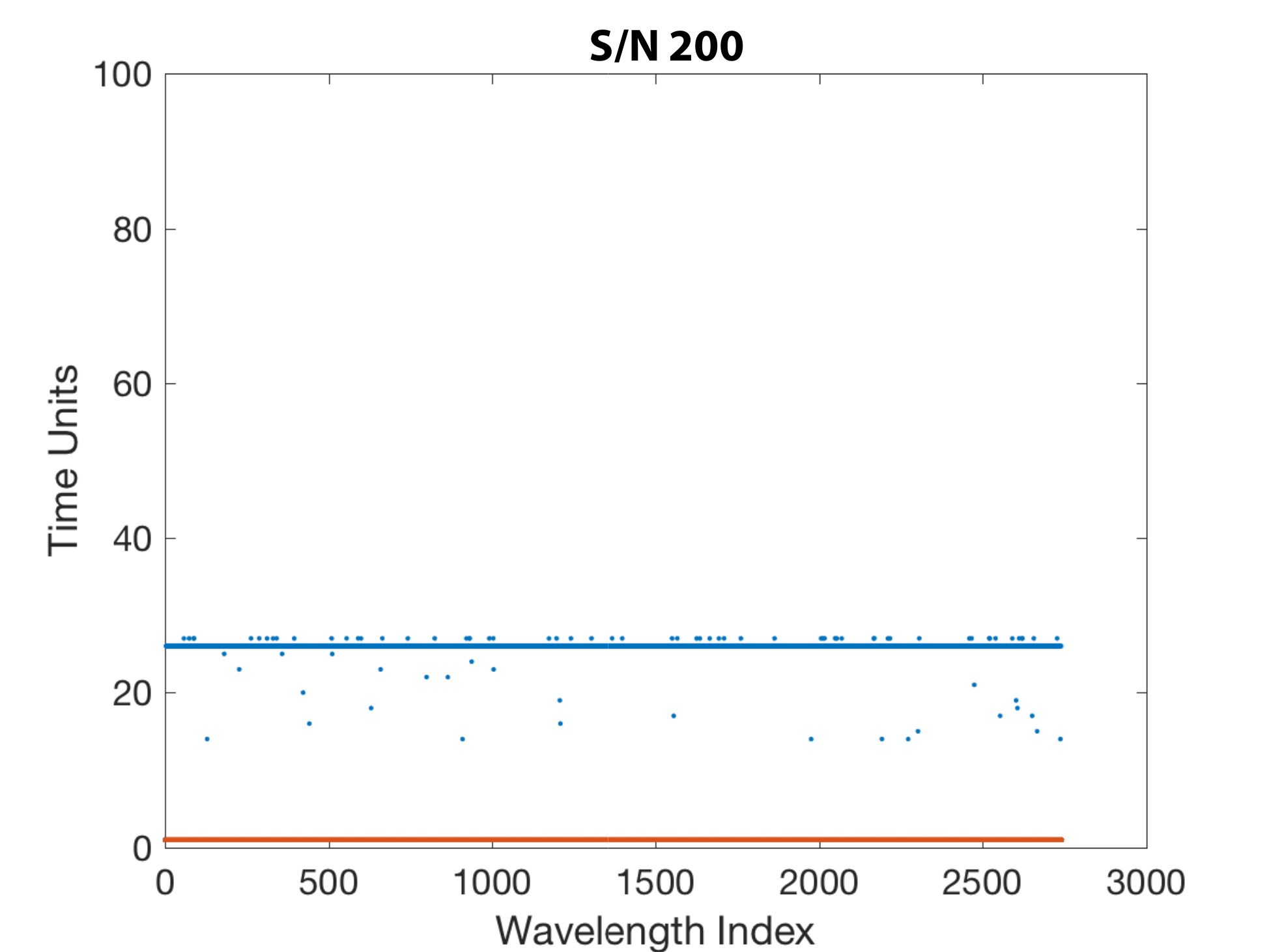}	
    \end{subfigure}%
    ~ 
    \begin{subfigure}[]
        \centering
        \includegraphics[trim = 15 0 20 0, clip, width = 0.3\textwidth]{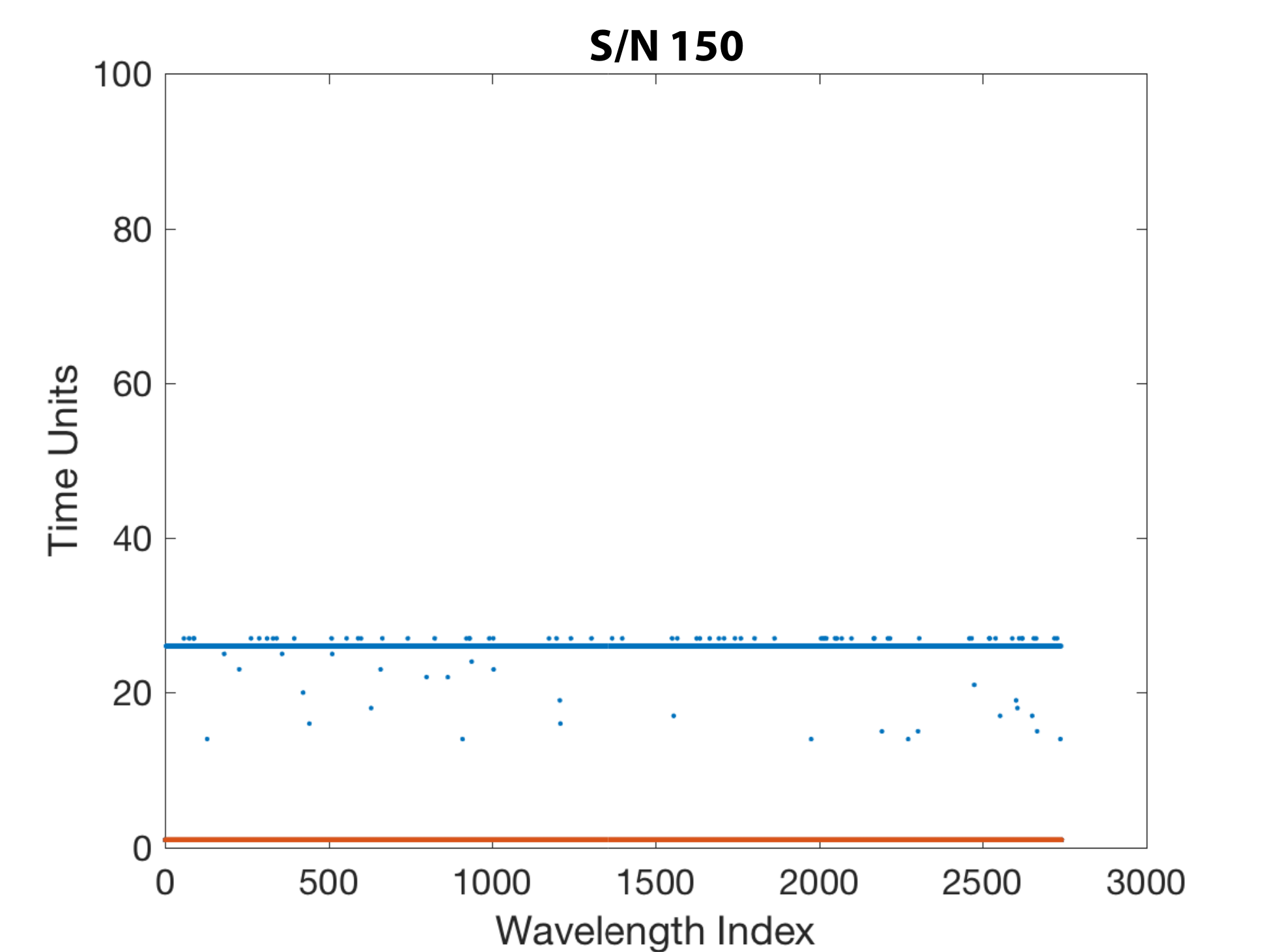}	
    \end{subfigure}
     \\
     \begin{subfigure}[]
        \centering
        \includegraphics[trim = 15 0 20 0, clip, width = 0.3\textwidth]{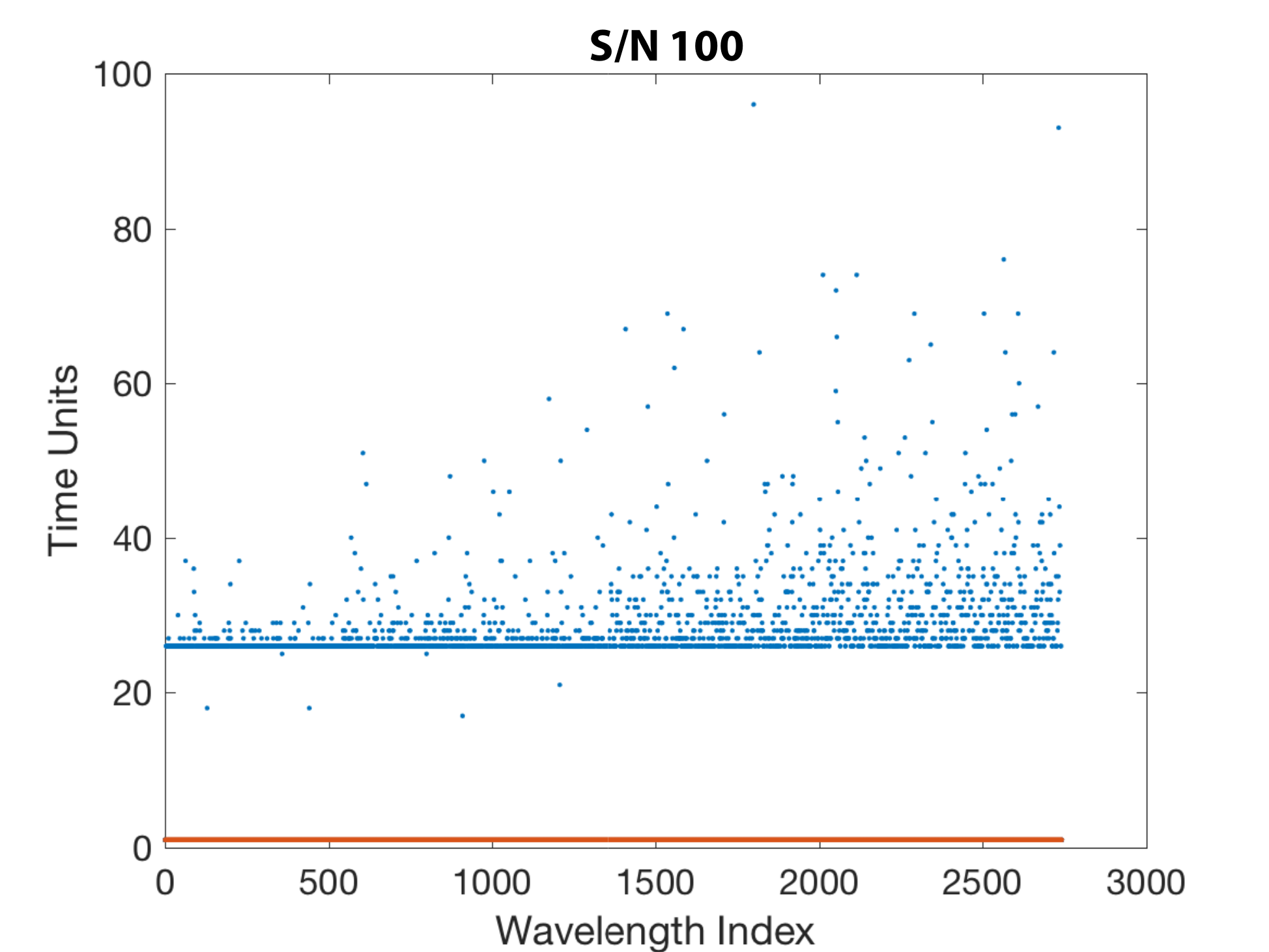}	
    \end{subfigure}%
    ~ 
    \begin{subfigure}[]
        \centering
        \includegraphics[trim = 15 0 20 0, clip, width = 0.3\textwidth]{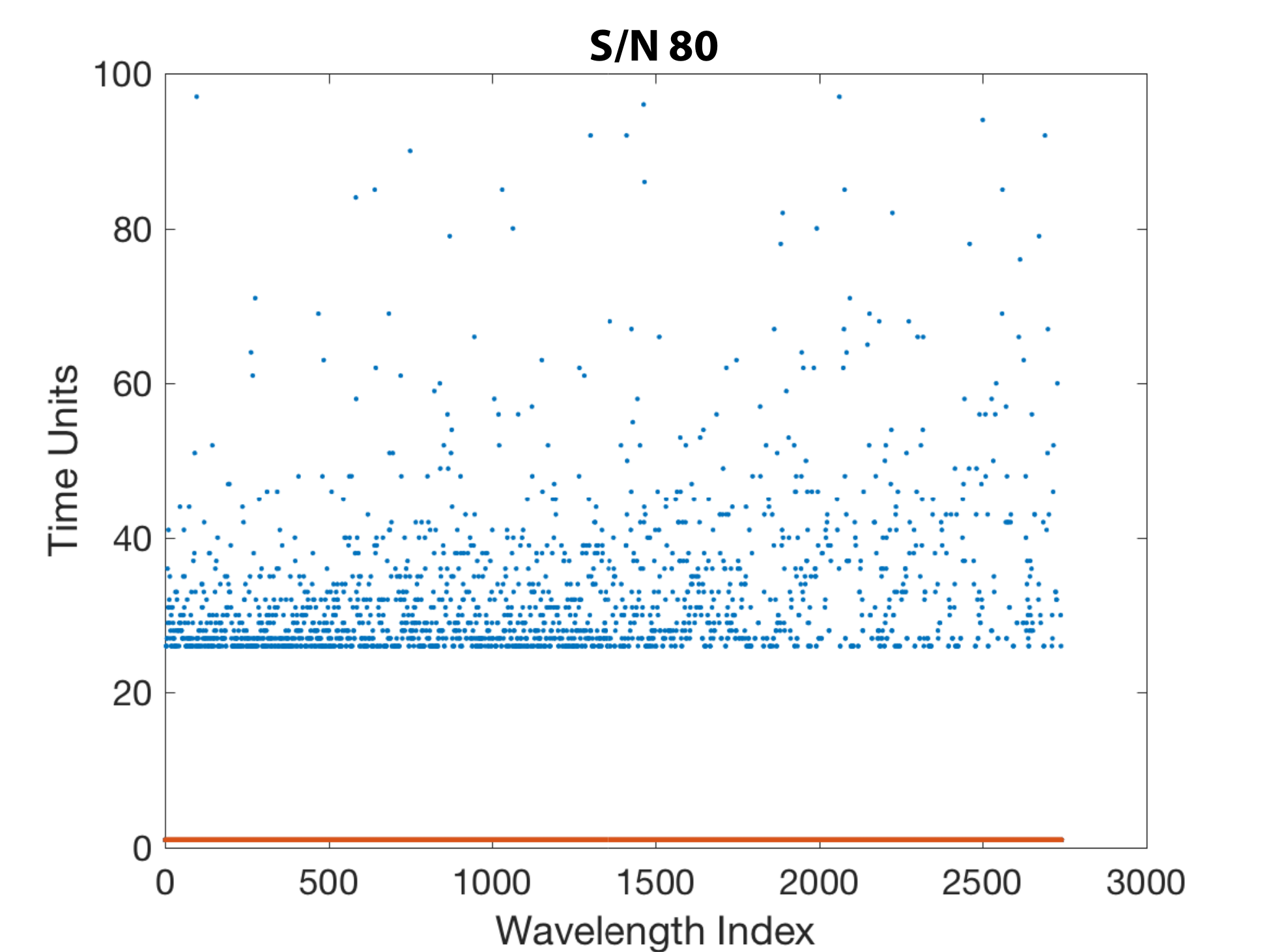}	
    \end{subfigure}%
    ~ 
    \begin{subfigure}[]
        \centering
        \includegraphics[trim = 15 0 20 0, clip, width = 0.3\textwidth]{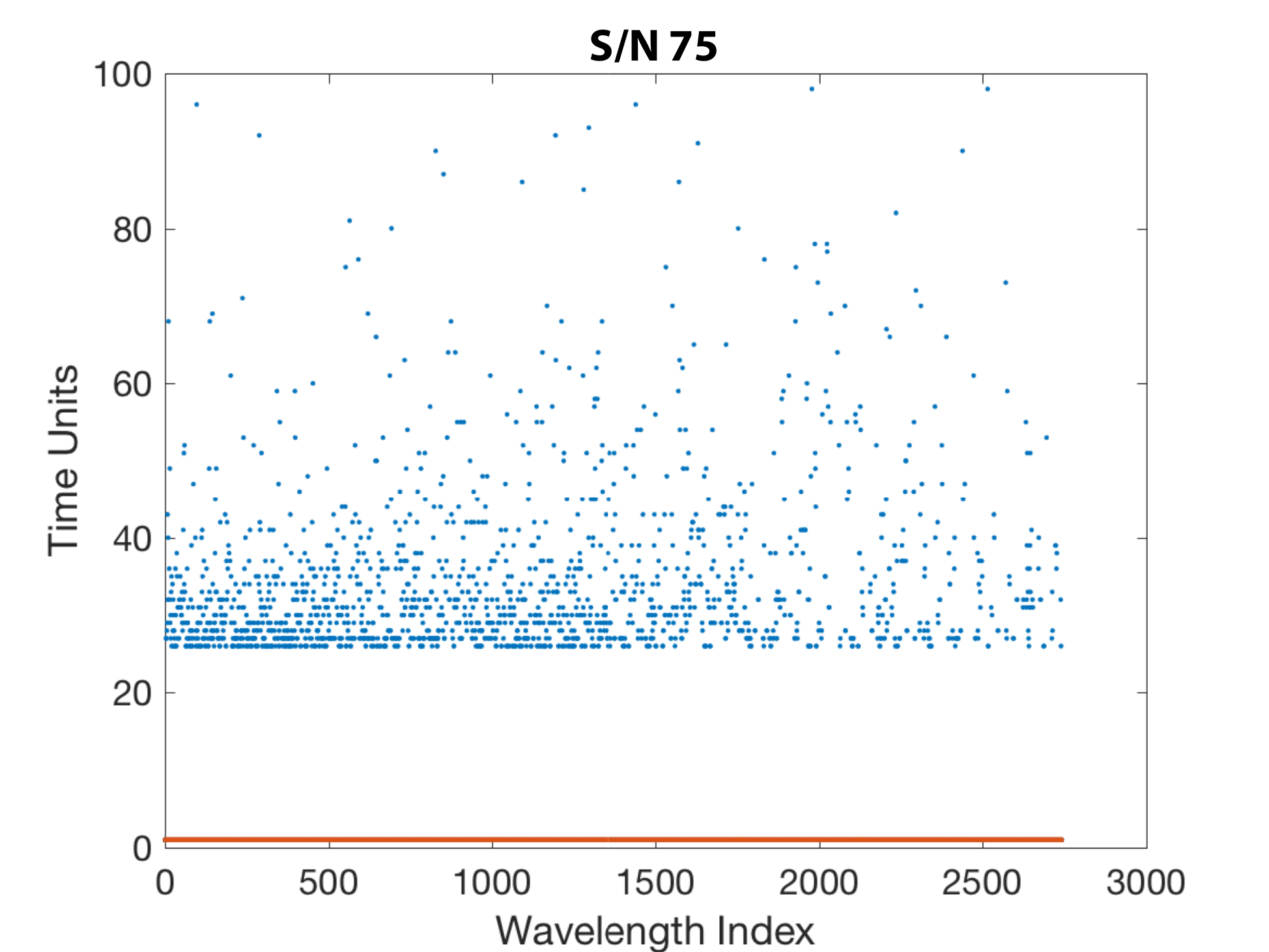}	
    \end{subfigure}
    \\
        \begin{subfigure}[]
        \centering
        \includegraphics[trim = 15 0 20 0, clip, width = 0.3\textwidth]{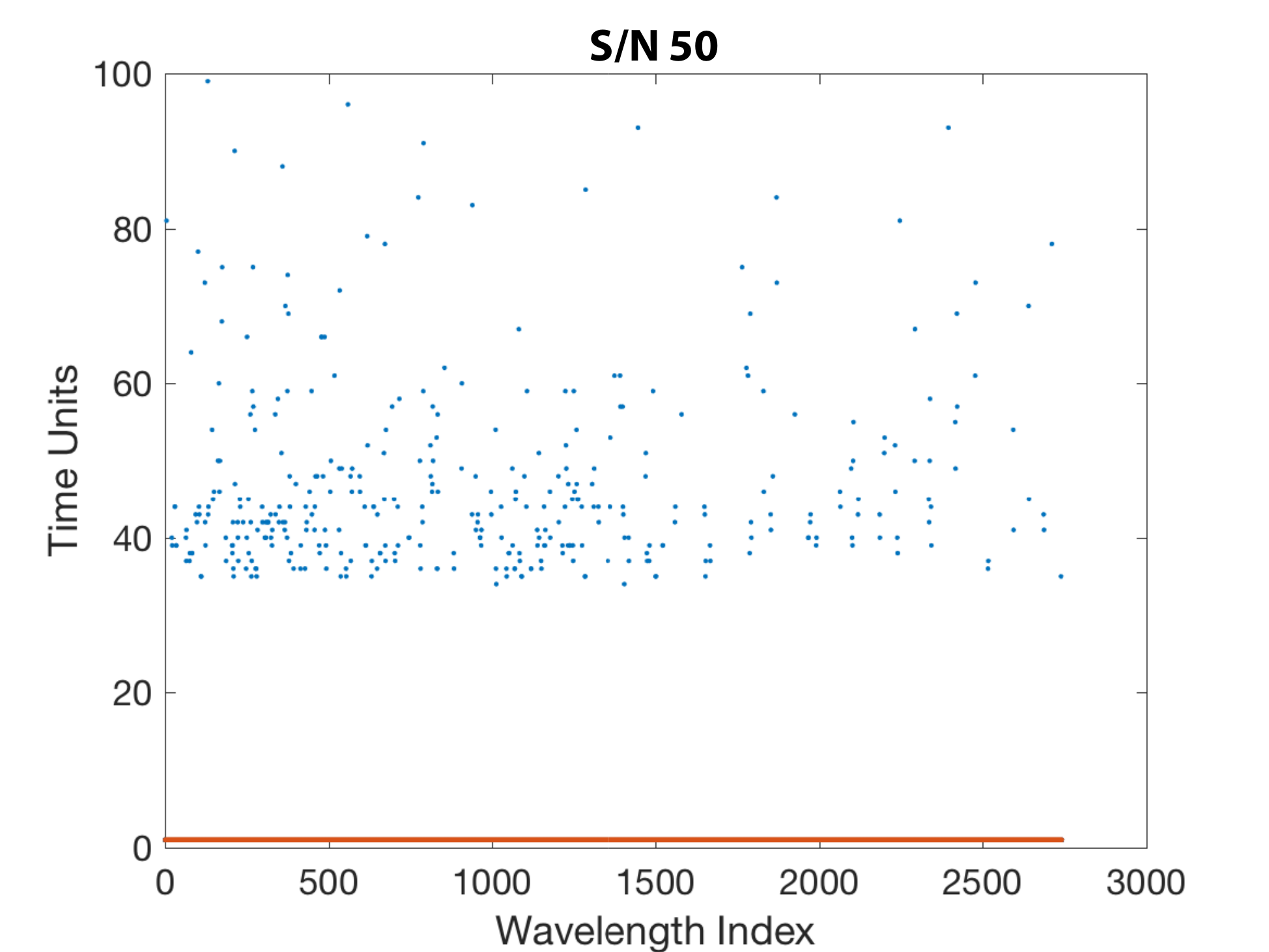}	
    \end{subfigure}%
    ~ 
    \begin{subfigure}[]
        \centering
        \includegraphics[trim = 15 0 20 0, clip, width = 0.3\textwidth]{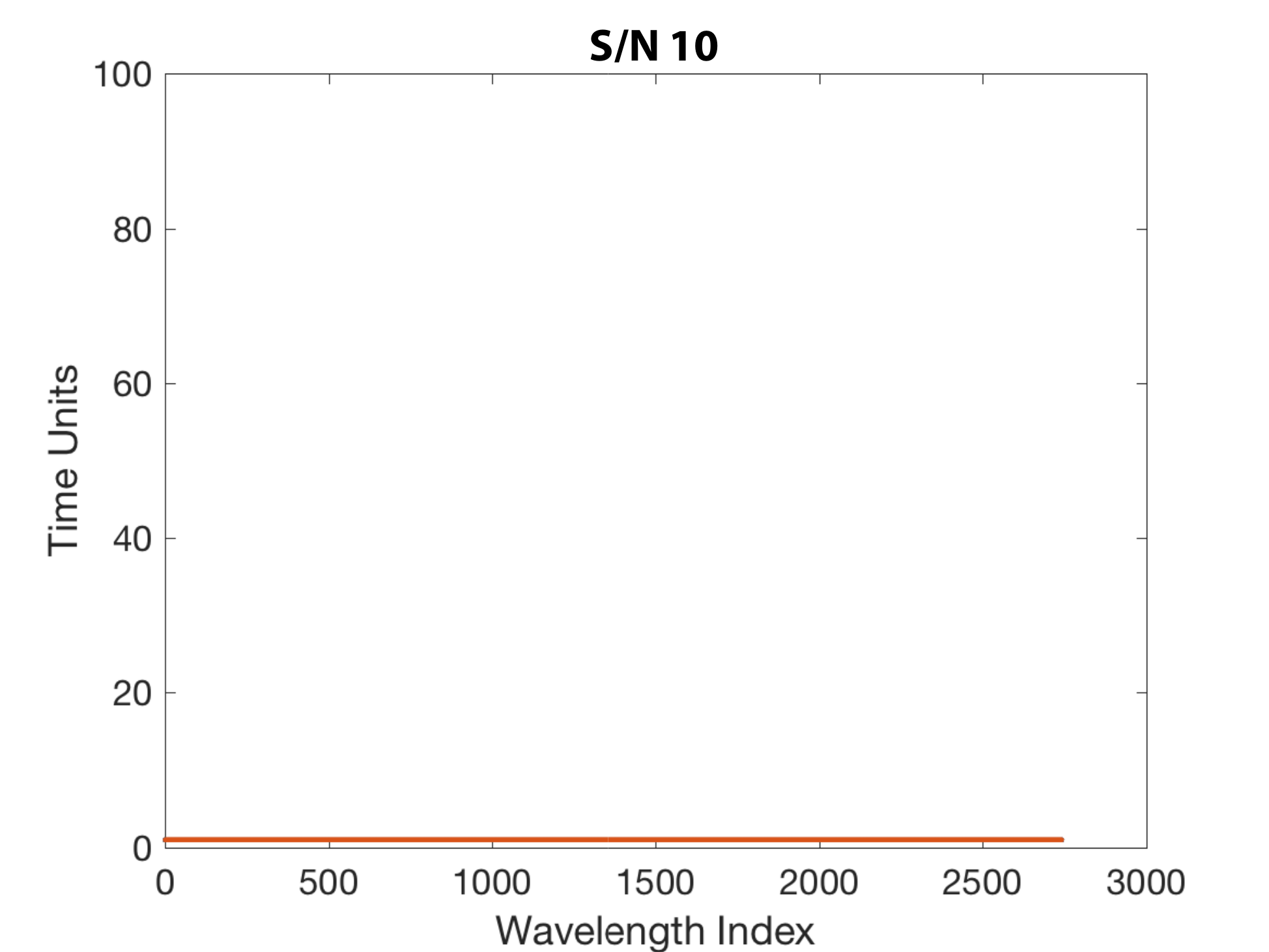}	
    \end{subfigure}%
    ~ 
    \begin{subfigure}[]
        \centering
        \includegraphics[trim = 15 0 20 0, clip, width = 0.3\textwidth]{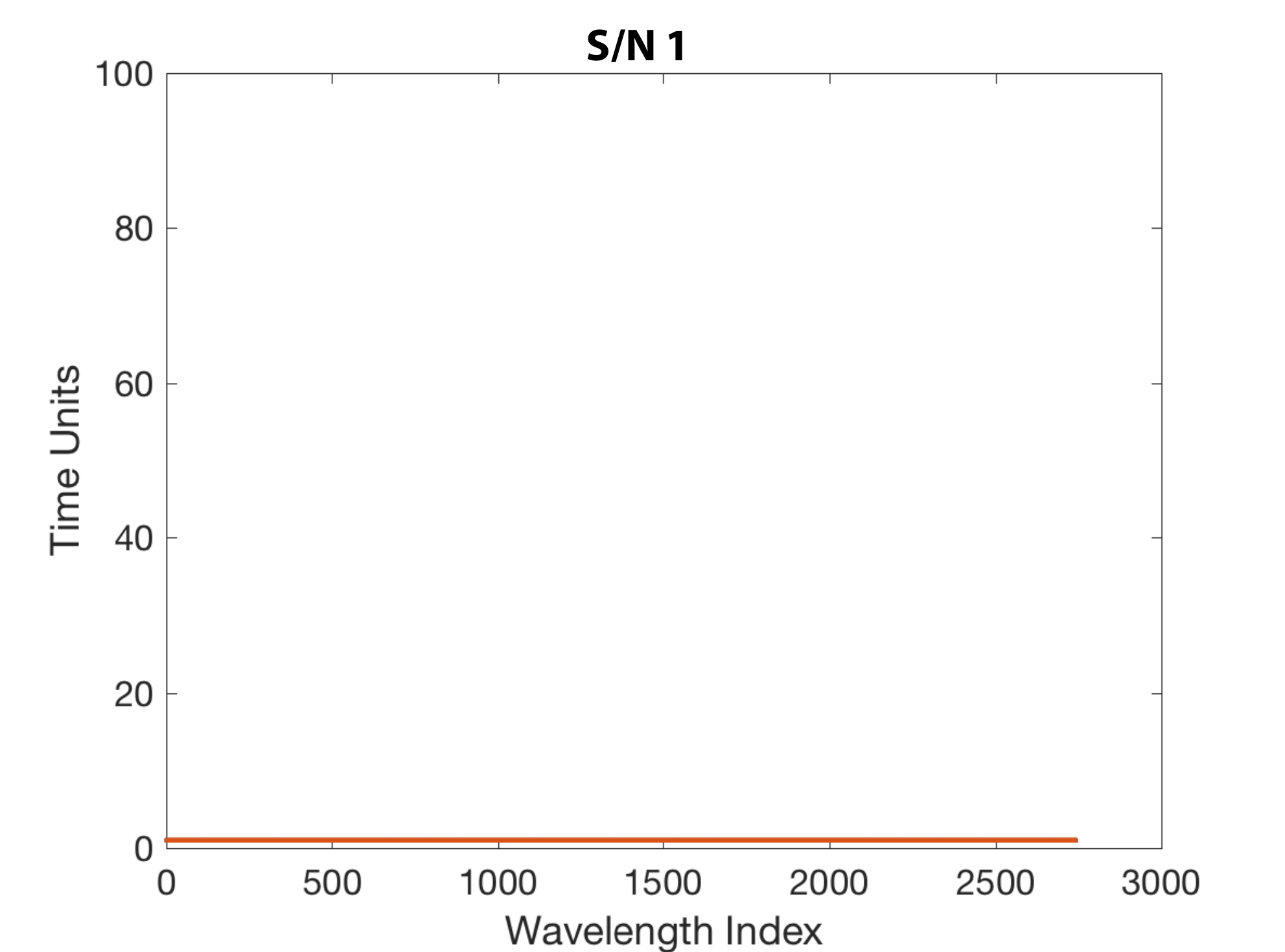}	
    \end{subfigure}      
\caption{The crossover times plotted for all wavelengths, with $C_{th} = 0.08$, for the simulated SOAP 2.0 spectra in presence of an orbiting planet. We show results for different S/N. (a) No Noise, (b) S/N = 1000, (c) S/N = 500, (d) S/N = 250, (e) S/N = 200, (f) S/N = 150, (g) S/N = 100, (h) S/N = 80, (i) S/N = 75, (j) S/N = 50, (k) S/N = 10, (l) S/N = 1.
\label{fig:SOAP_p2}}
\end{figure*}
\begin{figure*}[htbp]
    \centering
    \begin{subfigure}[]
        \centering
        \includegraphics[trim = 15 0 20 0, clip, width = 0.32\textwidth]{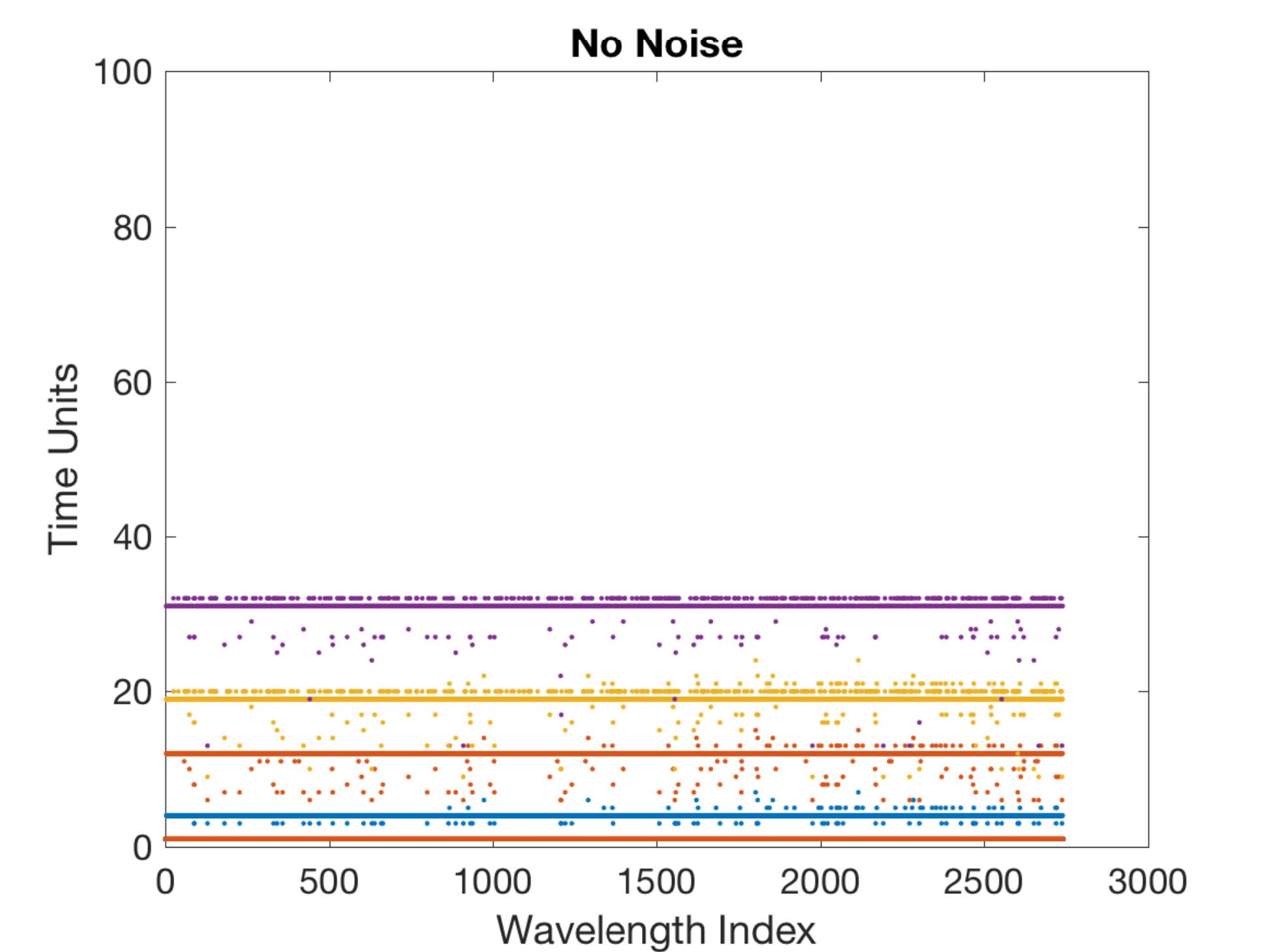}	
    \end{subfigure}%
    ~ 
    \begin{subfigure}[]
        \centering
        \includegraphics[trim = 15 0 20 0, clip, width = 0.32\textwidth]{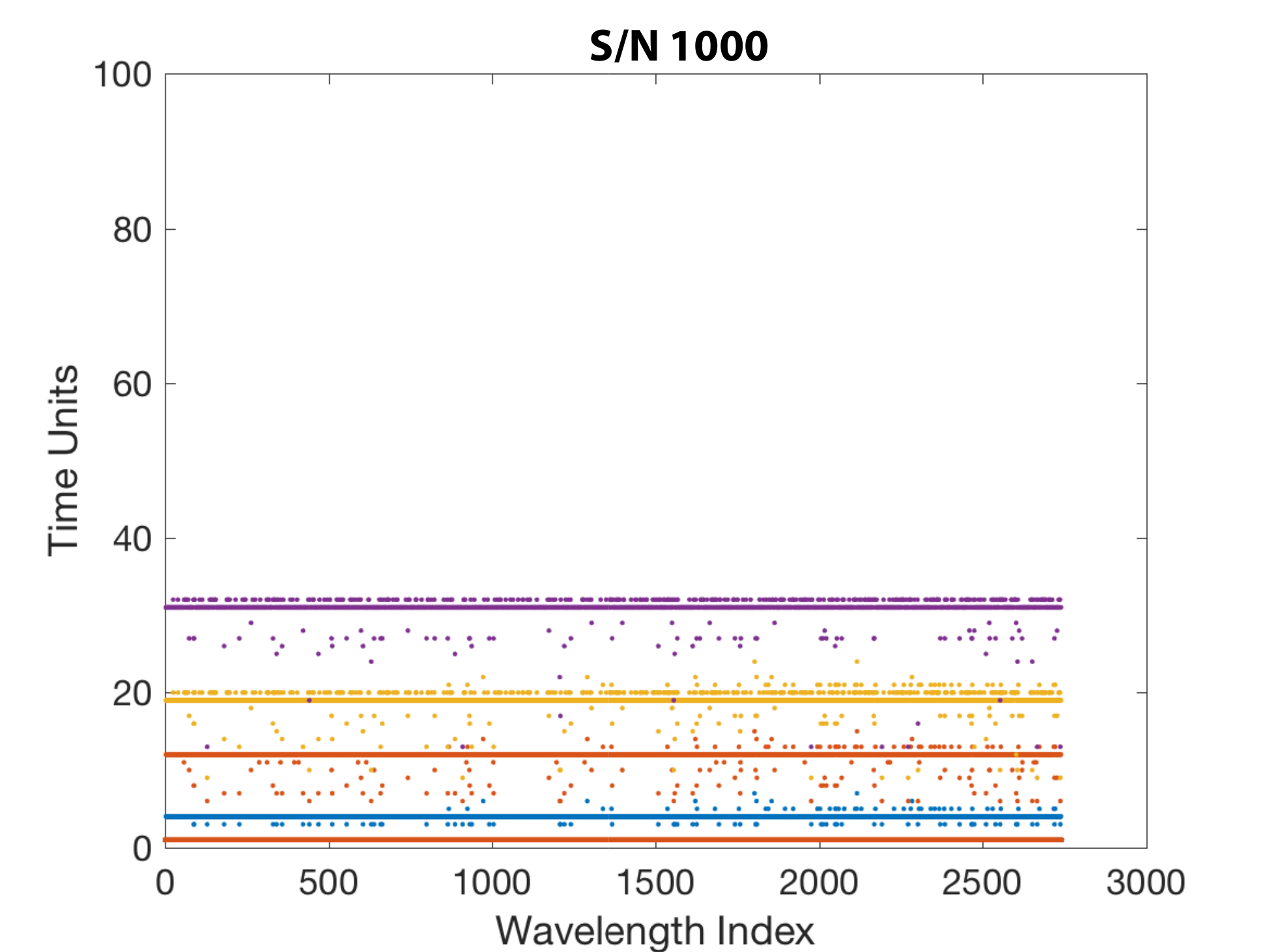}	
            \end{subfigure}
    ~
        \begin{subfigure}[]
        \centering
        \includegraphics[trim = 15 0 20 0, clip, width = 0.32\textwidth]{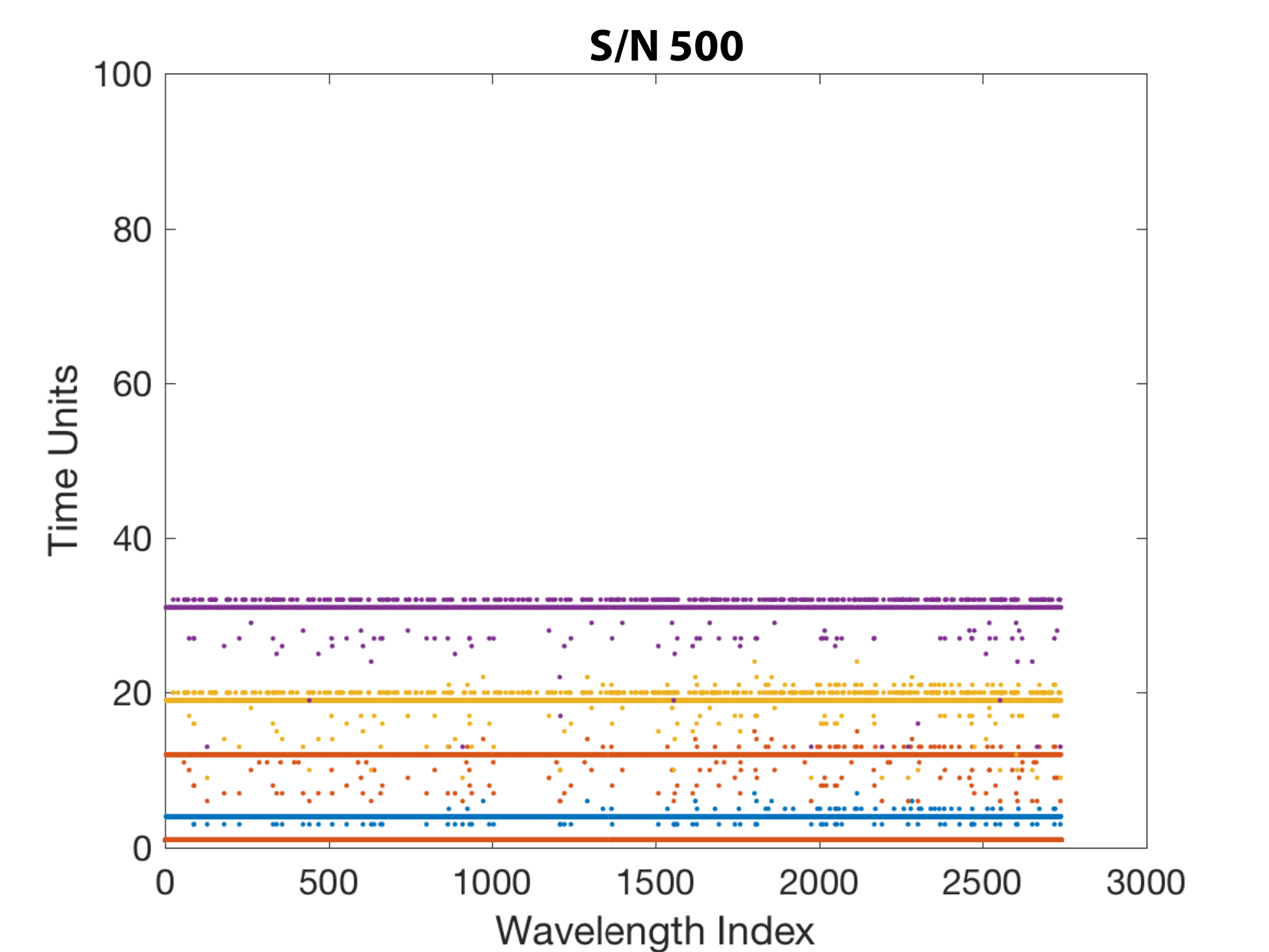}	
    \end{subfigure}%
    \\
    \begin{subfigure}[]
        \centering
        \includegraphics[trim = 15 0 20 0, clip, width = 0.32\textwidth]{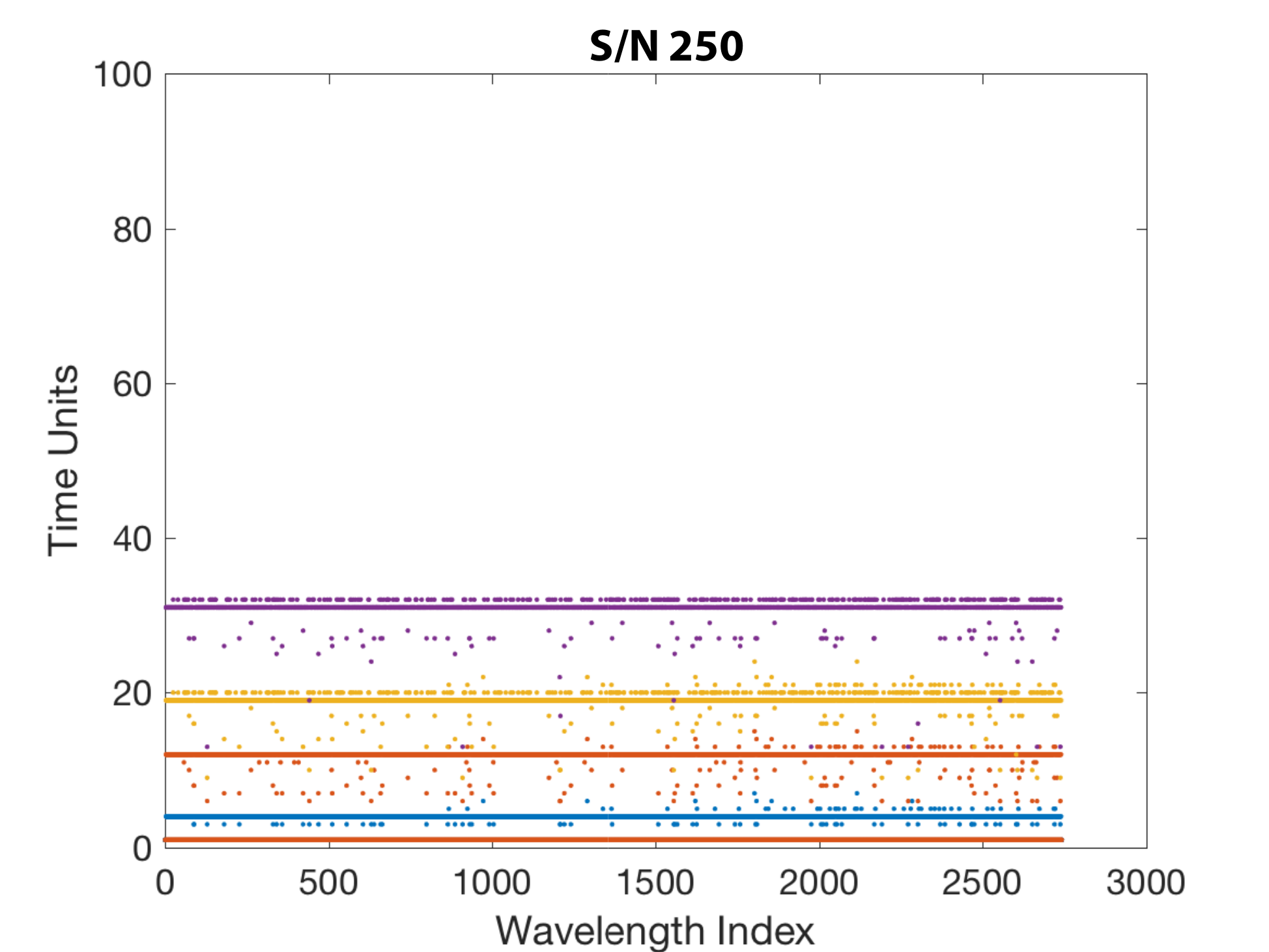}	
            \end{subfigure}
    ~
        \begin{subfigure}[]
        \centering
        \includegraphics[trim = 15 0 20 0, clip, width = 0.32\textwidth]{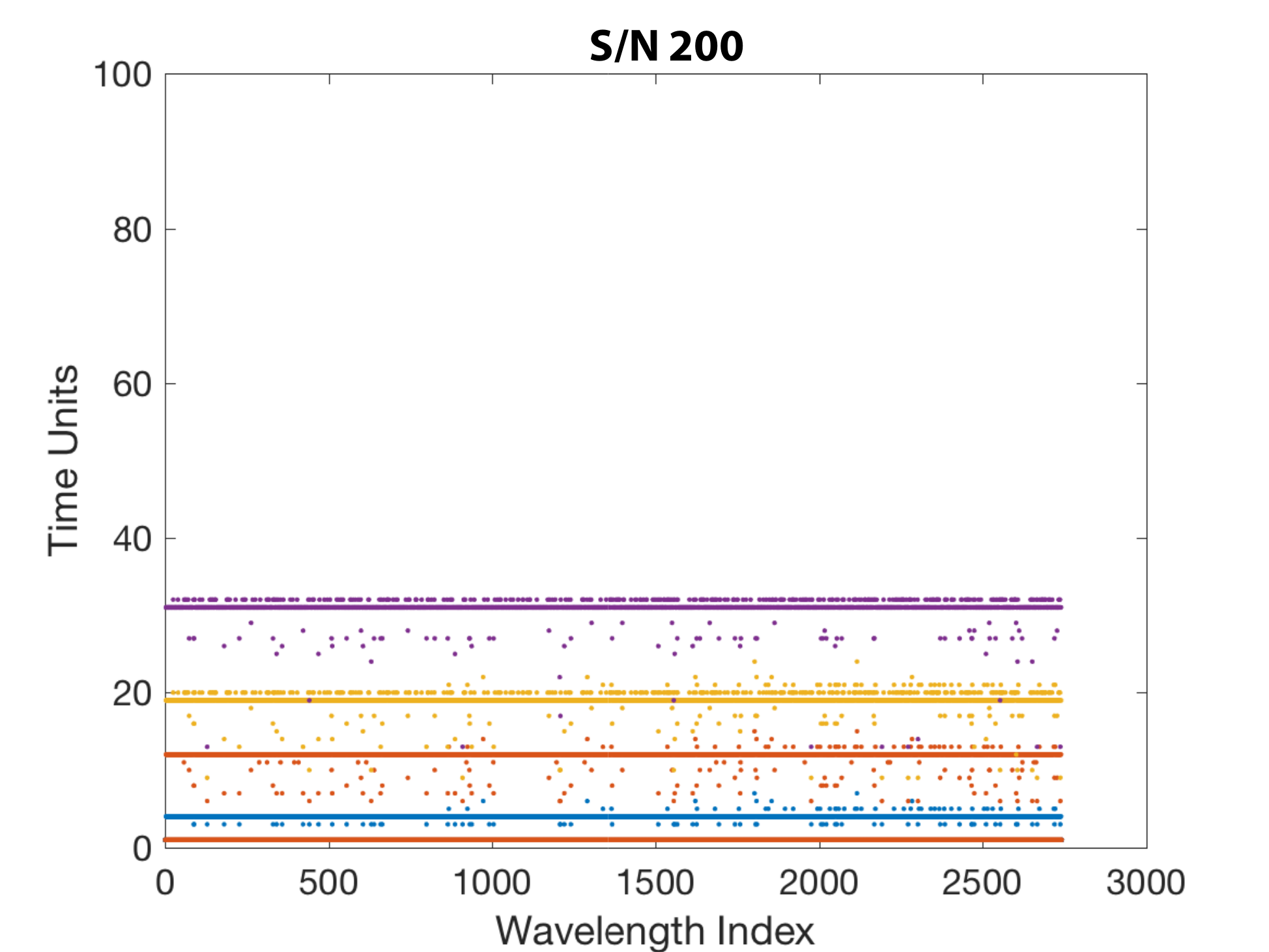}	
    \end{subfigure}%
    ~ 
    \begin{subfigure}[]
        \centering
        \includegraphics[trim = 15 0 20 0, clip, width = 0.32\textwidth]{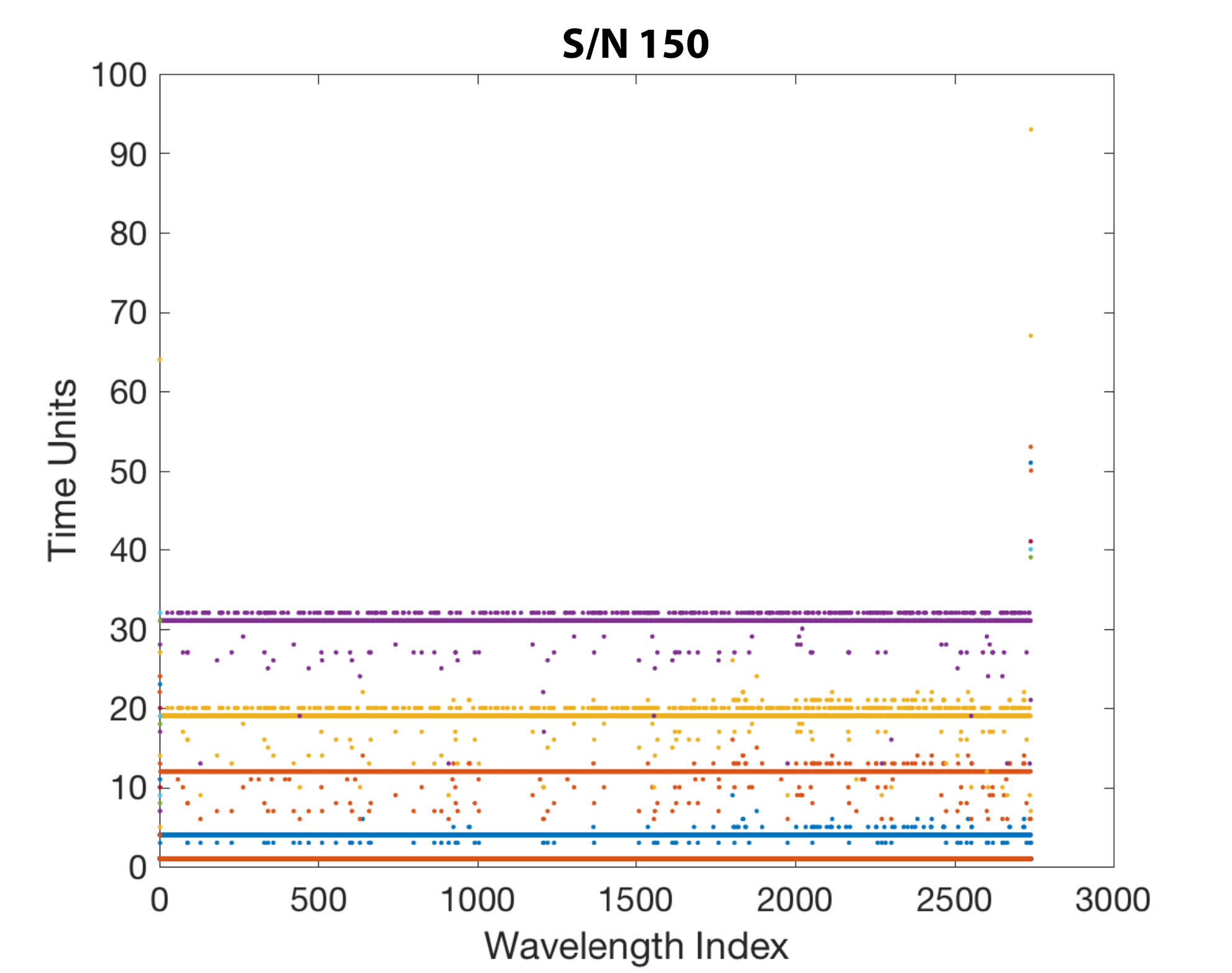}	
            \end{subfigure}
     \\       
     \begin{subfigure}[]
        \centering
        \includegraphics[trim = 15 0 20 0, clip, width = 0.32\textwidth]{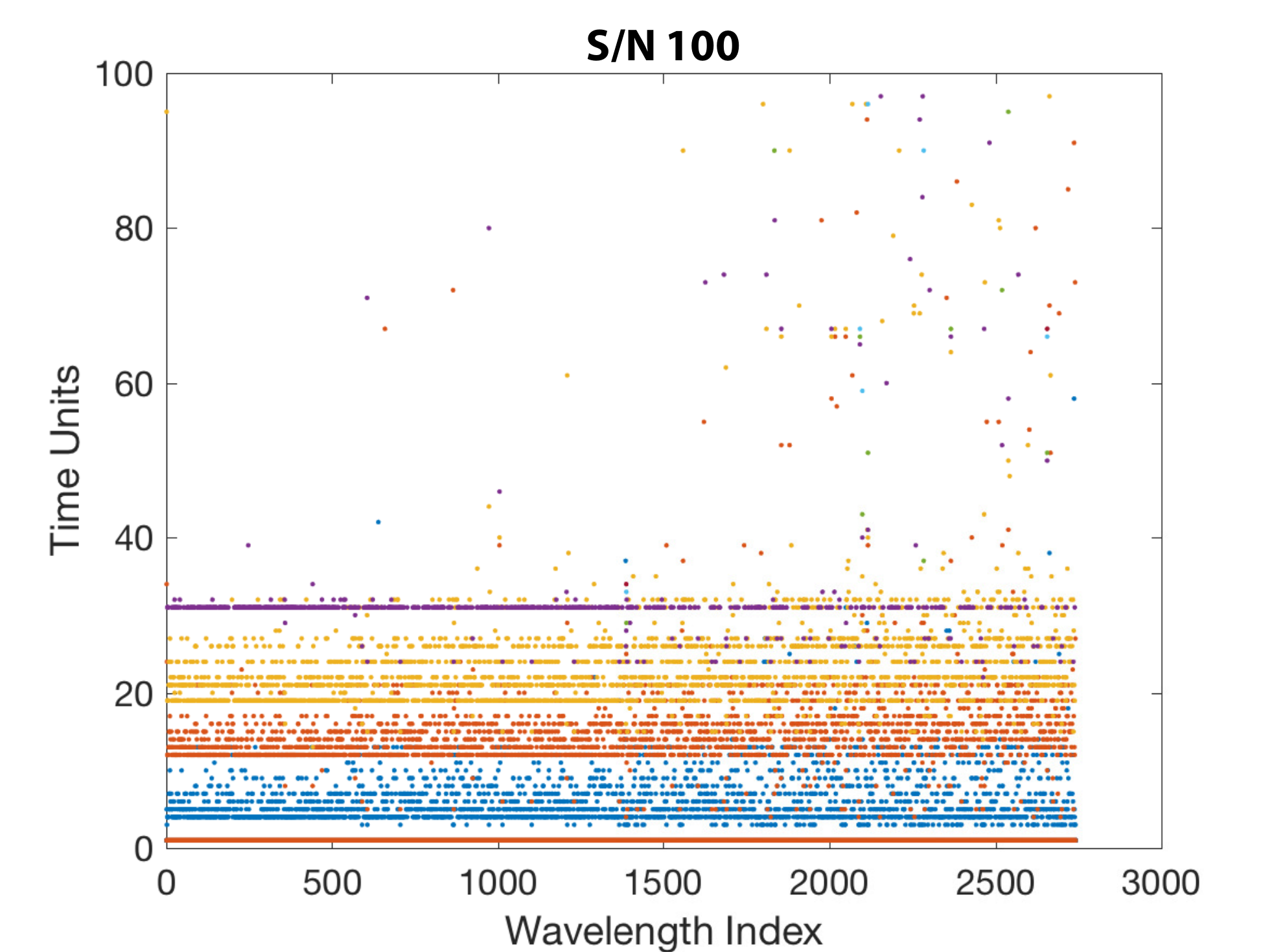}	
    \end{subfigure}%
    ~ 
    \begin{subfigure}[]
        \centering
        \includegraphics[trim = 15 0 20 0, clip, width = 0.32\textwidth]{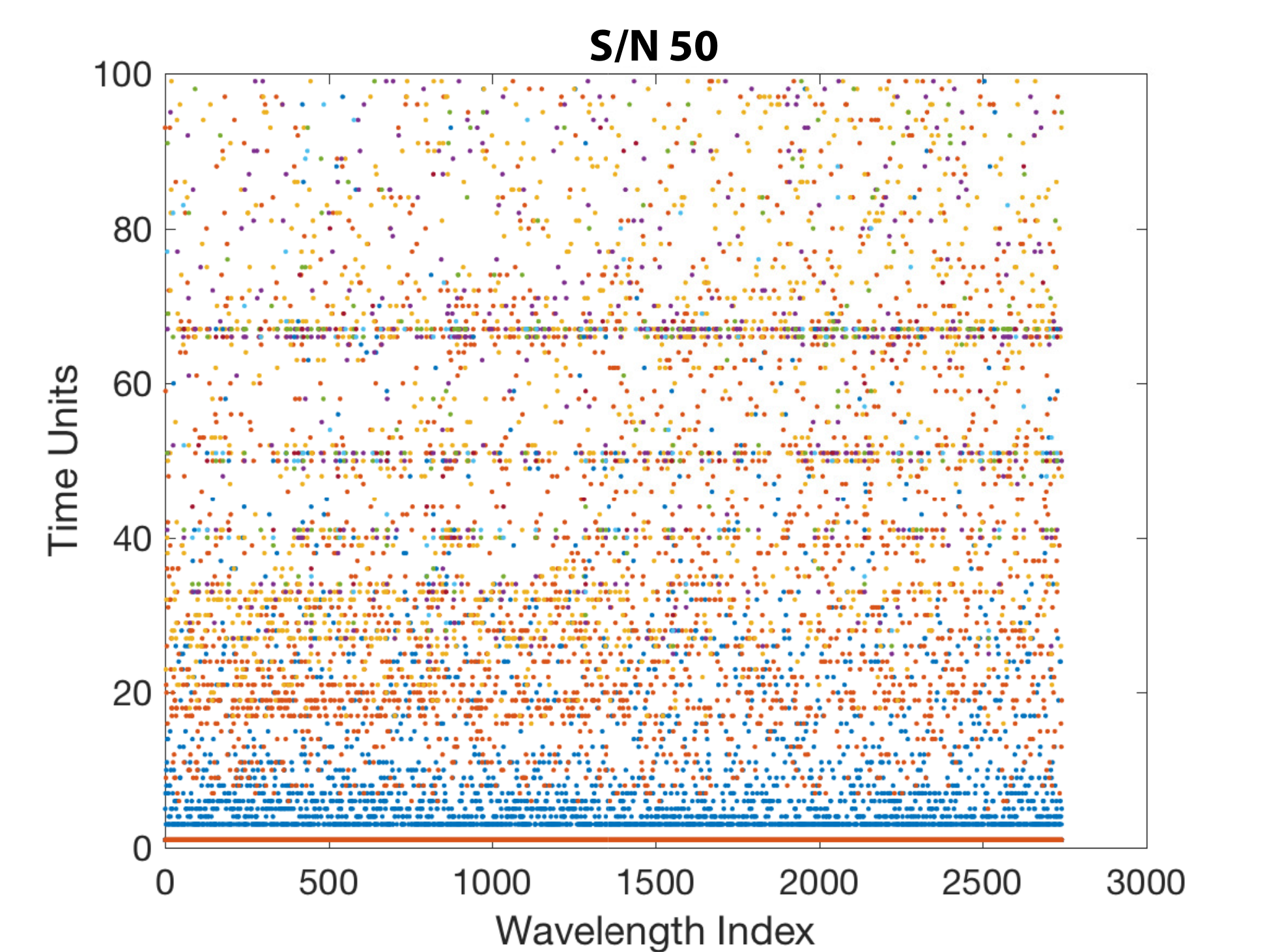}	
            \end{subfigure}
    \\
        \begin{subfigure}[]
        \centering
        \includegraphics[trim = 15 0 20 0, clip, width = 0.32\textwidth]{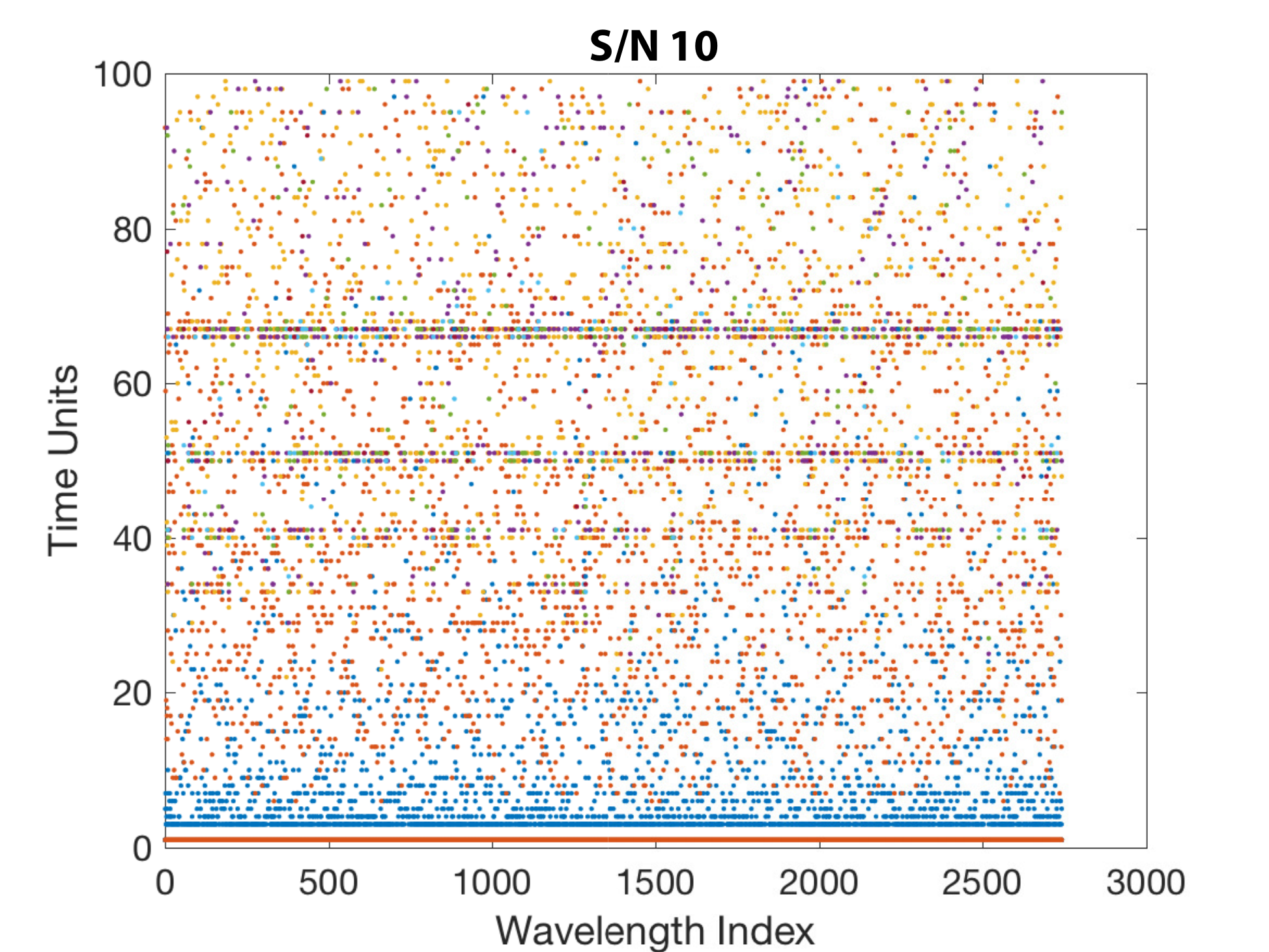}	
    \end{subfigure}%
    ~ 
    \begin{subfigure}[]
        \centering
        \includegraphics[trim = 15 0 20 0, clip, width = 0.32\textwidth]{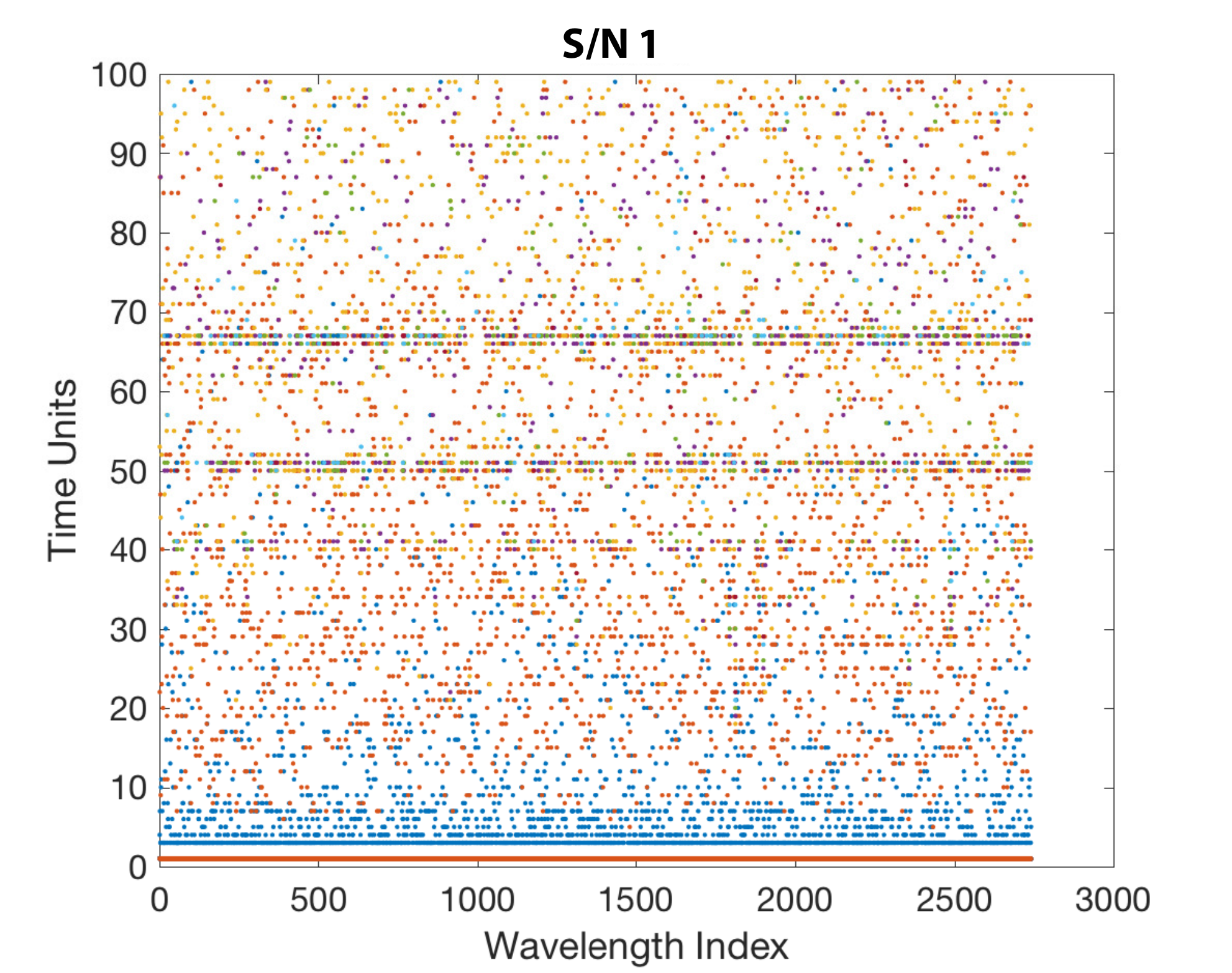}	
    \end{subfigure}
\caption{The crossover times plotted for all wavelengths, with $C_{th} = 0.01$, for the simulated SOAP 2.0 spectra in presence of an orbiting planet. We show results for different S/N. (a) No Noise, (b) S/N = 1000, (c) S/N = 500, (d) S/N = 250, (e) S/N = 200, (f) S/N = 150, (g) S/N = 100, (h) S/N = 50, (i) S/N = 10, (j) S/N = 1.
\label{fig:SOAP_e2_p2}}
\end{figure*}

Thus far, we have focused our analysis on data observed during a primary or secondary eclipse of an exoplanet. However, these measurements are not always available since only a small percentage of exoplanet transits occur between the observer and the host star. Instead, all the exoplanets produce an effect on the radial velocity of the host star. 

\begin{figure}[htbp!]
    \centering
    \includegraphics[trim = 0 0 20 10, clip, width = 0.5\textwidth]{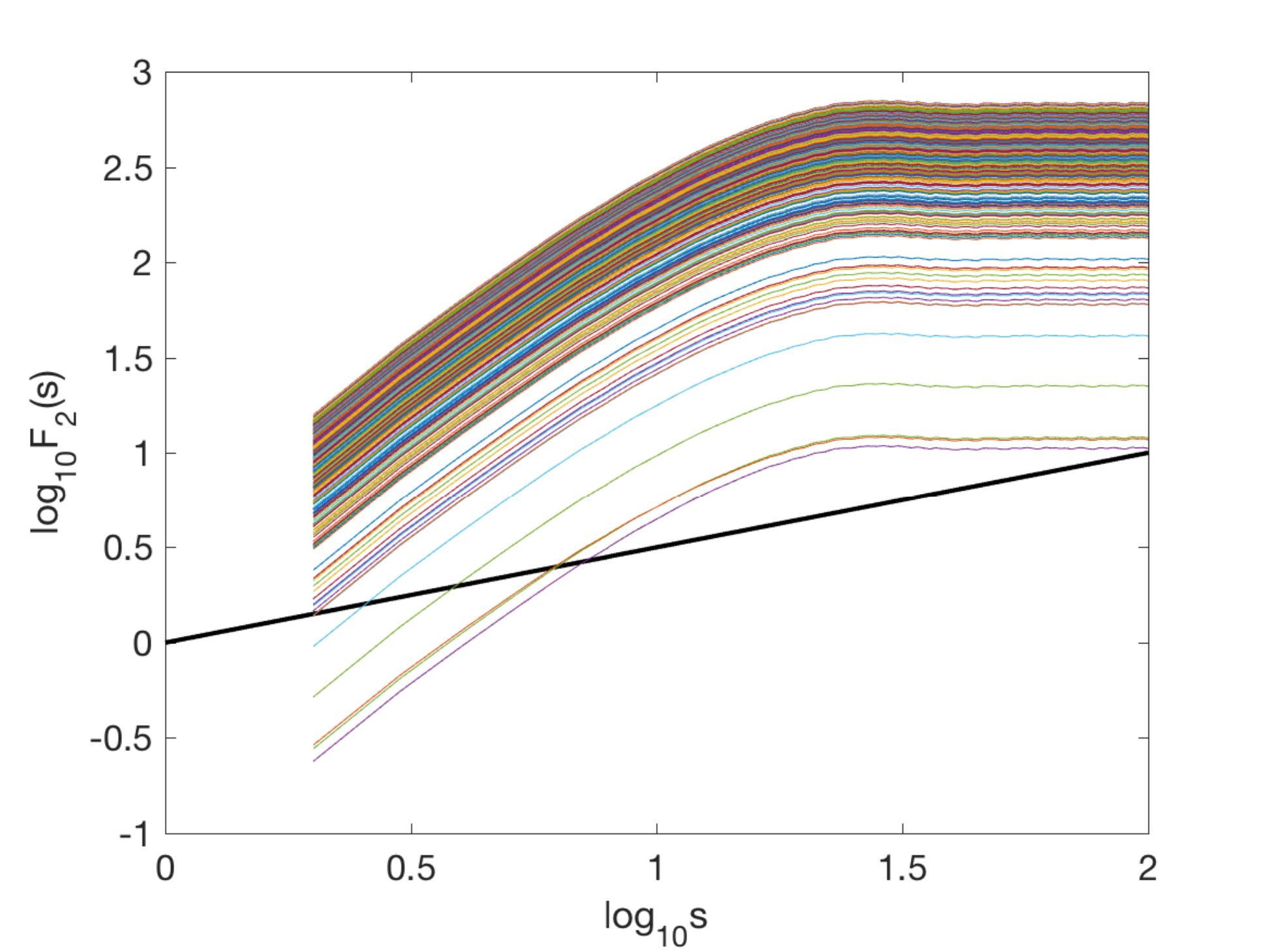}	  	
    \caption{The second moment of the fluctuation functions for all wavelengths for the simulated  SOAP 2.0 spectra with 5\% of the stellar disk covered with a spot, without noise in the spectra. The straight black line has a slope of 0.5, which denotes white noise dynamics.}
    \label{fig:Soap_MF_spot}
\end{figure}

Current technology only allows detection of the red and blue shifts of the spectrum when the motion of the star is sufficiently large, and although modern high-resolution spectrographs can attain a resolution of the order of 1 m s$^{-1}$, detections of a small planets often remain controversial.
Moreover, these data are contaminated by various sources of noise, such as instrumental noise, atmospheric noise, stellar noise, and telluric absorption lines.   Presently, the observations are fit to a radial velocity curve, which may or may not be robust and have a non-negligible false detection rate  \cite[see sect 4.3.2. of][]{Fischer:2016aa}.

\begin{figure*}[htbp]
    \centering
    \begin{subfigure}[]
        \centering
        \includegraphics[trim = 15 0 20 0, clip, width = 0.30\textwidth]{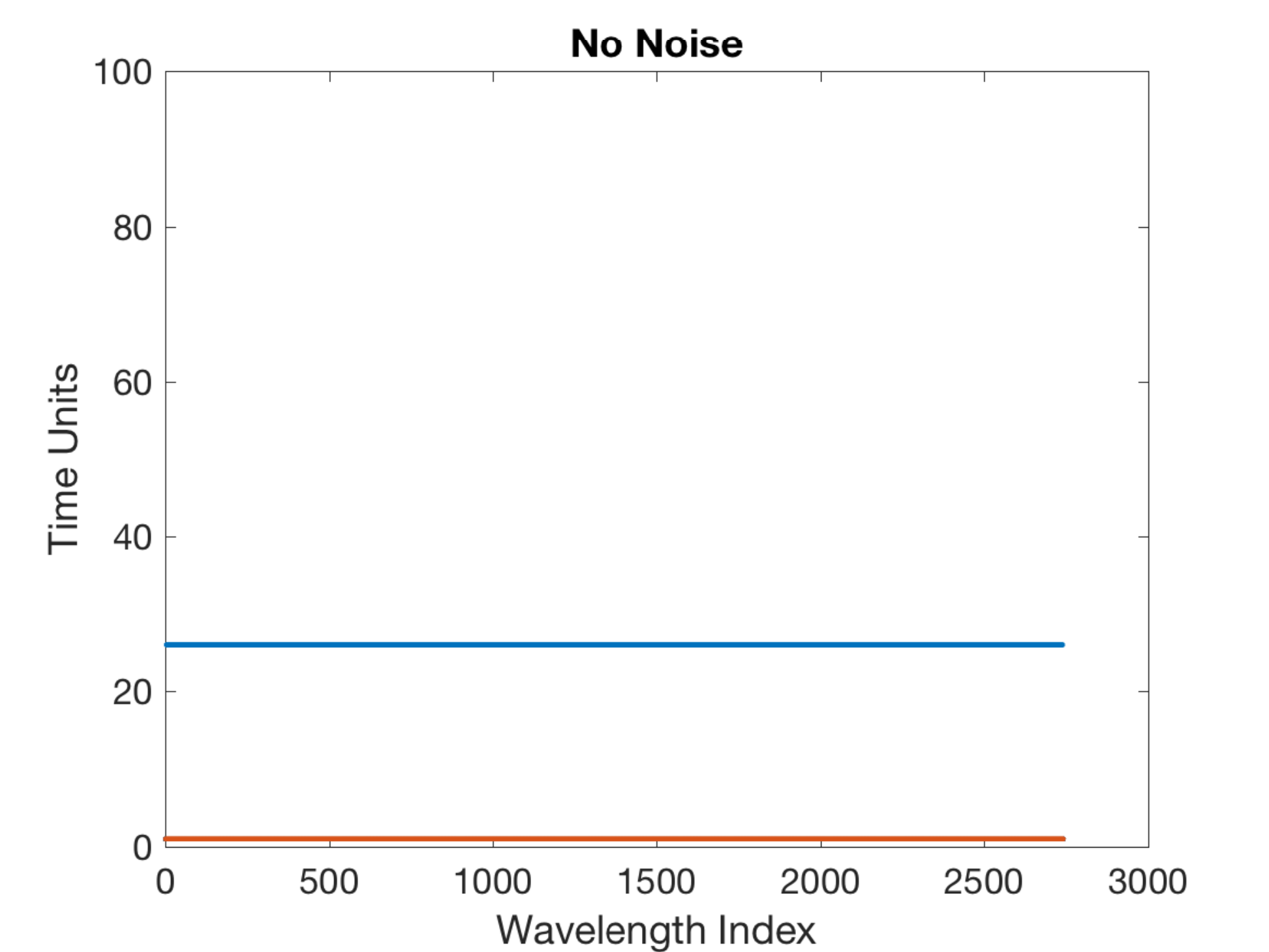}	
    \end{subfigure}%
    ~ 
    \begin{subfigure}[]
        \centering
        \includegraphics[trim = 15 0 20 0, clip, width = 0.30\textwidth]{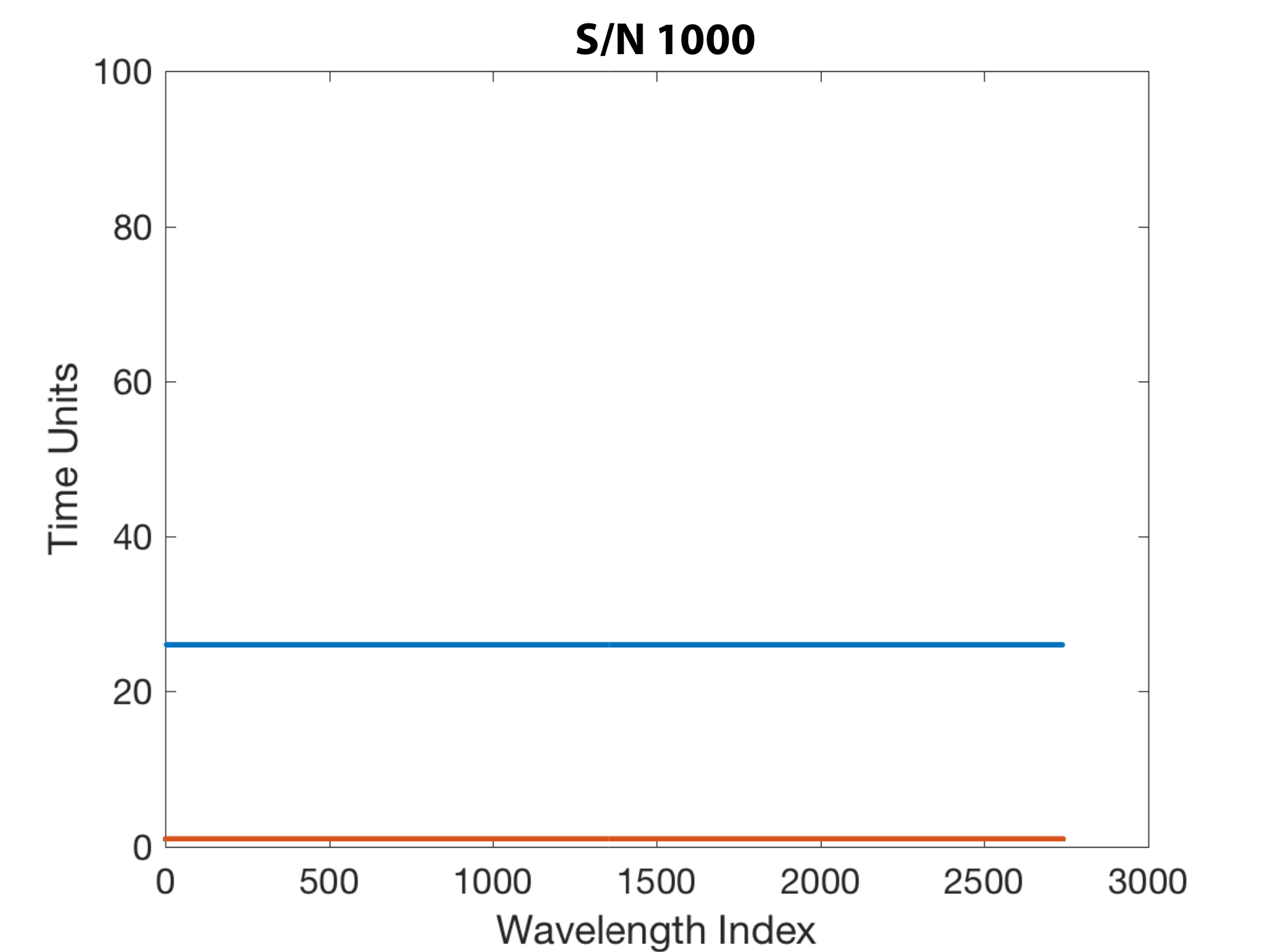}	
            \end{subfigure}
    ~
        \begin{subfigure}[]
        \centering
        \includegraphics[trim = 15 0 20 0, clip, width = 0.30\textwidth]{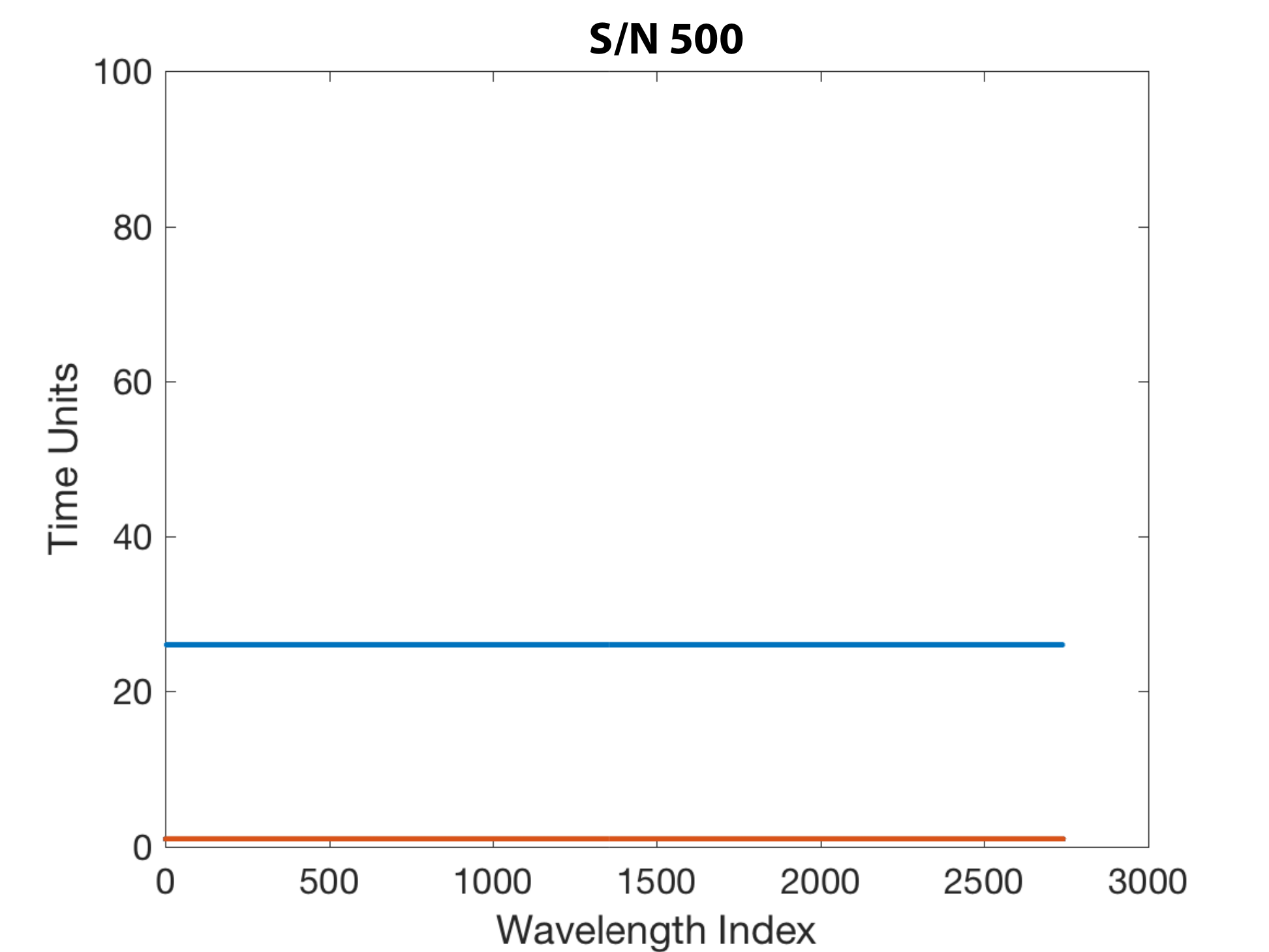}	
    \end{subfigure}%
    \\ 
    \begin{subfigure}[]
        \centering
        \includegraphics[trim = 15 0 20 0, clip, width = 0.30\textwidth]{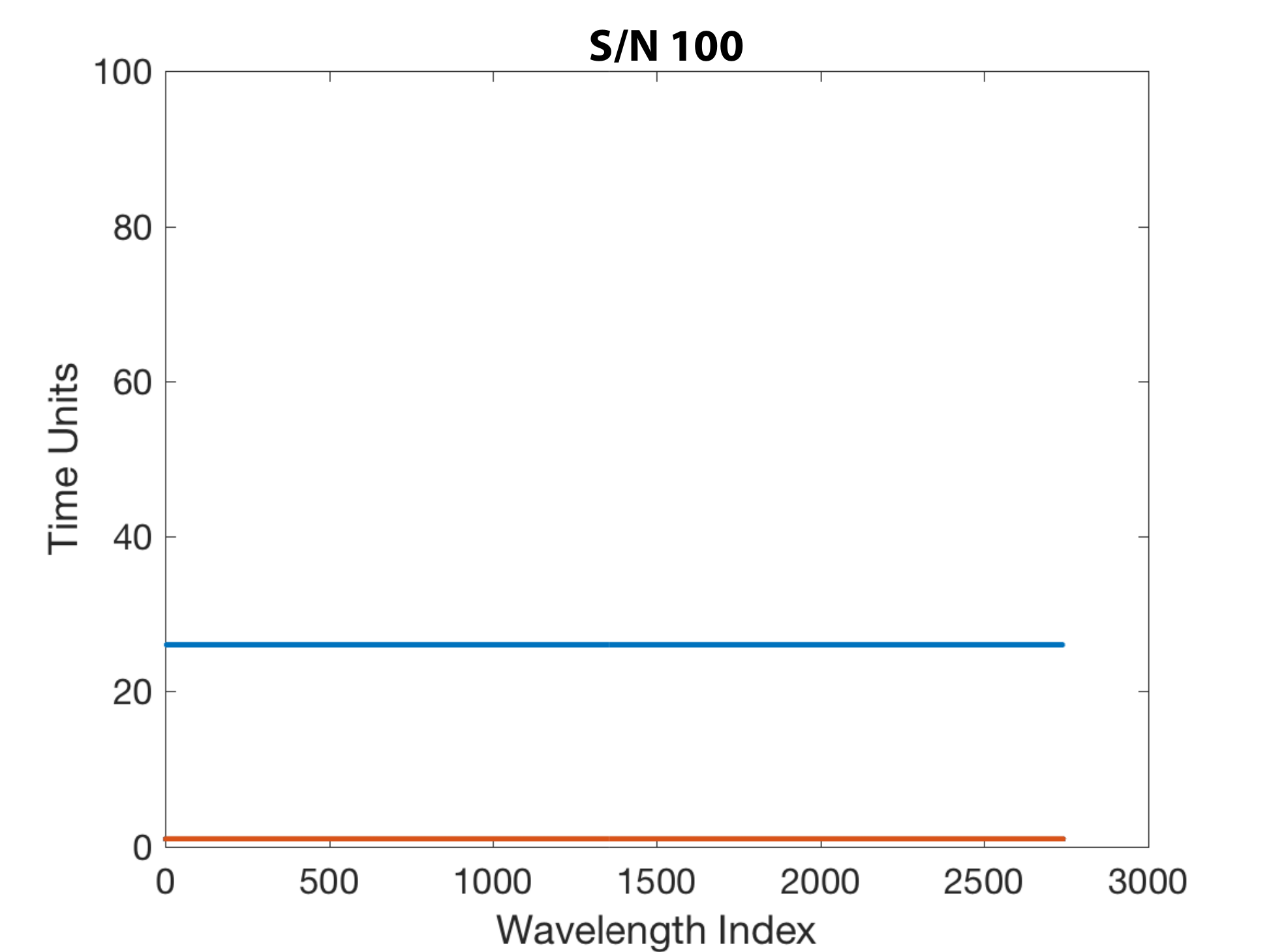}	
    \end{subfigure}
    ~
        \begin{subfigure}[]
        \centering
        \includegraphics[trim = 15 0 20 0, clip, width = 0.30\textwidth]{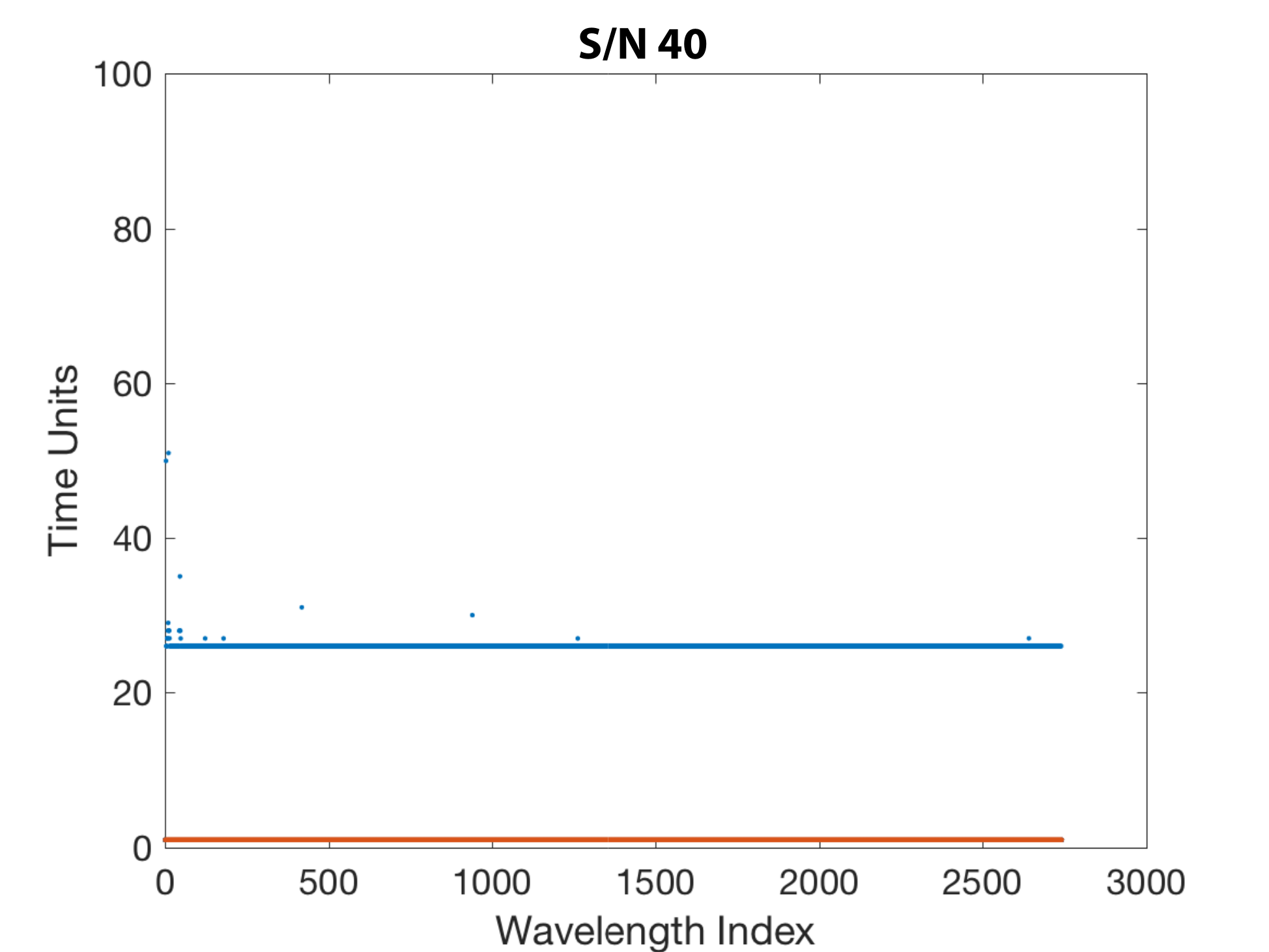}	
    \end{subfigure}%
    ~ 
    \begin{subfigure}[]
        \centering
        \includegraphics[trim = 15 0 20 0, clip, width = 0.30\textwidth]{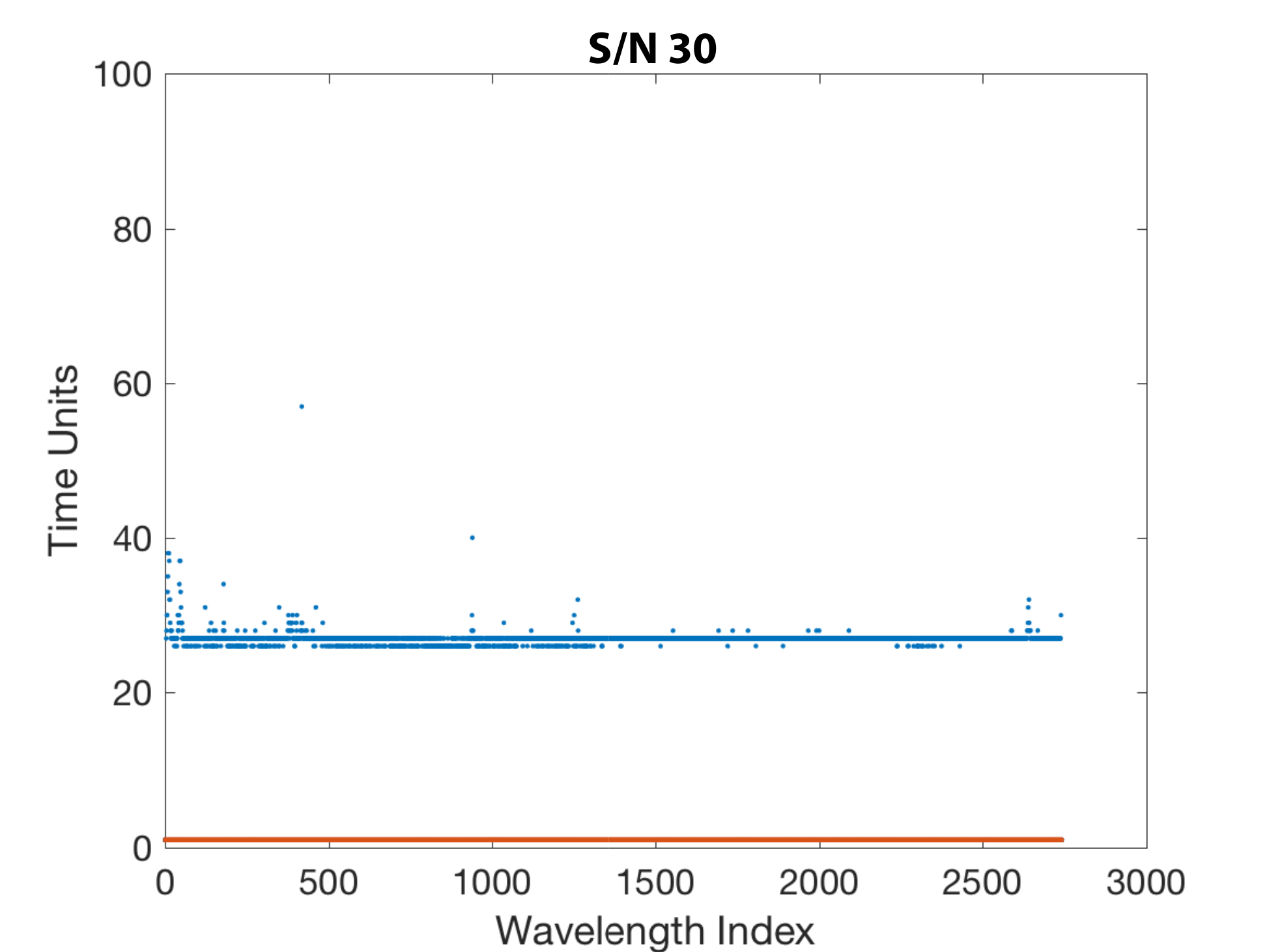}	
    \end{subfigure}
        \\ 
    \begin{subfigure}[]
        \centering
        \includegraphics[trim = 15 0 20 0, clip, width = 0.30\textwidth]{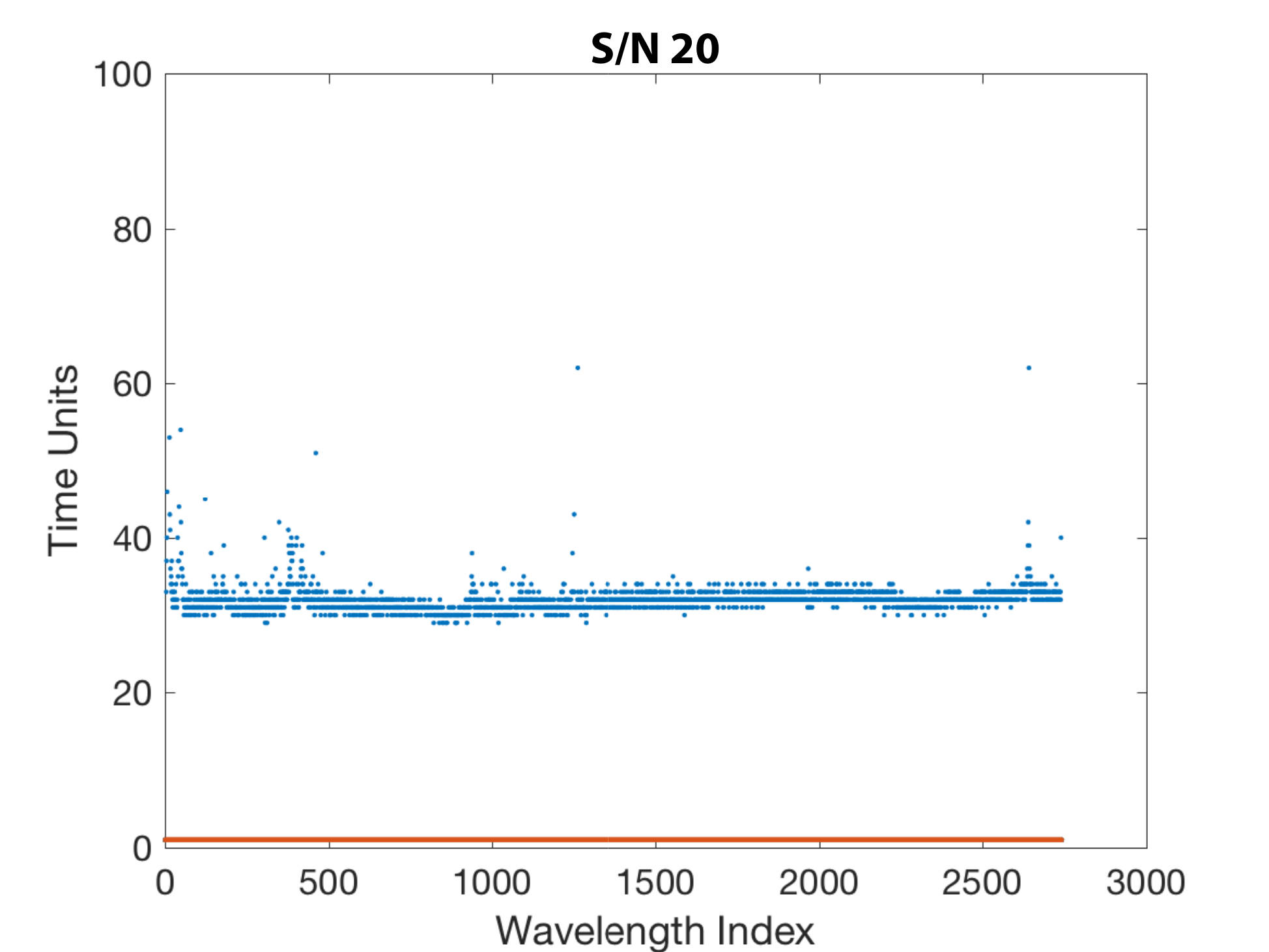}	
    \end{subfigure}
    ~
        \begin{subfigure}[]
        \centering
        \includegraphics[trim = 15 0 20 0, clip, width = 0.30\textwidth]{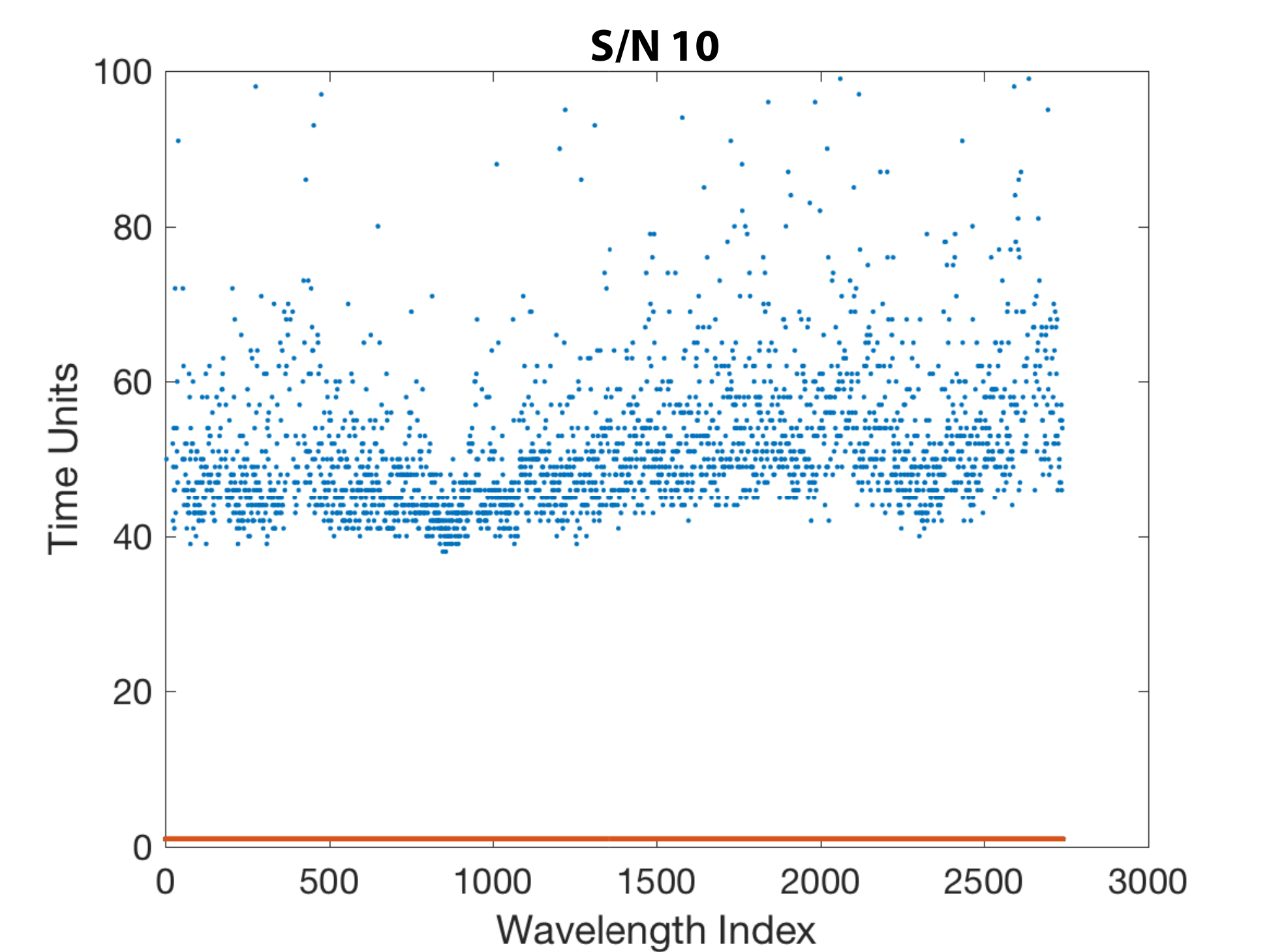}	
    \end{subfigure}%
    ~ 
    \begin{subfigure}[]
        \centering
        \includegraphics[trim = 15 0 20 0, clip, width = 0.30\textwidth]{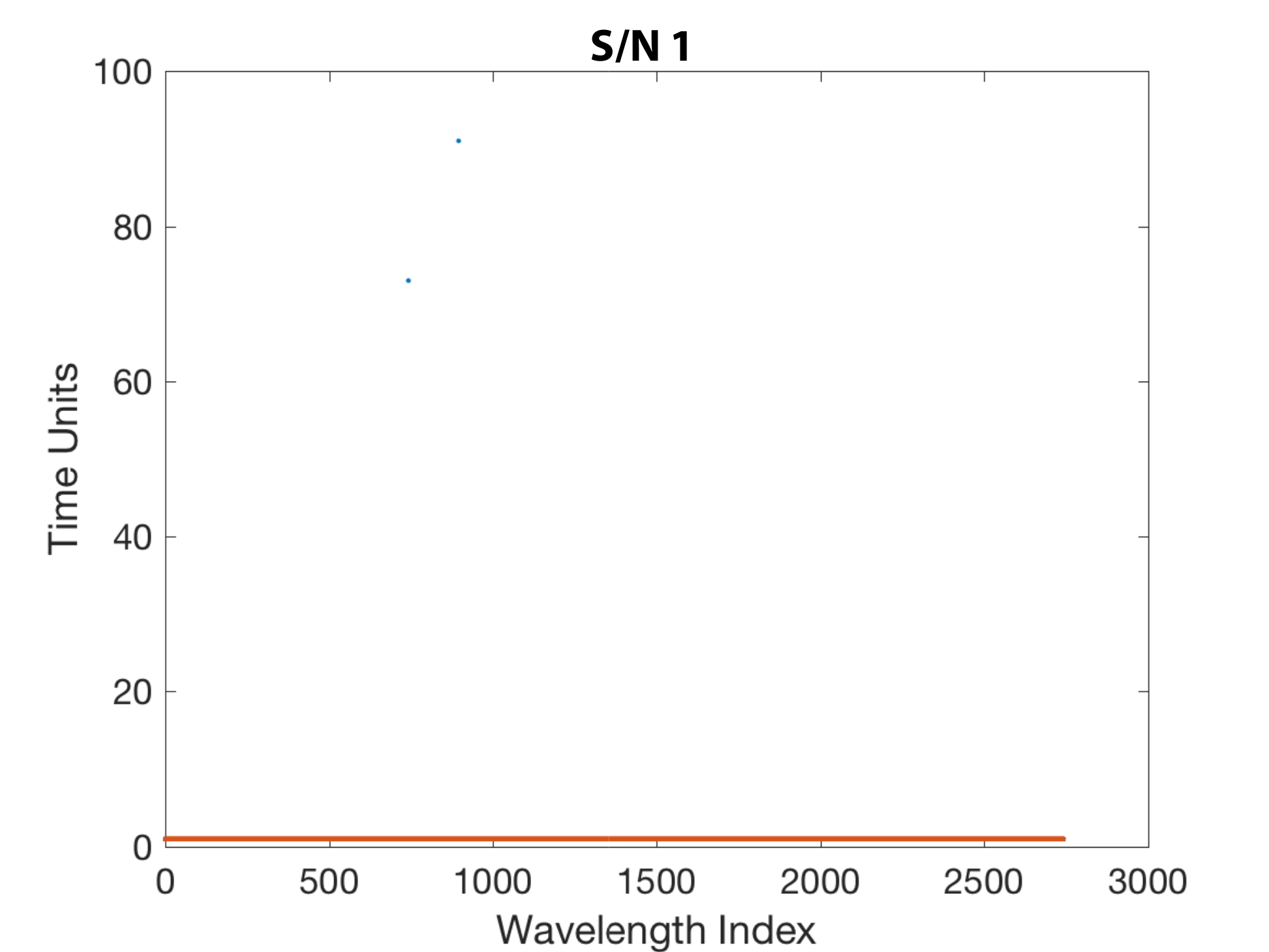}	
    \end{subfigure}
\caption{The crossover times plotted for all wavelengths, with $C_{th} = 0.08$, for the simulated SOAP 2.0 spectra with 5\% of the stellar disk covered with a spot, for different S/N. (a) No Noise, (b) S/N = 1000, (c) S/N = 500, (d) S/N = 100, (e) S/N = 40, (f) S/N = 30, (g) S/N = 20, (h) S/N = 10, (i) S/N = 1.
\label{fig:SOAP_s}}
\end{figure*}

\begin{figure*}[htbp]
    \centering
    \begin{subfigure}[]
        \centering
        \includegraphics[trim = 15 0 20 0, clip, width = 0.30\textwidth]{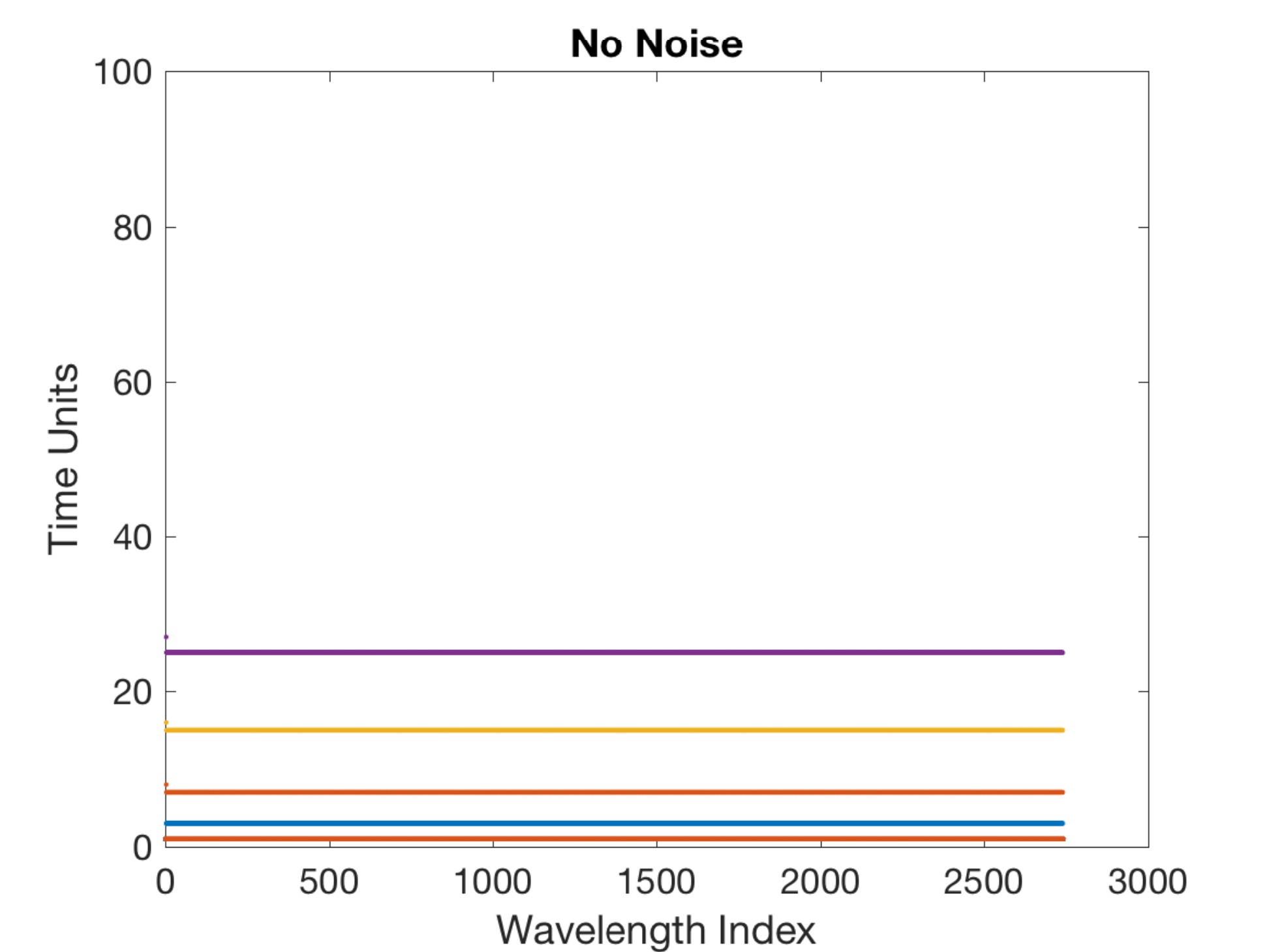}	
    \end{subfigure}%
    ~ 
    \begin{subfigure}[]
        \centering
        \includegraphics[trim = 15 0 20 0, clip, width = 0.30\textwidth]{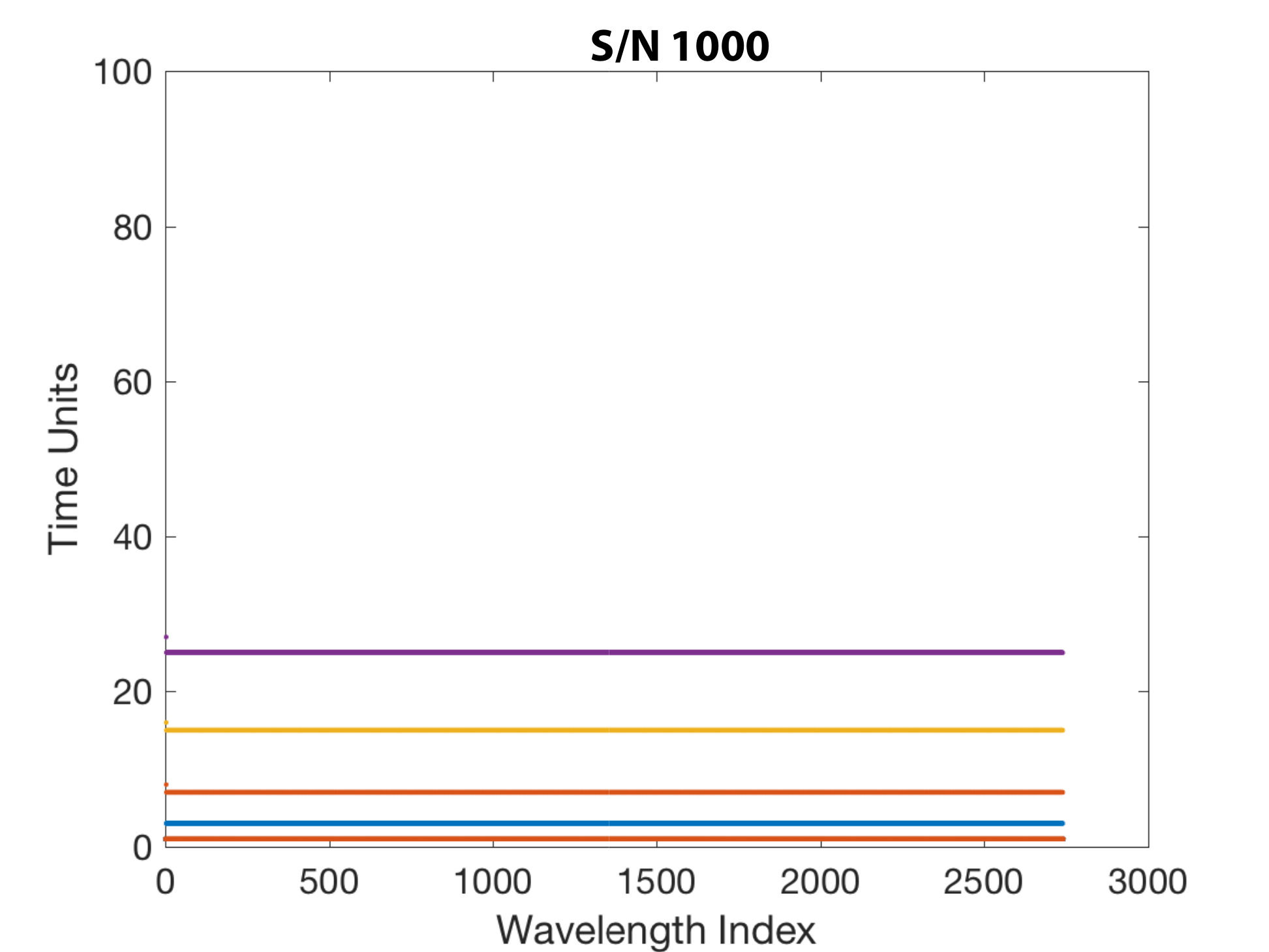}	
    \end{subfigure}
    ~
        \begin{subfigure}[]
        \centering
        \includegraphics[trim = 15 0 20 0, clip, width = 0.30\textwidth]{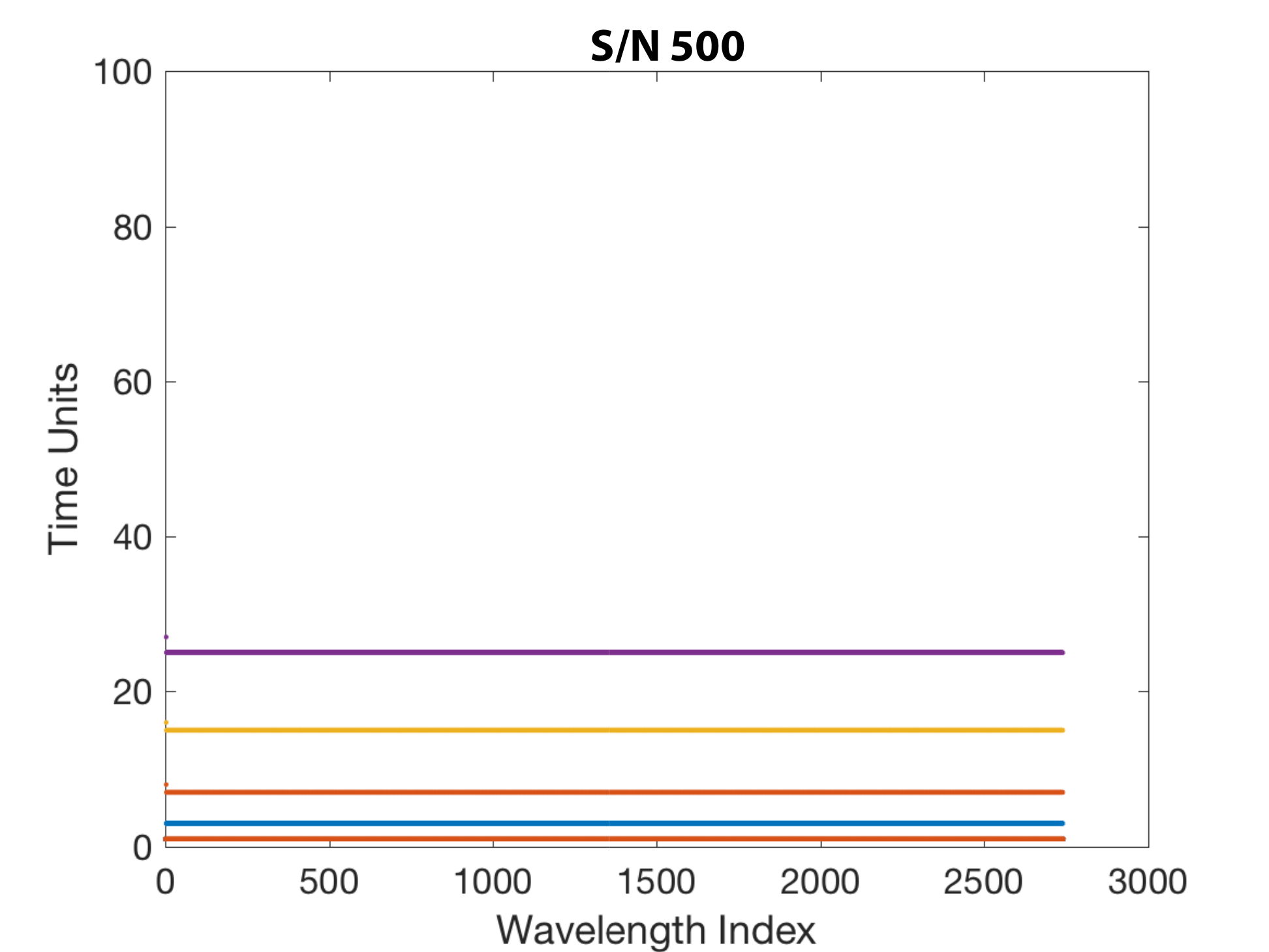}	
    \end{subfigure}%
    \\ 
    \begin{subfigure}[]
        \centering
        \includegraphics[trim = 15 0 20 0, clip, width = 0.30\textwidth]{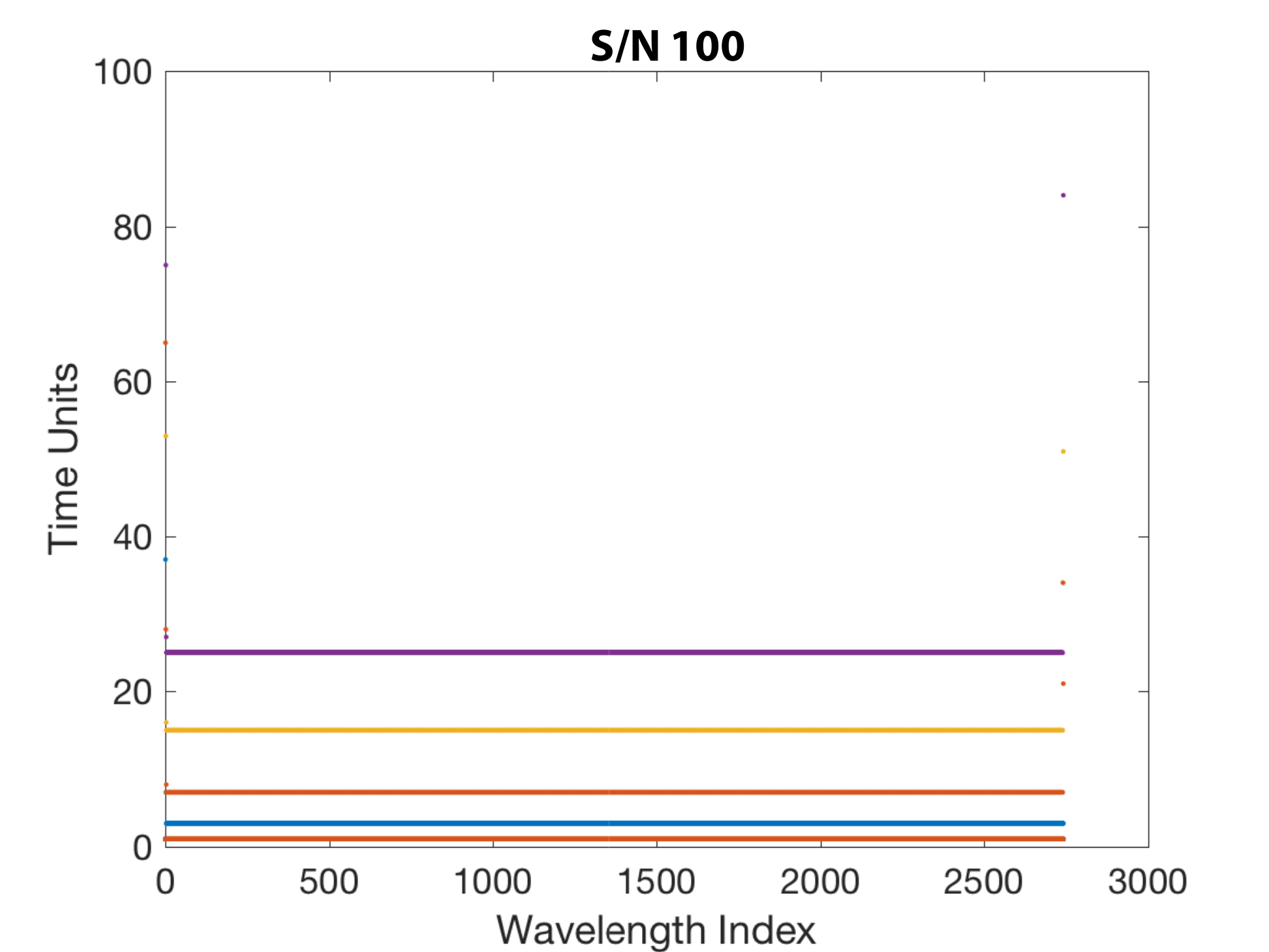}	
    \end{subfigure}
    ~
        \begin{subfigure}[]
        \centering
        \includegraphics[trim = 15 0 20 0, clip, width = 0.30\textwidth]{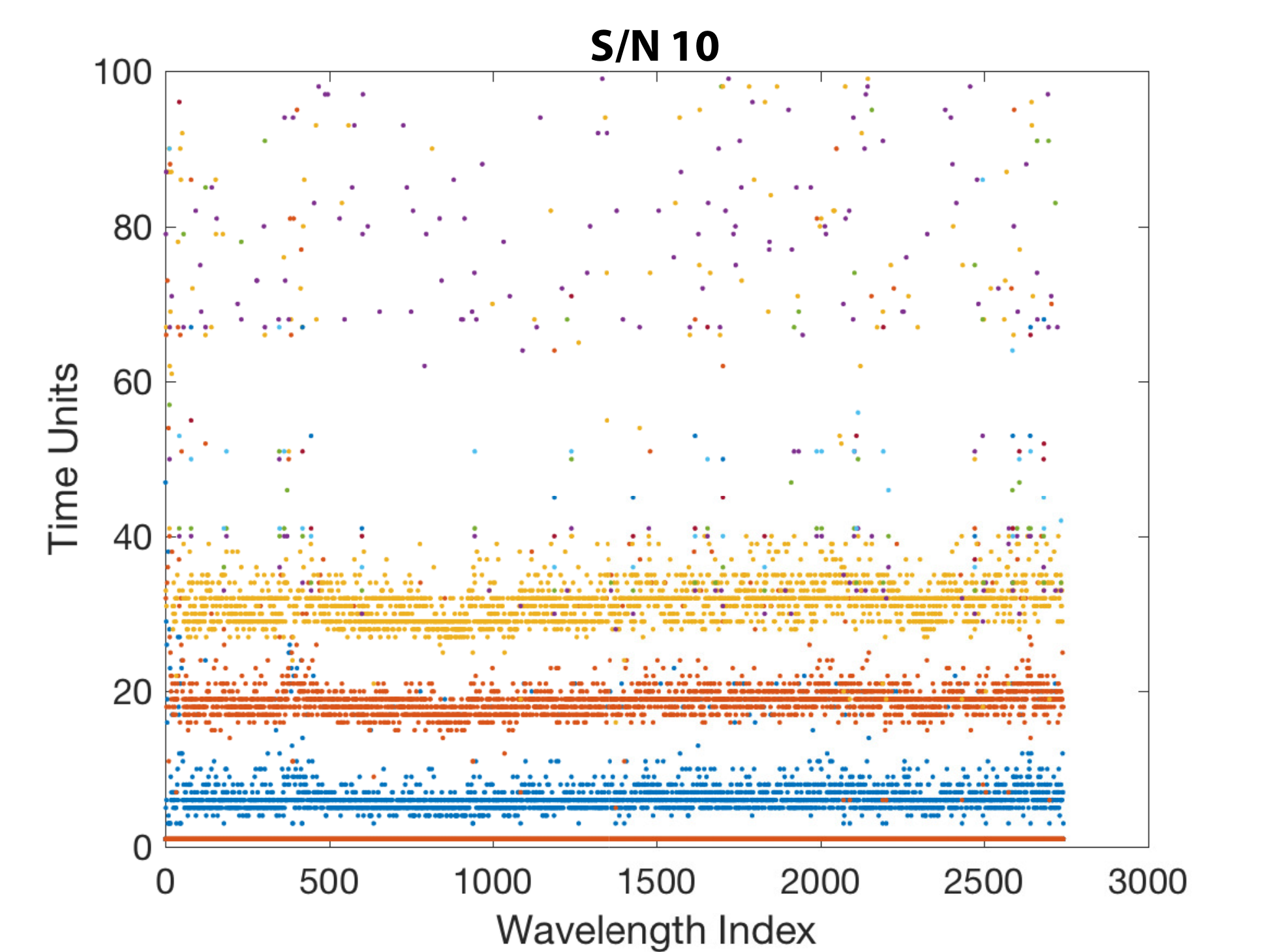}	
    \end{subfigure}%
    ~ 
    \begin{subfigure}[]
        \centering
        \includegraphics[trim = 15 0 20 0, clip, width = 0.30\textwidth]{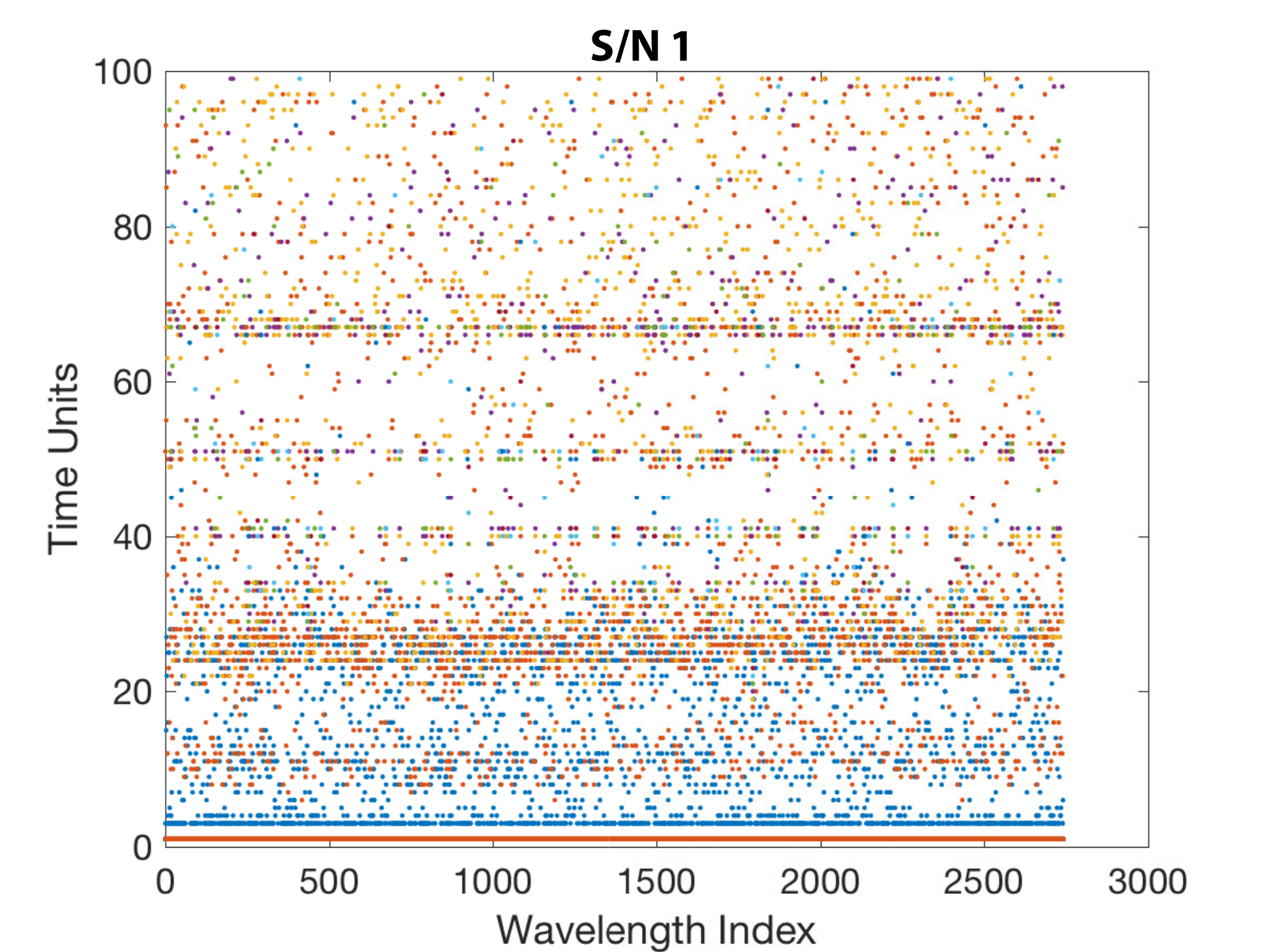}	
    \end{subfigure}
\caption{The crossover times plotted for all wavelengths, with $C_{th} = 0.01$, for the simulated SOAP 2.0 spectra with 5\% of the stellar disk covered with the spot, for different S/N. (a) No Noise, (b) S/N = 1000, (c) S/N = 500, (d) S/N = 100, (e) S/N = 10, (f) S/N = 1.
\label{fig:SOAP__e2_s}}
\end{figure*}

To address this problem, we test the detectability of time scales related to the radial velocity effect with our method. We start from the analysis of a well defined case using data from the SOAP 2.0 simulations.  This allows us to study the effect of noise as well as the robust estimation of crossover times, which we know to be present in the data set.

The original ("noise-free") data describe the shifts in the spectrum of the star with a radial velocity of amplitude 40 ms$^{-1}$, due to a planet orbiting around it. We add Gaussian white noise to the spectra at each time to get a specific S/N, and thereby obtain 12 such data sets for S/N of 1, 10, 50, 75, 80, 100, 150, 200, 250, 500, and 1000. Figure \ref{fig:Soap_MF_planet} shows the second moment of the fluctuation functions for all the wavelengths in the reduced spectra without noise, and subfigures in Figure \ref{fig:SOAP_p2} show the crossover time scales extracted for the different noise cases. 
As explained in Sec. \ref{Sec:method}, we evaluate these crossovers using a threshold value $C_{th}$ for the slope change.  The plots in Figure \ref{fig:SOAP_p2} use a threshold value $C_{th} = 0.08$.  Whence, we are able to extract the exact timescale for the orbital period of the exoplanet. Even for a $S/N = 150$, the methodology is efficient and robust against noise, and we extract the correct timescale. As the signal quality is degraded further, we see the noise starting to affect the calculated crossovers with scatter above the actual value.  Finally, as we further degrade  the signal quality the crossover times disappear altogether. 
As the noise begins to dominate, the signal becomes white. We know this quantitatively from our analysis, as fluctuation functions with a constant slope of 0.5 are, by definition, white noise processes (e.g. Figure \ref{fig:N1_complete}).

To examine how to capture timescales when the signal quality is poor, we decrease the threshold to $C_{th} = 0.01$ (Figure \ref{fig:SOAP_e2_p2}). Even this threshold is able to approximately capture (31 time units) the orbital period, but it also detects other multiple time scales, which may be spurious. The important point here is the ability to study how noise affects these timescales; as the S/N is decreased time scales remain robust, but become harmonics of the robust scales for a further increase in the S/N.  These may be due to a form of stochastic resonance between the threshold, radial velocity measurements and noise.  The value of threshold is thus crucial.  

In actual data, one cannot calculate this threshold for all the wavelengths separately and for each night, and hence one value is chosen in accordance with the observed noise characteristics. Presently, use of periodograms and fitting to sine curves is performed to model the radial velocity curves from the observations. The above analysis shows how noise can lead to spurious estimation of orbital periods, and thus potentially result in spurious detection of exoplanets.

In the same manner as above, we have analyzed the case in which no planet is orbiting around the star, but a spot covers 5\% of its surface. In a similar fashion we have analyzed 8 rotation periods of the star and reconstructed its rotational period (Figure \ref{fig:Soap_MF_spot} -- \ref{fig:SOAP__e2_s}). 
The use of this method to distinguish stellar features from planets will be subject of future work.

\section{Conclusion}\label{sec:conc}

We have presented and tested a new multi-fractal approach for the analysis of exo-planetary spectral observations.  The goal is to use a fit-free procedure to identify robust time scales associated with the exo-planetary orbital motion around a host star, as well as to detect time scales associated with stellar features.
With these timescales in hand, one can compute key system parameters such as the ratio of the size of the planet to that of the star and the latitude of transit, {\em without} use of stellar evolution models, data fitting, noise filtering, and the additional wide variety of other assumptions about the system that are typically made.  The concept of the approach is to take an agnostic (or model free) view of the observed spectral structure.  The method makes no {\em a priori} assumptions about the temporal structure in any observed spectra and makes use of only one number, the generalized Hurst exponent, the value of which underlies the identification of the key time scales. We have reconstructed the primary and secondary transit times of the exoplanet HD 189733b, using data from both ground-based (HARPS spectrograph) and space-based (\emph{Spitzer}) observations.

Using the SOAP 2.0 tool, which simulates a stellar spectrum and how it can be affected by the presence of a planet across one orbital period, our method is further tested in the context of measuring planetary orbital motion via Doppler shift detection.  Because the SOAP tool can also simulate the presence of a spot on the stellar surface across one rotational period of the star, we have demonstrated that our approach  reconstructs the planetary orbital period, as well as the rotation period of a spot covering 5\% of the stellar surface. Moreover, we have tested the analysis with a wide range of S/Ns.  We can reconstruct: (a) the period of a planet producing a 40 m s$^{-1}$ Doppler shift of the stellar spectrum, and  (b) the rotational period of the star, based on the presence of a spot on its surface, provided that the S/Ns are $\ge 75$ and $\ge 30$ respectively.  Importantly, to avoid introduction of errors arising from intrinsic irregularities in the analyzed time series, this method has the highest fidelity when observations 
are carried out at sufficiently high frequency that the time-difference between each measurement is less than the shortest relevant timescale that may be present in the system.

In conclusion this method based on Multi-fractal Temporally Weighted Detrended Fluctuation Analysis of time series is a robust way to measure planetary orbital motion. It provides a fertile framework to examine  
other data sets and to explore trying to systematically distinguish stellar noise from planetary motion.  
\acknowledgments

The authors thank the referee for comments that have helped to improve our paper, Xavier Dumusque for providing the SOAP 2.0 data used for this work and Debra Fischer and Allen Davis for discussions on this and other detection schemes.  
S.A. and J.S.W. acknowledge NASA Grant NNH13ZDA001N-CRYO for support.  F.D.S. acknowledges the Swedish Research Council International Postdoc fellowship for support.  J.S.W. acknowledges Swedish Research Council grant no. 638-2013-9243 and a Royal Society Wolfson Research Merit Award for support.  

%

\end{document}